\documentclass[a4paper,11pt]{article}
\pdfoutput=1 

\usepackage{jheppub} 
\usepackage{slashed}
\usepackage[normalem]{ulem}
\usepackage{braket}
\usepackage[T1]{fontenc} 
\usepackage{amsmath, amssymb}
\usepackage{xcolor}
\usepackage{tikz}
\usepackage{enumitem}
\usetikzlibrary{shapes.geometric, arrows.meta, positioning, calc}

\title{\boldmath BMN Spread Complexity Across Phase Transitions}

\author[a]{Dibakar Roychowdhury}

\affiliation[a]{Department of Physics, Indian Institute of Technology Roorkee,\\Roorkee 247667, Uttarakhand, India}

\abstract{
We investigate the dynamical behavior and spread complexity of quantum states within the mass-deformed BMN matrix model equipped with an emergent $U(1)$ global charge at finite temperature $\beta^{-1}$ and chemical potential $\nu$. Focusing on the strong-coupling regime ($\mu \gg 1$), we model the dynamics via a charged Thermofield Double (cTFD) state and trace the evolution of spread complexity across three distinct thermodynamic regimes. In the low-temperature gapped phase ($\beta \mu \gg 1$), discrete mass-gap bound states dominate, trapping wavepacket dispersion and producing non-chaotic, oscillatory early-time growth. Conversely, in the high-temperature continuum phase ($\beta \mu \ll 1$), thermal excitations overwhelm the mass gap, driving a transition to a continuous advection field that exhibits maximal chaotic scrambling with a Krylov Lyapunov exponent $\lambda_K = \pi / \beta$ that saturates the universal bound. In the intermediate temperature regime ($\beta \mu \sim 1, \beta \nu \sim 1$), the interplay between mass-gap bound states and the continuous thermal background induces a sub-leading correction to the Lanczos coefficients $b_n \sim \frac{\pi}{\beta} n + \gamma \sqrt{n}$, governing a continuous sub-exponential crossover before full chaotic thermalization. Technical derivations regarding KMS boundary conditions, grand canonical spectral moments, residue analysis, and time-reversal symmetry breaking in Krylov space are detailed in four dedicated appendices.
}
\begin{document} 
\maketitle
\flushbottom
\section{Introduction and General Idea}
Krylov and/or the spread complexity of states and operators \cite{Parker:2018yvk}-\cite{Caputa:2025ozd} and its various applications including chaos \cite{Hashimoto:2023swv}-\cite{Bhattacharjee:2024yxj}, holography \cite{Caputa:2024sux}-\cite{Erdmenger:2022lov} and black holes \cite{Kar:2021nbm} have garnered renewed attention in the recent years due to their several remarkable features\footnote{For a nice and comprehensive set of reviews on the subject see \cite{Baiguera:2025dkc}-\cite{Nandy:2024evd}.}. The purpose of this article is to extend the notion of spread complexity of states \cite{Parker:2018yvk}-\cite{Balasubramanian:2022tpr} in the context of BMN matrix models \cite{Berenstein:2002jq}-\cite{Roychowdhury:2026mpd} at finite chemical potential $\nu$ and temperature $\beta^{-1}$ \cite{Costa:2014wya}. In what follows, we consider the BMN model derived at a large mass $\mu\gg 1$ deformation from its parent BFSS matrix model \cite{Banks:1996vh} and therefore is non-conformal (or gapped).  

We argue that in the strong coupling limit $\mu \gg 1$, the BMN matrix model has an emergent $U(1)$ global symmetry at LO in the mass deformation $\mu$. The global $U(1)$ symmetry corresponds to a complex oscillator model whose energy levels are largely spaced (in units of $\mu$) at low temperatures. We use this symmetry to construct a charged thermo-field double (cTFD) \cite{Caputa:2013eka}-\cite{Chapman:2019clq} and compute the corresponding return amplitude and the spread complexity of states. The charged TFD (cTFD) is parametrized by the mass parameter $\mu \gg 1$ and a fixed chemical potential $\nu$. For a large mass gap, and fixed chemical potential, the BMN Hamiltonian can be \emph{approximated} at leading order as a pair of harmonic oscillators that can be combined to form a complex oscillator model \cite{Caputa:2025ozd}. 

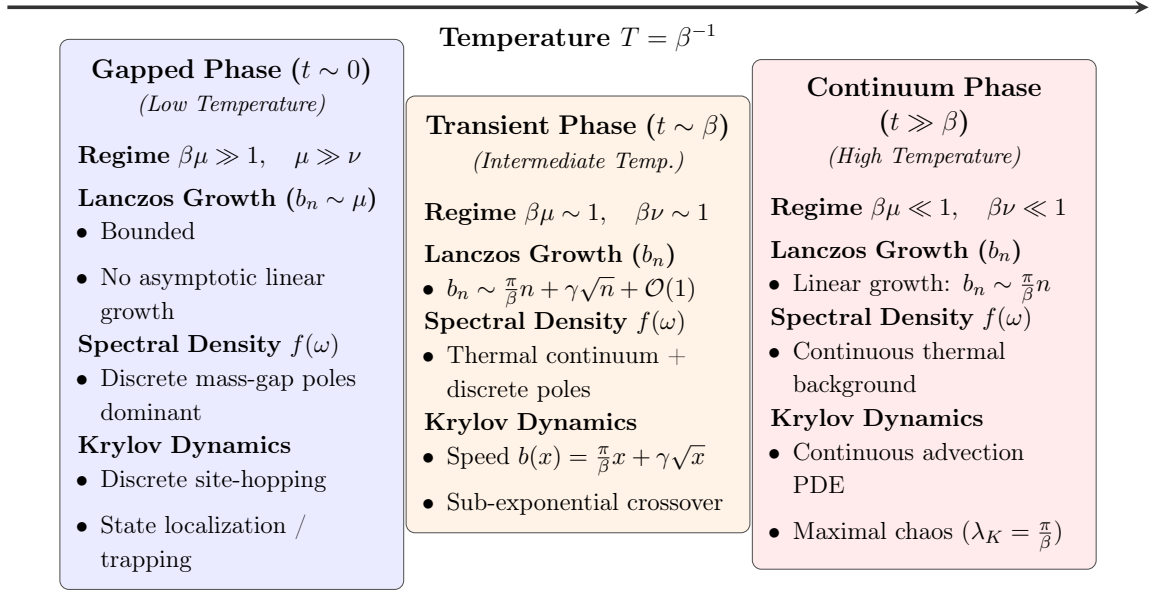
\begin{figure}[htbp]
\centering
\resizebox{\linewidth}{!}{%
\begin{tikzpicture}[
    box/.style={
        draw=black!70, 
        rounded corners=4pt, 
        text width=4.8cm, 
        inner sep=8pt, 
        minimum height=6.2cm, 
        align=left
    }
]

\draw[->, >=stealth, ultra thick, color=black!80] (-3.5, 4.8) -- (14.3, 4.8);
\node[above=5pt, font=\large\bfseries] at (5.4, 3.8) {Temperature $T = \beta^{-1}$};

\node[box, fill=blue!8] (box1) at (0,0) {
    \centering \textbf{\large Gapped Phase ($t \sim 0$)} \\
    \textit{\small (Low Temperature)} \\[8pt]
    \raggedright
    \textbf{Regime} $\beta\mu \gg 1, \quad \mu \gg \nu$ \\[4pt]
    \textbf{Lanczos Growth ($b_n \sim \mu$)}
    \begin{itemize}[leftmargin=10pt, itemsep=1pt, topsep=0pt]
        \item Bounded
        \item No asymptotic linear growth
    \end{itemize}
    \textbf{Spectral Density $f(\omega)$}
    \begin{itemize}[leftmargin=10pt, itemsep=1pt, topsep=0pt]
        \item Discrete mass-gap poles dominant
    \end{itemize}
    \textbf{Krylov Dynamics}
    \begin{itemize}[leftmargin=10pt, itemsep=1pt, topsep=0pt]
        \item Discrete site-hopping
        \item State localization / trapping
    \end{itemize}
};

\node[box, fill=orange!10] (box2) at (5.4,0) {
    \centering \textbf{\large Transient Phase ($t \sim \beta$)} \\
    \textit{\small (Intermediate Temp.)} \\[8pt]
    \raggedright
    \textbf{Regime} $\beta\mu \sim 1, \quad \beta\nu \sim 1$ \\[4pt]
    \textbf{Lanczos Growth ($b_n$)}
    \begin{itemize}[leftmargin=10pt, itemsep=1pt, topsep=0pt]
        \item $b_n \sim \frac{\pi}{\beta}n + \gamma \sqrt{n} + \mathcal{O}(1)$
    \end{itemize}
    \textbf{Spectral Density $f(\omega)$}
    \begin{itemize}[leftmargin=10pt, itemsep=1pt, topsep=0pt]
        \item Thermal continuum + discrete poles
    \end{itemize}
    \textbf{Krylov Dynamics}
    \begin{itemize}[leftmargin=10pt, itemsep=1pt, topsep=0pt]
        \item Speed $b(x) = \frac{\pi}{\beta}x + \gamma \sqrt{x}$
        \item Sub-exponential crossover
    \end{itemize}
};

\node[box, fill=red!8] (box3) at (10.8,0) {
    \centering \textbf{\large Continuum Phase ($t \gg \beta$)} \\
    \textit{\small (High Temperature)} \\[8pt]
    \raggedright
    \textbf{Regime} $\beta\mu \ll 1, \quad \beta\nu \ll 1$ \\[4pt]
    \textbf{Lanczos Growth ($b_n$)}
    \begin{itemize}[leftmargin=10pt, itemsep=1pt, topsep=0pt]
        \item Linear growth: $b_n \sim \frac{\pi}{\beta}n$
    \end{itemize}
    \textbf{Spectral Density $f(\omega)$}
    \begin{itemize}[leftmargin=10pt, itemsep=1pt, topsep=0pt]
        \item Continuous thermal background
    \end{itemize}
    \textbf{Krylov Dynamics}
    \begin{itemize}[leftmargin=10pt, itemsep=1pt, topsep=0pt]
        \item Continuous advection PDE
        \item Maximal chaos ($\lambda_K = \frac{\pi}{\beta}$)
    \end{itemize}
};

\end{tikzpicture}%
}
\caption{Schematic diagram summarizing the key characteristics of the three phases in the matrix model across temperature regimes, detailing the behavior of Lanczos coefficients $b_n$, spectral density $f(\omega)$, and Krylov dynamics.}
\label{fig:matrix_model_phases}
\end{figure}

Given the above state of the art, we treat temperature $\beta^{-1}$ as the tuning parameter and/or variable that ranges between low and high values. We categories our results for three distinct phases of the matrix model- (i) the low temperature gapped phase which is primarily governed by the mass scale $\mu\gg \beta^{-1}$ and $\mu \gg \nu$, (ii) the high temperature ($\beta \sim 0$) continuum phase where thermal fluctuations override the mass scale $\beta^{-1}\gg\mu$, $\beta^{-1}\gg \nu$ and (iii) intermediate temperature scale $\mu \sim \beta^{-1}$ and $\nu \sim \beta^{-1}$, where the chemical potential $\nu$ plays an important role (Fig. \ref{fig:matrix_model_phases}). Below, we summarize various key outcomes of the paper. 

Notice that the treatment of the Krylov site index $n$ is fundamentally different in the two extreme phases. For example, in the gapped low temperature phase $\beta \mu \gg 1$, the Lanczos coefficients $b_n$ do not exhibit linear growth. Instead, discrete mass-gap bound states produce localized, oscillatory, or power-law dynamics across discrete sites. The system remains strictly a discrete hopping problem on the Krylov chain, as wavepackets do not delocalize smoothly into a continuum. On the other hand, in the very high temperature (continuum) phase $\beta \mu \ll 1$, when evaluating operator growth or state complexity in the chaotic regime, the dominant physical processes occur at large Krylov site indices ($n \gg 1$) and the Lanczos coefficients grow asymptotically linearly ($b_n \sim \frac{\pi}{\beta} n$), reflecting fast thermalization and chaotic scrambling. At high $n$, the step size between adjacent sites ($\Delta n = 1$) becomes infinitesimally small relative to the overall scale of $n$ ($\frac{\Delta n}{n} = \frac{1}{n} \ll 1$), allowing the discrete wave amplitude $\psi_n(t)$ (or discrete Krylov site $\ket{K_n}$) to be approximated by a continuous field $\psi(x,t)$ governed by a continuous advection PDE.

In the intermediate temperature regime ($\beta\mu \sim 1$, $\beta \nu \sim 1$), the Krylov chain dynamics reflect a competition between the localized discrete bound states (characteristic of the gapped phase) and the smooth, chaotic thermal continuum (characteristic of the continuum phase). Rather than being strictly discrete or purely continuous, the spectral density $f(\omega)$ contains two coexisting components - (i) Thermal Continuum: High-frequency, high-energy states form a continuous thermal background that drives exponential advection at large $n$ and (ii) Mass-Gap Bound States: Low-energy isolated discrete poles (arising from mass scales $\mu$) survive alongside the continuum. The Lanczos coefficients $b_n$ do not follow a simple linear growth $b_n \sim \frac{\pi}{\beta} n$. Instead, they acquire a sub-leading correction term 
\begin{align}
    b_n \sim \frac{\pi}{\beta} n + \gamma \sqrt{n} + \mathcal{O}(1).
\end{align}

The primary term $\frac{\pi}{\beta} n$ comes from the continuous thermal background, preserving maximal chaos at asymptotic infinity. The $\gamma \sqrt{n}$ correction term comes from the mass-gap bound states, where $\gamma(\beta\mu, \beta\nu)$ depends directly on the ratio of discrete pole residues to the thermal continuum prefactor $C(\beta\mu, \beta\nu)$. In the continuum limit ($n \to x$), the evolution PDE includes an advection field $b(x) = \frac{\pi}{\beta} x + \gamma \sqrt{x}$. Solving for the trajectory $x(t)$ of a localized wavepacket $\psi(x,t)$ yields the following exponential growth controlled by the parameter $\gamma$ 
\begin{align}
    x(t) = \left[ \left( \sqrt{\epsilon} + \frac{\gamma \beta}{\pi} \right) e^{\frac{\pi t}{2 \beta}} - \frac{\gamma \beta}{\pi} \right]^2.
\end{align}
Asymptotically ($t \gg \beta/\pi$), the exponential growth is controlled by the Krylov exponent $\lambda_K = \frac{\pi}{\beta}$, which is governed purely by the thermal continuum limit. In the ``Transient Phase'' (early-to-intermediate times), the parameter $\gamma$ causes a sub-exponential acceleration or deceleration in the state growth profile, acting as a crossover bridge between non-chaotic bound-state oscillation and maximal thermal chaos.

Spread complexities in the three different phases can be expressed as\footnote{See Section \ref{sec3.3} for a detailed qualitative discussion.}
\begin{equation}
\label{e1.3}
C(t) \sim \begin{cases} \mu^2t^2+\mathcal{O}(t^4) & \text{(gapped phase, $t\sim 0$)} \\ \exp\left( \frac{\pi t}{\beta} \right) \left[ 1 + \frac{2\gamma \beta}{\pi \sqrt{\epsilon}} \left( 1 - e^{-\frac{\pi t}{2\beta}} \right) \right]+\mathcal{O}(\gamma^2) & \text{(intermediate phase, $t \sim \beta$)} 
\\ \exp\left( \frac{\pi t}{\beta} \right)  & \text{(continuum phase, $t\gg \beta$)}.\end{cases}
\end{equation}

Equation \eqref{e1.3} encapsulates the qualitative crossover in state complexity dynamics as the system transitions from a low-temperature localized regime to a high-temperature chaotic continuum. In the low-temperature gapped phase ($\beta\mu \gg 1$), thermal fluctuations are heavily suppressed by the large mass gap, causing the spread complexity $\mathcal{C}(t)$ to exhibit purely non-chaotic, quadratic short-time growth dominated by discrete mass-gap bound-state oscillations. Conversely, in the high-temperature continuum phase ($\beta\mu \ll 1$), thermal excitations wash out the mass gap and yield maximal quantum chaos, characterized by pure asymptotic exponential growth driven by the maximal Krylov Lyapunov exponent $\lambda_K = \pi/\beta$. The intermediate temperature regime ($\beta\mu \sim 1, \beta\nu \sim 1$) acts as a dynamical bridge between these two extremes, while the leading asymptotic behavior preserves maximal thermal scrambling ($\sim e^{\pi t/\beta}$), the sub-leading correction parameter $\gamma$ introduces a sub-exponential transient phase which reflects the underlying competition between the discrete mass-gap bound states and the continuous thermal background, effectively accelerating or decelerating the complex state trajectory before full chaotic thermalization sets in.

The organization of the rest of the paper is as follows. Section \ref{sec2} reviews the zero-temperature ($\beta^{-1}=0$) framework by examining the Fuzzy sphere model in the large mass deformation limit ($\mu \gg 1$) of the BMN Matrix Model, demonstrating how dimensional reduction generates non-linear effective potentials parameterized by $(k, l, n)$. Within this regime, constructing the Krylov basis $\ket{K_n}$ and evaluating the initial Lanczos coefficients reveals that the large mass gap $\mu \gg 1$ strictly confines early-time $t \sim 0$ wavepacket dispersion to the neighborhood of the initial lattice site. Section \ref{sec3} seamlessly extends this picture to finite temperatures ($\beta^{-1}$) and non-zero chemical potentials ($\nu$) by elevating single-mode dynamics to a continuous thermal ensemble via the charged Thermofield Double state (cTFD). Incorporating continuous thermal background states alongside discrete mass-gap bound-state corrections modifies the high-$n$ Lanczos spectrum into $b_n \sim \frac{\pi}{\beta} n + \gamma \sqrt{n}$. Taking the continuum limit ($n \to x$) maps the discrete Krylov chain onto a continuous advection field $b(x)$, enabling an analytical derivation of the wavepacket trajectory $x(t)$ and directly predicting the three-phase spread complexity scaling behavior detailed in equation \eqref{e1.3}. Finally, we draw our conclusion in Section \ref{sec4} with potential future remarks.

The technical details and supporting derivations are organized into four appendices as follows:
Appendix~\ref{appenA} details the exact evaluation of the continuous spectral function $f_{\text{cont}}(\omega)$ and its leading-order pole structure, establishing the analytic continuation leading to the KMS boundary conditions and thermal sum rules.
Appendix~\ref{appenB} provides the explicit step-by-step derivation of the $2n$-th spectral moments $\mathcal{M}_{2n}$ under grand canonical conditions, detailing the asymptotic expansion and contour integration techniques that yield the characteristic factorial growth $(2n)!$.
Appendix~\ref{appenC} presents the detailed residue analysis at the primary KMS singularity, isolating the interpolating amplitude $C(\beta\mu, \beta\nu)$ and quantifying the interplay between chemical potential enhancement and mass-gap suppression.
Finally, Appendix~\ref{appenD} details how time-reversal symmetry ($\mathcal{T}$) and Krylov parity operator ($\mathcal{P}_{\text{Krylov}}$) constraints govern the Krylov space geometry, dictating the vanishing of diagonal Lanczos coefficients $a_n$ and the positivity of off-diagonal hopping amplitudes $b_n$.
\section{Fuzzy sphere model and spread complexity: A review}
\label{sec2}
The purpose of this Section is to briefly review the Fuzzy sphere model and spread complexity of states along the lines of \cite{Roychowdhury:2026vzq}-\cite{Roychowdhury:2026mpd}. We begin by reviewing the basics of the Fuzzy sphere model \cite{Asano:2015eha} at large mass deformation, where we discuss the emergent scaling symmetries of the quantum mechanical model. We construct the Krylov basis and outline the derivation of the first few Lanczos coefficients $a_n$ and $b_n$ in the limit of large mass deformation $\mu \gg 1$. We outline how to obtain the wave function (or the probability amplitude) $\psi_n(t)$ associated with the $n$th Krylov site $\ket{K_n}$ \cite{Balasubramanian:2022tpr}. It is possible to show that at short time scales ($t \sim 0$), these wavefunctions allow a polynomial expansion in powers of $\mu t$, where the product $\mu t$ is kept \emph{fixed} in the limit $\mu \gg 1$ \cite{Roychowdhury:2026mpd}. In other words, at large deformation, the Krylov spread of states is controlled by the value of the mass parameter $\mu$.

The BMN or Plane Wave Matrix Model \cite{Berenstein:2002jq} is defined as\footnote{Throughout this paper, we choose to work with natural units where we set $\hbar =c=l=1$.} \cite{Asano:2015eha}
\begin{align}
\label{e2.1}
    S_{PWMM}&=\frac{1}{g^2}\int dt \text{Tr}\Big[ \frac{1}{2}(D_t X^r)^2+\frac{1}{4}[X^r,X^s]^2 -\frac{1}{2}\Big(\frac{\mu}{3}\Big)^2X^2_i \nonumber\\
    &-\frac{1}{2}\Big(\frac{\mu}{6}\Big)^2X^2_a -\frac{i\mu}{3}\epsilon_{ijk}X^i X^j X^k\Big]+\text{Fermions}.
\end{align}

The action \eqref{e2.1} can be obtained following a systematic reduction of $\mathcal{N}=4$ SYM on three sphere $S^3$. Here $\mu$ is the mass deformation parameter (or the relevant coupling of the theory), which is considered to be large enough in our computation. Given the above coupling, the matrix model \eqref{e2.1} could be thought of as the (relevant) deformation of the BFSS matrix model \cite{Banks:1996vh}. Here, $X^r (r=1,\cdots , 9)$ are $N \times N$ gauge invariant matrices which satisfy the following set of equations of motion \cite{Asano:2015eha}
\begin{align}
\label{e2.2}
    &\ddot{X}^i+[X^r,[X^r,X^i]]+\Big(\frac{\mu}{3}\Big)^2 X^i+i \mu \epsilon^{ijk}X_j X_k =0\\
    \label{e2.3}
    &\ddot{X}^a+[X^r,[X^r,X^a]]+\Big(\frac{\mu}{6}\Big)^2 X^a =0.
\end{align}

The above set of equations \eqref{e2.2}-\eqref{e2.3} are supplemented with the Gauss law constraint 
\begin{align}
\label{e2.4}
    [X^r,\dot{X^r}]=0.
\end{align}

Notice that $i,j=1,2,3$ correspond to the $SU(2)$ indices and $a,b=4,\cdots ,9$ are the $SO(6)$ indices. We consider several consistent reductions of the matrix model \cite{Asano:2015eha} and in particular focus on $N=2,4$ representations of the Fuzzy sphere model that satisfy the required $SU(2)$ algebra. The corresponding Hamiltonian density can be expressed as
\begin{align}
\label{e2.5}
    \mathcal{H}_{BMN}=\mathcal{H}=\frac{1}{2}(p_x^2+p_y^2)+V(x,y).
\end{align}

The potential function has the generic expression of the form \cite{Asano:2015eha}, \cite{Roychowdhury:2026igc}
\begin{align}
\label{e2.6}
    V(x,y)=\frac{\mu^2}{2}(k x^2+y^2)+\frac{1}{2}(x^4+y^4)+l x^2 y^2-\mu y^3-n \mu x^2 y.
\end{align}

The parameters $k,l,n$ characterize different representations of the Fuzzy sphere. For example, setting $k=\frac{1}{4}$, $l=1$ and $n=0$ one finds the pulsating fuzzy sphere \cite{Asano:2015eha} corresponding to the $N=2$ representation of the $SU(2)$ algebra. On the other hand, for the choice of the parameters $k=1$, $l=3$, $n=3$, the resulting theory leads to the integrable fuzzy sphere model, which corresponds to the $N=4$ representation of the $SU(2)$ algebra. 

The Hamiltonian \eqref{e2.5} has an emergent scaling symmetry \cite{Amore:2024ihm} of the form
\begin{align}
     x \rightarrow \lambda x~;~y \rightarrow \lambda y~;~t \rightarrow t/\lambda ~;~\mu\rightarrow \lambda \mu
\end{align}
which results in an overall scaling of the Hamiltonian as $\mathcal{H}\rightarrow \lambda^4 \mathcal{H}$. Clearly, the minimum energy configuration would correspond to setting $\lambda \rightarrow 0$. Keeping the re-scaled $\mu$ finite in this scaling limit, would naturally lead to setting $\mu \gg 1$. In summary, the ground state of the matrix model in the limit of the large mass gap $\mu$ would correspond to a configuration of localized harmonic oscillators that is characterized by the potential function
\begin{align}
\label{e2.8}
   V(x,y)=\frac{\mu^2}{2}(k x^2+y^2)+\mathcal{O}(\mu^{-2})  
\end{align}
where both $\mu x$ and $\mu y$ are kept fixed in the limit $\mu \gg 1$.

Given the above state of the art, we choose the Krylov basis at $t=0$ as \cite{Roychowdhury:2026sgg}
\begin{align}
\label{e2.18}
    &\ket{K_0}=\ket{\Psi_0}=\int dx dy \Psi_0(x,y)\ket{x,y}\\
    & \Psi_0(x,y)=\sqrt{\frac{2 \mu}{\pi}}e^{-\mu(x^2+y^2)}
    \label{e2.19}
\end{align}
where, $\Psi_0(x,y)$ is the Gaussian wavefunction of the ground state of the composite system and $\ket{x,y}$ is the eigen basis in the position representation.

For generic $n$, the Krylov basis can be expressed as
\begin{align}
\label{e2.11}
    \ket{K_{n+1}}=\ket{\Psi_{n+1}}-c_n \ket{K_n}-d_n \ket{K_{n-1}}~;~n=0,1,2,\cdots.
\end{align}

The coefficients $c_n$ and $d_n$ are fixed from the orthonormality criteria $\braket{K_n|K_m}=\delta_{nm}$, for details, see \cite{Roychowdhury:2026vzq}-\cite{Roychowdhury:2026mpd}. The $n$th state $\ket{\Psi_n}$ can be expressed as $\ket{\Psi_n}=\mathcal{H}^n\ket{\Psi_0}$. The Krylov basis elements $\{\ket{K_n}\}$ satisfy the Krylov chain criterion \cite{Balasubramanian:2022tpr}
\begin{align}
\label{e2.12}
    \mathcal{H} \ket{K_n}=a_n \ket{K_n}+b_n \ket{K_{n-1}}+b_{n+1}\ket{K_{n+1}}.
\end{align}

The Lanczos coefficients $a_n$ and $b_n$ can be expressed as
\begin{align}
    a_n=\braket{K_n|\mathcal{H}|K_n}~;~b_n = \braket{K_{n-1}|\mathcal{H}|K_n}.
\end{align}

The quantum state at any given time $t$ has a generic expression of the form
\begin{align}
\label{e2.14}
     \ket{\Psi(t)}=\sum_{n=0}^{\infty}\frac{(-it)^n}{n!}\mathcal{H}^n \ket{\Psi_0}
     =\sum_{n=0}^{\infty}\frac{(-it)^n}{n!}\ket{\Psi_n}=\ket{\Psi_0}-it \ket{\Psi_1}+\cdots.
\end{align}

The same quantum state can be expanded in a Krylov basis as
\begin{align}
\label{e2.15}
\ket{\Psi(t)}=\sum_{n=0}^{\infty}\psi_n(t)\ket{K_n}=\psi_0(t)\ket{K_0}+\psi_1(t)\ket{K_1}+\cdots  
\end{align}
where the coefficients $|\psi_n(t)|^2$ are the probability amplitudes of the state to be found at the lattice site $n$. The wavefunctions $\psi_n(t)$ can be obtained by comparing \eqref{e2.14} and \eqref{e2.15}, which satisfy the discrete Schrodinger equation of the form
\begin{align}
\label{e2.16}
     i  \frac{\partial}{\partial t}\psi_n(t)=a_n \psi_n (t)+b_{n+1}\psi_{n+1}(t)+b_n \psi_{n-1}(t).
\end{align}

The spread complexity of states is defined as the average position on the Krylov chain
\begin{align}
    C(t)=\sum_{n=0}^\infty n | \psi_n(t)|^2.
\end{align}

Analytically solving the first few orthonormal wave functions $\psi_n(t)(n=0,1,2)$, the spread complexity of states can be expressed as
\begin{align}
\label{e2.25}
    C(t)|_{t\sim 0}=b_1^2 t^2+\mathcal{O}(t^4)
\end{align}
where the LO coefficient $b_1 \sim \mu$ controls the spread complexity growth of states.\\\\
\uline{\textbf{Physical insight into early-time complexity growth:}}\\\\
Let us summarize the take home message from the above analysis. It appears that physically the mass gap parameter $\mu$ acts as an \emph{effective} harmonic confining potential in the state space representation. In the limit when the mass gap becomes large enough, that is, $\mu \gg 1$, the steep energy cost associated with populating higher excited states strictly suppresses quantum fluctuations away from the reference state $\ket{K_0}$. Consequently, the initial wave packet remains dynamically trapped within the immediate neighborhood of the initial lattice site ($n=0$) on the Krylov chain, significantly delaying state expansion and preventing rapid operator growth into higher Krylov modes $n \ge 1$.

The quadratic early-time growth of the spread complexity of states, $C(t)|_{t \sim 0} \sim \mu^2 t^2+\mathcal{O}(t^4)$, reflects the universal perturbative onset of quantum state spreading along a one-dimensional Krylov lattice pertaining to the BMN matrix model. Here, the leading-order coefficient $b_1$ measures the intrinsic rate of state dispersion into adjacent Krylov sites. The $t^2$ scaling of the spread complexity of states at early time scales $t \sim 0$ is a universal feature. However, it is worth noting that, because the coefficient $b_1 \propto \mu$, the mass gap $\mu$ sets the natural energy scale and an intrinsic clock governing the transition rates $b_n$ between two neighboring lattice sites, for example, $n$ and $n-1$ (or $n+1$). In other words, the mass gap $\mu$ controls how fast the state can jump from one Krylov state to the next nearest neighbor. A larger mass gap thus effectively rescales the physical time $t \to \mu t$, meaning that it speeds up system's ``internal clock'', effectively replacing normal time $t$ with a faster one $\mu t$. As a result, the initial growth rate scales as $\mu^2$, although the overall state dynamics remains strongly gapped to lower Krylov dimensions over microscopic time scales as explained above.

So far we were concerned about a single localized state in the matrix model. In reality, one could think of adding temperature $\beta^{-1}$ and chemical potential $\nu$ to the system. While the localized single-state dynamics above provides clear intuition for single oscillator modes, realistic matrix model dynamics at finite temperature $\beta^{-1}$ and non-zero chemical potential $\nu$ require accounting for continuous mode superpositions. In the next section, we generalize this framework by constructing the charged Thermofield Double state (cTFD), where global $U(1)$ charge symmetry resolution introduces additional charge-dependent chemical potential factors that non-trivially alter the Krylov return amplitude and complexity evolution.

\section{Charged thermofield double and spread complexity of states}
\label{sec3}
Until now we focused on a single isolated quantum state at zero temperature and its evolution in a Krylov basis. 
Having established the state localization and early-time complexity dynamics for isolated oscillator modes in the large mass limit, we now turn our attention to the full thermodynamic and finite-temperature regime of the theory. At finite temperatures, the initial state $\ket{\Psi_0}$ is not just a single mode, rather it is a superposition of many thermal states continuously interacting. In other words, in a realistic matrix model, like BMN, a comprehensive description of spread complexity of states at non-zero temperature requires elevating single-state dynamics to continuous thermal mode superpositions. 

In physical terms, at zero temperature, the system resides strictly in a single quantum mode where state dispersion is easily tracked. At non-zero temperatures $\beta^{-1}$, thermal fluctuations are important and they excite multiple energy channels simultaneously. Spread complexity must therefore account not only for quantum transitions within a single channel, but for the collective, statistical spreading of information across a thermal ensemble.

To achieve this, we introduce the notion of the charged Thermofield Double state (cTFD) \cite{Caputa:2013eka}-\cite{Chapman:2019clq} in the presence of an emergent global $U(1)$ charge symmetry. The resolution of global symmetries plays a central role in probing fine-grained quantum information properties, for example, the decomposition of the Hamiltonian $\mathcal{H}$ into super-selection sectors \cite{Caputa:2025ozd}. Moreover, the introduction of a global charge $Q$ and its associated chemical potential $\nu$ has a direct impact on the corresponding energy levels, which non-trivially deforms the thermal spectrum and directly influences the Krylov return amplitude (survival probability) and the resulting spread complexity of states \cite{Caputa:2025mii}-\cite{Caputa:2025ozd}, as we will see below. 

Physically, turning on a chemical potential $\nu$ acts as a selective potential barrier across charge sectors. It biases the statistical weights towards specific conserved charge configurations, thereby reshaping the Krylov geometry and effectively suppressing or enhancing the speed at which quantum information spreads depending on the charge sector $Q$.
\subsection{The complex oscillator model}
The fundamental premise of our construction relies on the fact that for sufficiently large mass deformation, that is, $\mu \gg 1$, the non-linear interactions in the matrix model effectively decouple, allowing the system to be mapped onto a set of localized quantum harmonic oscillators anchored around their respective ground states. In this regime, the high restoring force restricts matrix fluctuations to small oscillations near the classical potential minima \eqref{e2.8}. Consequently, non-linear interactions vanish at leading order, simplifying the non-integrable matrix dynamics into an \emph{integrable} set of non-interacting quantum harmonic oscillators where exact analytical control over the Krylov basis becomes tractable. In what follows, we leverage this asymptotic oscillator representation to systematically evaluate the moments $\mathcal{M}_n$, derive the exact form of the return amplitude, and analyze the interplay between symmetry resolution and the spread complexity growth of states.

The key idea is based on the fact that at large mass deformation the BMN matrix model could be thought of as localized harmonic oscillators around their respective ground states. The corresponding Hamiltonian follows from \eqref{e2.6} (with $k=1$)
\begin{align}
\label{e3.1}
    \mathcal{H}_{BMN}=\frac{1}{2}(p_x^2+p_y^2)+\frac{\mu^2}{2}(x^2+y^2)-\mu y (nx^2+y^2)+\mathcal{O}(\mu^{-4})
\end{align}
where $\mu x$ and $\mu y$ are kept fixed in the large deformation limit $\mu \gg 1$.

One can schematically express the above Hamiltonian \eqref{e3.1} as 
\begin{align}
\label{ee3.2}
    \mathcal{H}_{BMN}=\mathcal{H}_{SHO}+\mathcal{H}_{deform}
\end{align}
where the leading term corresponds to the Hamiltonian of a simple harmonic oscillator. All the sub-leading terms correspond to deformation which we treat as perturbation to the harmonic oscillator model in the limit $\mu \gg 1$ where $\mu x <1$ and $\mu y <1$ are kept fixed.

To begin with, we switch off the perturbation and obtain the return amplitude and moments for the harmonic oscillator potential
\begin{align}
\label{e3.2}
    \mathcal{H}_{SHO}=\frac{1}{2}(p_x^2+p_y^2)+\frac{\mu^2}{2}(x^2+y^2).
\end{align}

Clearly, the Hamiltonian \eqref{e3.2} is a sum of decoupled oscillators, and the matrix model has an \emph{emergent} global $U(1)$ symmetry in the large mass deformation. To see this explicitly, we introduce the following change of variables \cite{Caputa:2025ozd}
\begin{align}
    \Pi = \frac{1}{\sqrt{2}}(p_{x}-ip_{y})~;~\Phi = \frac{1}{\sqrt{2}}(x+i y)
\end{align}
which yields the Hamiltonian in the form
\begin{align}
\label{e3.5}
    \mathcal{H}_{SHO}=\Pi \Pi^\dagger+\mu^2\Phi \Phi^\dagger.
\end{align}

The Hamiltonian has a $U(1)$ global symmetry
\begin{align}
    \Pi \rightarrow e^{i \theta}\Pi~;~\Phi \rightarrow e^{i\theta}\Phi
\end{align}
which is broken at NLO in the $1/\mu$ expansion.

Next, we introduce the bosonic creation and annihilation operators 
\begin{align}
\label{e2.52}
    \Pi=i\sqrt{\frac{\mu}{2}}(d^\dagger-c)~;~\Phi=\frac{1}{\sqrt{2 \mu}}(d+c^\dagger).
\end{align}

Using \eqref{e2.52}, we find the following Hamiltonian
\begin{align}
\label{e3.8}
    \mathcal{H}_{SHO}=\mu(N_d+N_c+1)
\end{align}
where we have used commutation relations between creation and annihilation operators as
\begin{align}
    [d,d^\dagger]=1~;~[c,c^\dagger]=1.
\end{align}

Here, we introduce a set of number operators as $N_d=d^\dagger d$ and $N_c = c^\dagger c$ such that
\begin{align}
   & \mathcal{H}_{SHO}\ket{n_d,n_c}=\mathcal{E}_{(n_d,n_c)}\ket{n_d,n_c}\\
   &\mathcal{E}_{(n_d,n_c)} = \mu (n_d+n_c+1)
\end{align}
where $n_d$ and $n_c$ are the respective occupation numbers of the state.

Finally, we note the $U(1)$ charge associated with the global symmetry
\begin{align}
    Q=N_d-N_c
\end{align}
which acts on a charge eigen-state as follows
\begin{align}
    Q \ket{n_d,n_c}=(n_d-n_c)\ket{n_d,n_c}=q_{(n_d,n_c)}\ket{n_d,n_c}.
\end{align}

Next, we introduce the charged TFD (cTFD) for the BMN matrix model at leading order \eqref{e3.8} in the strong coupling $\mu \gg 1$ as
\begin{align}
\label{e3.14}
    \ket{TFD(t)}^{(0)}_{BMN}=\sum_{n_d,n_c}\frac{e^{-\frac{\beta}{2}\Big[\mu(n_d+n_c+1)+\nu (n_d-n_c) \Big]-i \mu t(n_d+n_c+1)}}{\sqrt{Z_{BMN}(\beta, \nu)}}\ket{n_d,n_c}\otimes \ket{n_d,n_c}
\end{align}
where $\nu$ is the chemical potential associated with the global charge $q_{(n_d,n_c)}$. The above expression\footnote{The superscript refer to the fact that we are considering the undeformed BMN Hamiltonian \eqref{e3.2}.} \eqref{e3.14} has to be understood as the \emph{zeroth} order solution in a perturbative expansion \eqref{e3.2}. Notice that in order for the sum \eqref{e3.14} to be convergent, one must satisfy the condition $\nu \leq \mu$. It is worth highlighting that, in contrast to previous analyses \cite{Caputa:2025ozd}, \cite{Caputa:2013eka}-\cite{Chapman:2019clq}, the coefficient in the infinite series \eqref{e3.14} has to be understood in the limit of the large mass gap $\mu$, where we retain only the leading order term in the expansion \eqref{e3.5}, after we perform the infinite sum over the (quantum) harmonic oscillator modes $n_d$ and $n_c$.\\\\
\uline{\textbf{Remarks on dual interpretation:}}\\\\
It is entirely possible to construct such a pair of maximally \emph{entangled} states \eqref{e3.14} for the BMN matrix model, which exists in the limit of large mass deformation $\mu \gg 1$. The holographic counterpart of \eqref{e3.14} should correspond to a pair of eternal black holes in type IIA supergravity that are maximally entangled. From a gauge-gravity perspective, the charged Thermofield Double (cTFD) state represents two asymptotic boundaries of an eternal charged black hole connected by an Einstein-Rosen bridge (wormhole). The chemical potential $\nu$ maps to the electric potential at the horizon, directly governing the late-time growth of the wormhole's interior volume, which serves as the dual gravitational measure of spread complexity of states in the matrix model at large mass gap.

Let us add a few more comments on the dual gravitational interpretation of the BMN model. At high temperatures $\beta \mu \ll 1$, the matrix model undergoes a ``continuum transition'' where matrix fluctuations are not constrained. This refers to a non-extremal D0 brane in 10d type IIA flux background \cite{Costa:2014wya} with an unbroken $SO(9)$ symmetry. 

\paragraph{Non-Extremal D0-Brane Background.}
The 10-dimensional Type IIA supergravity solution (in the string frame) representing a non-extremal D0-brane black hole carrying $N$ units of Ramond-Ramond (RR) 0-brane charge is given by the following line element
\begin{align}
\label{ee3.15}
ds_{10,\text{string}}^2 &= - H(r)^{-1/2} f(r) \, dt^2 + H(r)^{1/2} \left[ f(r)^{-1} dr^2 + r^2 d\Omega_8^2 \right], \\
e^{\Phi(r)} &= g_s \, H(r)^{3/4}, \\
A_1(r) &= \left( \frac{1 - H(r)^{-1}}{\coth \alpha} \right) dt,
\end{align}
where $d\Omega_8^2$ is the line element on the unit 8-sphere ($S^8$) preserving the $SO(9)$ rotational symmetry of the unconstrained matrix fluctuations in the continuum phase. The harmonic function $H(r)$ and non-extremality thermal emittance function $f(r)$ are defined as
\begin{equation}
H(r) = 1 + \frac{r_0^7 \sinh^2\alpha}{r^7}, \quad \text{and} \quad f(r) = 1 - \frac{r_0^7}{r^7}.
\end{equation}
Here, $r_0$ denotes the non-extremal horizon radius, $\alpha$ is a charge parameter related to the D0-brane charge $N$, and $g_s$ is the string coupling constant. 

At the event horizon $r = r_0$, $f(r_0) = 0$, giving rise to a finite Hawking temperature dual to the matrix model temperature $T = \beta^{-1}$ \cite{Costa:2014wya}
\begin{equation}
T_H = \frac{7}{4\pi r_0 \cosh\alpha}.
\end{equation}
In the zero-temperature / BPS limit ($r_0 \to 0$ with $r_0^7 \sinh^2\alpha \equiv R^7$ kept fixed), $f(r) \to 1$, recovering the extremal supersymmetric D0-brane metric.

As the transition temperature drops down, the effects due to massive deformation $\mu$ become important that breaks $SO(9)\rightarrow SO(6)\times SO(3)$. At low temperatures $\beta \mu \gg 1$, the black hole solution becomes sub-dominant and evaporates. The resulting geometry smoothly transits into LLM geometry \cite{Lin:2004nb}-\cite{Lin:2004kw} supported by background fluxes.

\paragraph{Lin-Lunin-Maldacena (LLM) Geometry.}
In the low-temperature gapped phase ($\beta \mu \gg 1$), the 10-dimensional Type IIA/M-theory supergravity dual smoothly transits into a horizonless LLM geometry \cite{Lin:2004nb}-\cite{Lin:2004kw} specified by a boundary coloring function $z(x_1, x_2, 0)$. In the string frame, the metric preserving $SO(6) \times SO(3)$ symmetry can be expressed as
\begin{equation}
\label{eq:LLM_metric_10d}
ds_{10,\text{string}}^2 = - \frac{h^{-2}}{\sqrt{1 - z^2}} \, dt^2 + h^2 \sqrt{1 - z^2} \left[ dr^2 + r^2 d\Omega_2^2 + y^2 d\Omega_5^2 \right] + \frac{h^{-2}}{\sqrt{1 - z^2}} \left( dx_1^2 + dx_2^2 \right),
\end{equation}
where $d\Omega_2^2$ and $d\Omega_5^2$ are the line elements on the unit 2-sphere ($S^2$) and 5-sphere ($S^5$) corresponding to the $SO(3)$ and $SO(6)$ isometry groups, respectively. The warp factor $h(x_1, x_2, y)$ and function $z(x_1, x_2, y)$ are determined by a single electrostatic potential $V(x_1, x_2, y)$, whose explicit form is given by 
\begin{align}
z(x_1, x_2, y) &= -\frac{1}{2} y \, \frac{\partial V}{\partial y}, \\
h^{-2}(x_1, x_2, y) &= 2 y \, e^{-G} \sinh G, \quad \text{with } e^{2G} = \frac{1 + z}{1 - z}.
\end{align}
The potential $V(x_1, x_2, y)$ obeys Laplace's equation in three virtual spatial dimensions $(x_1, x_2, y)$ with Dirichlet boundary conditions at $y = 0$:
\begin{equation}
z(x_1, x_2, 0) = \begin{cases} +\frac{1}{2} & \text{(black regions / flux droplets)} \\ -\frac{1}{2} & \text{(white regions / spacetime background).} \end{cases}
\end{equation}
Because $z^2 = 1/4$ everywhere on the boundary $y = 0$, the metric is everywhere non-singular and completely free of horizons, reflecting the absence of Hawking radiation and thermal dissipation in the low-temperature phase of the BMN matrix model.

In summary, in the high temperature ($\beta \mu \ll 1$) disconnected phase, the cTFD corresponds to a two sided eternal black hole, where two copies of the BMN matrix model ($L$ and $R$) live on the two exterior boundaries connected by a smooth spatial Einstein-Rosen bridge. On the other hand, in the low temperature ($\beta \mu \gg 1$) gapped phase, the geometry transits into two disconnected copies of smooth Lin-Maldacena geometries \cite{Lin:2004nb}-\cite{Lin:2004kw} supported by background fluxes. The wormhole pinches off, reflecting the fact that the matrix degrees of freedom are gapped and lack the large-$N^2$ entanglement entropy required to hold open a semi-classical Einstein-Rosen bridge connecting these two geometries. Therefore, given the state of the art, in the following, we divide our results into two categories. In the first example, we discuss results near the low temperature ($\beta \mu \gg 1$) gapped phase. Finally, we discuss results for the high temperature ($\beta \mu \ll 1$) continuum phase.\\\\
\uline{A word of caution:}\\\\
Given the charged TFD (cTFD) state \eqref{e3.14}, in the following we carry out a ``preliminary'' analysis of return amplitude and moments. However, these results are further reshaped (while keeping the qualitative features intact) in Section \ref{sec3.2} by imposing constraints due to time reversal or equivalently a parity operation on the 1D Krylov chain.
\\\\
\uline{\textbf{Return amplitude in the gapped phase:}}\\\\
We now return to the discussion of the cTFD and the return amplitude for the low temperature gapped phase corresponding to $\beta \mu \gg 1$ where we fix the mass gap $\mu \gg 1$ and also set the chemical potential $\nu \ll \mu$ to a small value comparable to $\mu$. The mass gap $\mu$ is such that one could still consider the deformation \eqref{ee3.2} as a perturbation.

The first step is to estimate the partition function, which is given by
\begin{align}
\label{e3.15}
    Z_{BMN}(\beta, \nu)=\sum_{n_d,n_c}e^{-\beta\Big[\mu(n_d+n_c+1)+\nu (n_d-n_c) \Big]}.
\end{align}

After performing the sum over the oscillator modes, it yields the following expression
\begin{align}
\label{e3.16}
  Z_{BMN}(\beta, \nu)=\frac{1}{2}\frac{1}{\cosh(\beta \mu)-\cosh(\beta \nu)}.
\end{align}

Taking into account an expansion in the limit $\beta \mu \gg 1$ and $\beta \nu \gg 1$, further yields
\begin{align}
\label{e3.17}
    Z_{BMN}(\beta,\nu)=\frac{e^{-\beta \mu}}{1-e^{-\beta (\mu -\nu)}}~;~\mu \gg \nu
\end{align}
which is the partition function of the BMN model at small temperatures and low density.

Next, we compute the return amplitude, which turns out to be
\begin{align}
    \mathcal{R}^{(0)}_{BMN}(t)&=\braket{TFD(t)|TFD(0)}^{(0)}_{BMN}\nonumber\\
    &=\frac{1}{Z_{BMN}(\beta,\nu)}\sum_{n_d,n_c}e^{-\beta\Big[\mu(n_d+n_c+1)+\nu (n_d-n_c) \Big]+i \mu t(n_d+n_c+1)}.
\end{align}

After performing the sum over quantum oscillator modes, one finds the following
\begin{align}
\label{e3.19}
    \mathcal{R}^{(0)}_{BMN}(t)=\frac{\cosh(\beta \mu)-\cosh(\beta \nu)}{\cosh\Big[\mu (\beta -i t) \Big]-\cosh(\beta \nu)}.
\end{align}

Next, we perform an expansion in the limit $\beta \mu \gg 1$ and $\beta \nu \gg 1$, which finally reveals
\begin{align}
\label{e3.20}
    \mathcal{R}^{(0)}_{BMN}(t)=\frac{1-e^{-\beta (\mu -\nu)}}{e^{-i\mu t}-e^{-\beta (\mu -\nu)}}=e^{i \mu t}+\mathcal{O}(e^{-\beta(\mu- \nu)}).
\end{align}\\\\
\uline{\textbf{Remarks on possible factorization:}}\\\\
We now wish to comment on the possibilities of the factorization of the cTFD \eqref{e3.14} and the return amplitude \eqref{e3.20}, which should follow directly from the expansions \eqref{e3.17} and \eqref{e3.19}. However, before doing any computation, it is worth mentioning that one should not expect any emergent $SL(2,R)$ symmetry \cite{Caputa:2025ozd} to exist. One of the prime reasons behind this is that the system has a large mass gap, parametrized by $\mu$, which makes the theory non-conformal. This is further confirmed by the holographic construction, which reveals that the dual geometry does not have any $AdS_2$ counterpart \cite{Lin:2005nh}. 

Looking at \eqref{e3.17}, one finds that the partition function does not allow any decomposition along the lines of \cite{Caputa:2025ozd}. The root cause of this lies in the fact that in the limit of low temperatures $ \beta \mu \gg 1$ and \emph{finite} density $ \beta \nu \gg 1$, we approximate the partition function \eqref{e3.16} by dropping terms $e^{-\beta \mu}\ll 1$ and $e^{-\beta \nu}\ll 1$ compared to their accompanying terms $e^{\beta \mu}$ and $e^{\beta \nu}$, which finally leads to \eqref{e3.17}. Similar remarks hold for the return amplitude.\\\\
\uline{\textbf{Moments:}}\\\\
In principle, moments $\mathcal{M}^{(0)}_n$ can be calculated using the first principle \cite{Balasubramanian:2022tpr}, that is, knowledge of the return amplitude \eqref{e3.20}. We should think of these moments $\mathcal{M}^{(0)}_n$ as representing the \emph{composite} system in an effective description. Given the above state of the art, one should understand that these moments $\mathcal{M}^{(0)}_n$ are strictly evaluated in the large $\mu$ limit, while considering only the simple harmonic contribution in the BMN Hamiltonian \eqref{e3.2}. 

For example, using \eqref{e3.19} and setting $n=1$, we find the first moment
\begin{align}
    \mathcal{M}^{(0)}_1 (\beta ,\nu)= \frac{d}{dt}\mathcal{R}^{(0)}_{BMN}(t)\Big|_{t=0}=\frac{i \mu  \sinh (\beta  \mu )}{\cosh (\beta  \mu )-\cosh (\beta  \nu )}.
\end{align}

Considering an expansion in the low temperature domain further reveals
\begin{align}
    \mathcal{M}^{(0)}_1(\beta,\nu) = \frac{i \mu}{1-e^{-\beta (\mu -\nu)}}=i\mu +\mathcal{O}(e^{-\beta (\mu -\nu)}).
\end{align}

On a similar note, for $n=2$, we obtain the moment at second order
\begin{align}
  &\mathcal{M}^{(0)}_2 (\beta ,\nu)= \frac{d^2}{dt^2}\mathcal{R}^{(0)}_{BMN}(t)\Big|_{t=0}\nonumber\\
  &= -\frac{\mu ^2 }{2 (\cosh (\beta  \mu )-\cosh (\beta  \nu ))^2}\Bigg[2 \cosh (\beta  \mu ) \cosh (\beta  \nu )+\cosh (2 \beta  \mu )-3 \Bigg].
\end{align}

After performing an expansion in the low temperature regime, one finds
\begin{align}
    \mathcal{M}^{(0)}_2 (\beta ,\nu)=-\mu^2\frac{(1+e^{-\beta (\mu - \nu)})}{(1-e^{-\beta (\mu -\nu)})^2}=-\mu^2+\mathcal{O}(e^{-\beta (\mu -\nu)}).
\end{align}

Next, considering $n=3$, we find the moment at order three
\begin{align}
   &\mathcal{M}^{(0)}_3 (\beta ,\nu)= \frac{d^3}{dt^3}\mathcal{R}^{(0)}_{BMN}(t)\Big|_{t=0}\nonumber\\
   &=  -\frac{i \mu ^3 \sinh (\beta  \mu ) }{2 (\cosh (\beta  \mu )-\cosh (\beta  \nu ))^3}\Bigg[8 \cosh (\beta  \mu ) \cosh (\beta  \nu )+\cosh (2 \beta  \mu )\nonumber\\
   &+\cosh (2 \beta  \nu )-10 \Bigg].
\end{align}

After performing an expansion in the low temperature regime, one finds
\begin{align}
    \mathcal{M}^{(0)}_3 (\beta ,\nu)=-i \mu^3 \frac{(1+e^{-\beta (\mu -\nu)})^2}{(1-e^{-\beta(\mu -\nu)})^3}=-i \mu^3+\mathcal{O}(e^{-\beta (\mu -\nu)}).
\end{align}

The above calculation can be repeated for higher values of $n \geq 4$ that generalizes the above structure in the following form
\begin{align}
\label{e3.27}
   \mathcal{M}^{(0)}_n (\beta ,\nu) =(i \mu)^n \frac{(1+e^{-\beta (\mu -\nu)})^{n-1}}{(1-e^{-\beta (\mu -\nu)})^{n}}=(i \mu)^n+\mathcal{O}(e^{-\beta(\mu-\nu)})~;~n = 1,2,3 \cdots.
\end{align}\\\\
\uline{\textbf{Return amplitude and moments in the continuum phase:}}\\\\
We now explore the return amplitude and moments in the continuum phase that corresponds to a high temperature regime $\beta \mu \ll 1$ and $\beta \nu \ll 1$. 

As a first step, we expand the partition function \eqref{e3.16}, which yields
\begin{align}
\label{ee3.37}
    Z(\beta , \nu)=-\frac{1}{\beta ^2 \nu ^2}+\beta ^2 \mu ^2 \left(-\frac{1}{\beta ^4 \nu ^4}+\frac{1}{6 \beta ^2 \nu ^2}-\frac{11}{720}\right)+\cdots
\end{align}
where $\mu > \nu$ is understood.

The return amplitude \eqref{e3.19} can be expanded in a similar fashion. However, a careful analysis reveals that a meaningful expression can be obtained first by expanding at early time scales $t \sim t_{\text{ref}}$ and thereby expanding in the high temperature limit. The emphasize on the expansion in early times $t \sim t_{\text{ref}}$ essentially reflects to the fact that it is the early time contribution to the zeroth order wavefunction $\psi_0(t)$ \eqref{e2.15} that determines the return amplitude and the moments. After some algebra, one finally obtains\footnote{As we elaborate later in Section \ref{sec3.3}, for each phase of the matrix model there is a reference time $t_{\text{ref}}$ (also known as the initial time $t_{\text{initial}}$) such that an early expansion is always referred to an expansion in the domain $t \sim t_{\text{ref}}$ and the argument of the return amplitude should be thought of as the difference $\Delta t = t-t_{\text{ref}}\sim 0$ for an early time expansion. The reference time for the gapped phase is taken to be $t_{\text{ref}}=0$ without any loss of generality. On the other hand, for the thermal phase, it is set by the temperature $\beta^{-1}$.}
\begin{align}
\label{ee3.38}
    \mathcal{R}^{(0)}_{BMN}(t \sim 0)=1-\frac{\mu ^2 t^2}{\beta ^2 \nu ^2}-\beta  \mu  \left(\frac{2 i \mu  t}{\beta ^2 \nu ^2}-\frac{i \mu  t}{6}\right)+\frac{\mu ^2 t^2}{12}+\cdots.
\end{align}

A further simplification can be achieved in the limit $\nu \rightarrow \mu$, which yields
\begin{align}
\label{e3.30}
    \mathcal{R}^{(0)}_{BMN}(t \sim 0)&=1-\frac{t^2}{\beta ^2}-\beta  \mu  \left(\frac{2 i t}{\beta ^2 \mu }-\frac{i \mu  t}{6}\right)+\frac{\mu ^2 t^2}{12}+\cdots\nonumber\\
    &=1-\frac{1}{\beta^2}\Bigg[t^2+ 2i \beta t - \frac{i}{6} \beta^3 \mu^2 t -\frac{1}{12}\mu^2 \beta^2 t+\cdots\Bigg]\nonumber\\
    &=1-\frac{1}{\beta^2}\Bigg[t^2+\mathcal{O}(\beta t)\Bigg].
\end{align}
It is trivial to show that the sub-leading corrections in \eqref{e3.30} are $\mathcal{O}(1/T)$ and are therefore suppressed in the high temperature limit, $T=\beta^{-1}\gg \mu$.

The moments can be computed following standard definition. Using \eqref{e3.30}, one finds
\begin{align}
\label{e3.31}
    &\mathcal{M}^{(0)}_1 (\beta ,\mu) =\beta  \mu  \left(\frac{i \mu }{6}-\frac{2 i}{\beta ^2 \mu }\right)+\cdots \\
    & \mathcal{M}^{(0)}_2 (\beta ,\mu) =\frac{\mu ^2}{6}-\frac{2}{\beta ^2}+\cdots.
    \label{e3.32}
\end{align}\\\\
\uline{\textbf{Physical Interpretation and Comparison of the Two Phases:}}\\\\
We now make a comparative analysis of the above results and discuss their physical interpretations.  The stark mathematical contrast between the entities in the gapped phase expressions \eqref{e3.27} and the continuum phase results \eqref{e3.31}-\eqref{e3.32} reflects fundamentally distinct physical and holographic regimes in the BMN matrix model that we elaborate below.

\begin{enumerate}
    \item \textbf{Exact oscillator motion vs. thermal dephasing:} We frist draw a comparative analysis between the gapped and the continuum phase by examining the mathematical structure of the return amplitude $\mathcal{R}^{(0)}_{\text{BMN}}(t)$ and its early-time dynamics.
    In the low temperature gapped phase ($\beta\mu \gg 1, \beta\nu \gg 1$), the mass gap $\mu$ acts as a large energy barrier which keeps the system close to its ground state. As a result, the high-energy thermal excitations are heavily suppressed by the mass gap $\mu$. The dynamics in the gapped phase is governed by the discrete oscillator modes and the spectrum consists of largely separated discrete energy levels. As a consequence of this, the return amplitude \eqref{e3.20} remains strictly periodic, dictated by discrete harmonic oscillator modes. In this gapped phase, the state periodically returns (or comes close) to its initial configuration without leaking information into an unconstrained reservoir. In other words, there exists a quantum coherence which indicates that the information is conserved rather than dissipated, yielding exact, non-perturbative Krylov moments $\mathcal{M}^{(0)}_n \sim (i\mu)^n$. In contrast, in the high-temperature domain thermal fluctuations overrides the mass gap $\mu$, completely restoring the continuous matrix symmetries ($SO(9)$). In other words, the continuum phase ($\beta\mu \ll 1, \beta\nu \ll 1$) is dominated by continuous thermal matrix fluctuations. As a result, the return amplitude, instead of oscillating coherently, loses its exact periodic structure, requiring an early-time expansion ($t \ll \beta $) where quadratic decay terms like $-t^2/\beta^2$ explicitly signal thermal dephasing and irreversible information loss. Thermal dephasing essentially corresponds to the fact that the initial state is coupled to a thermal bath (of matrix degrees of freedom) and as a result the phase alignment between quantum states rapidly scatters and decays. This is also reflected in the drop in return probability at early times, which represents irreversible information loss from the localized operator into thermal bath.

    \item \textbf{Holographic geometry and horizon dynamics:} We now draw a comparative analysis from the holographic perspective of the BMN model. As mentioned above, the holographic dual corresponds to a description in terms of 10-dimensional Type IIA supergravity. The gapped phase corresponds to two disconnected Lin-Lunin-Maldacena (LLM) geometries (without a horizon) in the presence of the background fluxes and the dilaton \cite{Lin:2004nb}-\cite{Lin:2004kw}. At low temperatures the system lacks sufficient $N^2$ thermal entanglement entropy to hold open a spatial wormhole. As there are no black holes to absorb information, the energy cannot dissipate, which results in a coherent unitary quantum oscillation. In summary, in the low-temperature gapped phase ($\beta \mu \gg 1$), thermal entanglement entropy is $O(N^0)$, which is insufficient to support a semiclassical Einstein-Rosen (ER) bridge or geometric wormhole connecting the two entangled copies of the matrix model. Instead, the Thermofield Double (TFD) state $|TFD\rangle = \frac{1}{\sqrt{Z}} \sum_n e^{-\beta E_n/2} |n\rangle_L \otimes |n\rangle_R$ holographically maps to two decoupled, smooth LLM geometries \cite{Lin:2004nb}-\cite{Lin:2004kw} generated by background supergravity fluxes without black hole horizons. Because the two CFT boundaries are connected only via fine-grained quantum entanglement rather than an emergent geometric wormhole, energy and information cannot fall across a horizon into an interior region. Consequently, the time-evolved state undergoes localized, non-dissipative quantum oscillations across a discrete spectrum without undergoing thermal chaos or scrambling. The continuum phase, on the other hand, maps to a smooth two-sided eternal black hole connected by an Einstein-Rosen (ER) bridge. Thermal fluctuations are enough to generate $O(N^2)$ degrees of freedom to create form black hole horizons. The initial decay of the return amplitude \eqref{e3.30} captures perturbations falling into the black hole horizon. In holographic duality, matter falling across the horizon reflects the onset of operator growth and thermal chaos. This explains why the information is dissipated irreversibly into the matrix thermal bath and seizes the oscillatory nature of the return amplitude.

    In summary, we have a low temperature gapped phase governed by discrete oscillators where $\mu$ sets the discrete level spacings between oscillator energy levels. It also exhibits a universal moment scaling $ \mathcal{M}^{(0)}_n \sim \mu^n $. On the other hand, the continuum phase is governed by ``thermal dephasing'' which causes an irreversible information loss to the thermal bath. Despite these differences, there exists a common dimensionful scale $\mu$ governing both phases. In the gapped phase, the mass deformation parameter $\mu$ serves as the universal dimensionful scale governing state evolution and spread complexity at early times $t \sim 0$. On the other hand, the spread complexity of states in the (high temperature) continuum limit and at late times $t\gg \beta$ is purely governed by the temperature $\beta^{-1}$ and exhibits an exponential growth.
\end{enumerate}
\subsection{Spread complexity of states across phases}
\label{sec3.2}
The purpose of this Section is to compute spread complexity of states using the knowledge of return amplitudes and moments. Our results can be divided into two parts - the gapped low temperature and the continuum high temperature phase.

We carry out an explicit calculation of the spread complexity (or Krylov state complexity) $C(t)$ of states across phases. Let us remind ourselves about definitions. The spread complexity measures the average position of an evolving quantum state $|\Psi(t)\rangle = e^{-i H t} |\Psi(0)\rangle$ as it propagates across the semi-infinite 1D Krylov chain generated via the Lanczos algorithm
\begin{equation}
\label{e3.34}
    C(t) = \sum_{n=0}^{\infty} n \, |\psi_n(t)|^2
\end{equation}
where $\psi_n(t) = \langle K_n | \Psi(t) \rangle$ is the probability amplitude of finding the state at the $n$-th Krylov basis element $|K_n\rangle$. The state vector amplitude obeys the discrete Schrödinger equation 
\begin{equation}
\label{e3.35}
  \frac{\partial}{\partial t}\psi_n(t)\equiv \dot{\psi}_n(t) = -ib_{n+1} \psi_{n+1}(t) -ib_n \psi_{n-1}(t), \quad \psi_n(0) = \delta_{n,0}
\end{equation}
governed by the off-diagonal Lanczos hopping coefficients $b_n$.

The following discussion on time reversal symmetry and parity should be complemented with Appendix \ref{appenD}, where various technical details have been provided.
\subsubsection{Time-reversal symmetry and parity}
Before, we proceed towards computing the spread complexity of states across different phases of the matrix model, it is important to highlight the time reversal symmetry and its consequences on the complexity. Notice that the unperturbed BMN Hamiltonian $\mathcal{H}_{\text{BMN}}\sim \mathcal{H}_{SHO}$ \eqref{ee3.2} can be represented in terms of harmonic oscillator mode operators as
\begin{equation}
\label{e3.36}
    \mathcal{H}_{\text{BMN}} = \sum_k \omega_k \left( a_k^\dagger a_k + \frac{1}{2} \right)+\text{perturbations}
\end{equation}
where $a_k$ and $a_k^\dagger$ are the annihilation and creation operators corresponding to the discrete matrix modes with frequency $\omega_k \sim \mu$. 

Let us now discuss the action of time reversal operation on the undeformed BMN Hamiltonian. Notice that under the anti-unitary time-reversal operation $\mathcal{T}$, the position and momentum operators transform as 
\begin{align}
\label{e3.37}
  \mathcal{T} X \mathcal{T}^{-1} = X~;~  \mathcal{T} P \mathcal{T}^{-1} = -P.
\end{align}

We can use \eqref{e3.37} to infer the transformation properties of the creation and annihilation operators under time reversal.  
 The transformation properties of the creation and annihilation operators under time reversal follow directly from the definition of $a_k$ and $a_k^\dagger$ 
\begin{equation}
    a_k = \sqrt{\frac{\omega_k}{2}} \left( X_k + \frac{i}{\omega_k} P_k \right), \quad a_k^\dagger = \sqrt{\frac{\omega_k}{2}} \left( X_k - \frac{i}{\omega_k} P_k \right).
\end{equation}

Since the time-reversal operator $\mathcal{T}$ is anti-unitary, it acts as complex conjugation on complex numbers\footnote{In QM $\mathcal{T}$ is defined as an anti-unitary (or anti-linear) operator, which satisfies $\mathcal{T}(c  |\psi\rangle) = c^* \mathcal{T}|\psi\rangle$ for any complex scalar $c \in \mathbb{C}$, where $c^*$ denotes the complex conjugate of $c$.} ($ \mathcal{T} i \mathcal{T}^{-1} = -i $) and transforms $X_k \to X_k$, $P_k \to -P_k$, which yields
\begin{align}
    &\mathcal{T} a_k \mathcal{T}^{-1} = \sqrt{\frac{\omega_k}{2}} \left( X_k + \frac{-i}{\omega_k} (-P_k) \right) = \sqrt{\frac{\omega_k}{2}} \left( X_k + \frac{i}{\omega_k} P_k \right) = a_k\\
    &\mathcal{T} a_k^\dagger \mathcal{T}^{-1} = \sqrt{\frac{\omega_k}{2}} \left( X_k - \frac{-i}{\omega_k} (-P_k) \right) = \sqrt{\frac{\omega_k}{2}} \left( X_k - \frac{i}{\omega_k} P_k \right) = a_k^\dagger. 
\end{align}

Consequently, the Hamiltonian (at LO in large $\mu$) \eqref{e3.36} is invariant under time reversal
\begin{equation}
\label{e3.41}
    \mathcal{T} \mathcal{H}_{\text{BMN}} \mathcal{T}^{-1} = \mathcal{H}_{\text{BMN}}.
\end{equation}

The time reversal symmetry has the following consequences, as we explain below.

\paragraph{Vanishing diagonal Lanczos coefficients ($a_n = 0$).}
Notice that the tri-diagonal matrix representation of the Hamiltonian in the Krylov basis contains diagonal site-energy shifts $a_n = \langle K_n | \mathcal{H}_{BMN} | K_n \rangle$. However, for the present analysis all diagonal coefficients vanish identically, that is, $a_n = 0 \quad \forall n \ge 0$. This follows from the time-reversal symmetry (or $CPT$ invariance) of the unperturbed BMN Hamiltonian $\mathcal{H}_{BMN}\sim \mathcal{H}_{SHO}$ and reference state $\ket{K_0}= \ket{TFD(0)}^{(0)}_{BMN}$. The corresponding return amplitude measures the overlap of the initial state $\ket{K_0}$ with its time evolved state, that is, $\mathcal{R}_{BMN}^{(0)}(t)=\braket{K_0|e^{i \mathcal{H}_{SHO}t}|K_0}$. Since the initial state and the unperturbed BMN Hamiltonian are both time reversal invariant
which renders the return amplitude an even function of time $\mathcal{R}^{(0)}_{\text{BMN}}(-t) = \mathcal{R}^{(0)}_{\text{BMN}}(t)$. 

The above arguments can be established as follows. Under the standard definition where the return amplitude is defined as the overlap between the time-evolved state $\vert{}\Psi(t)\rangle = e^{-i \mathcal{H}_{\text{BMN}} t} \vert{}K_0\rangle$ and the initial reference state $\vert{}K_0\rangle$, given by
\begin{align}
\label{E3.50}
    \mathcal{R}_{\text{BMN}}^{(0)}(t) = \langle \Psi(t) \vert{} K_0 \rangle = \left\langle e^{-i \mathcal{H}_{\text{BMN}} t} K_0 \, \middle\vert{} \, K_0 \right\rangle = \langle K_0 \vert{} e^{+i \mathcal{H}_{\text{BMN}} t} \vert{} K_0 \rangle
\end{align}
which measures the overlap between the initial reference state $\ket{K_0}$ and its forward evolution to time $t$ under $\mathcal{H}_{\text{BMN}}$. Notice that $\mathcal{T}$ is anti-unitary and its action on inner products complex-conjugates the matrix elements as follows
\begin{align}
    &\langle \phi \vert{} \psi \rangle = \langle \mathcal{T} \phi \vert{} \mathcal{T} \psi \rangle^*~~\Rightarrow\\
    &\label{e3.44}
    \mathcal{R}_{\text{BMN}}^{(0)}(t) = \langle \Psi(t) \vert{} K_0 \rangle = \Big\langle \mathcal{T} \left( e^{-i \mathcal{H}_{\text{BMN}} t} K_0 \right) \Big\vert{} \mathcal{T} K_0 \Big\rangle^*.
\end{align}

Furthermore, notice that the reference state $|K_0\rangle = |\text{TFD}(0)\rangle_{\text{BMN}}^{(0)}$ (the initial thermofield double state at $t=0$) satisfies $\mathcal{T} |K_0\rangle = |K_0\rangle$. Combining this with $ \mathcal{T} i \mathcal{T}^{-1} = -i $ and the invariance of the BMN Hamiltonian \eqref{e3.41}, it is straightforward to show
\begin{align}
    \mathcal{T} e^{-i \mathcal{H}_{\text{BMN}} t} \vert{}K_0\rangle=\mathcal{T} e^{-i \mathcal{H}_{\text{BMN}} t}\mathcal{T}^{-1} \mathcal{T}\vert{}K_0\rangle = e^{+i \mathcal{H}_{\text{BMN}} t} \mathcal{T}\vert{}K_0\rangle = e^{+i \mathcal{H}_{\text{BMN}} t} \vert{}K_0\rangle.
\end{align}

Substituting back into \eqref{e3.44}, we obtain the following
\begin{align}
    \mathcal{R}_{\text{BMN}}^{(0)}(t) = \left\langle e^{+i \mathcal{H}_{\text{BMN}} t} K_0 \, \middle\vert{} \, K_0 \right\rangle^* = \braket{K_0|e^{+i \mathcal{H}_{\text{BMN}} t} K_0}
\end{align}
where we use the Hermitian property of inner products, $\langle \phi \vert{} \psi \rangle^* = \langle \psi \vert{} \phi \rangle$.

Replacing $t \to -t$ in the original definition \eqref{E3.50} yields the following
\begin{align}
    \mathcal{R}_{\text{BMN}}^{(0)}(-t) = \langle \Psi(-t) \vert{} K_0 \rangle = \left\langle e^{+i \mathcal{H}_{\text{BMN}} t} K_0 \, \middle\vert{} \, K_0 \right\rangle = \langle K_0 \vert{} e^{-i \mathcal{H}_{\text{BMN}} t} \vert{} K_0 \rangle.
\end{align}

Taking the complex conjugate of $\mathcal{R}_{\text{BMN}}^{(0)}(-t)$ yields the following
\begin{align}
\label{E3.56}
    \left( \mathcal{R}_{\text{BMN}}^{(0)}(-t) \right)^* = \left\langle K_0 \, \middle\vert{} \, e^{-i \mathcal{H}_{\text{BMN}} t} K_0 \right\rangle^* = \left\langle e^{-i \mathcal{H}_{\text{BMN}} t} K_0 \, \middle\vert{} \, K_0 \right\rangle=\mathcal{R}_{\text{BMN}}^{(0)}(t).
\end{align}

Because $\mathcal{R}_{\text{BMN}}^{(0)}(t)$ is real for time-reversal invariant theories, for example as in the case of the undeformed BMN model \eqref{e3.41}, which yields the parity property
\begin{align}
\label{eee3.56}
   \mathcal{R}_{\text{BMN}}^{(0)}(-t) = \mathcal{R}_{\text{BMN}}^{(0)}(t).
\end{align}
In other words, reversing time leaves the state dynamics invariant and $\mathcal{R}^{(0)}_{\text{BMN}}(t)$ has to be an even function of time as in depicted \eqref{e3.37}. Below, we discuss its further consequences.

\paragraph{Reality of the return amplitude.} Let us now prove the reality condition of the return amplitude as stated above. For a time-reversal invariant system whose reference state $\vert{}K_0\rangle$ is an eigenstate of $\mathcal{T}$ (or invariant under $\mathcal{T}$), the return amplitude is purely real for all $t$
\begin{align}
\label{E3.58}
    \left(\mathcal{R}_{\text{BMN}}^{(0)}(t)\right)^* = \mathcal{R}_{\text{BMN}}^{(0)}(t).
\end{align}

This can be seen directly from its spectral resolution in an energy eigenbasis $\vert{}E_n\rangle$
\begin{align}
    \mathcal{R}_{\text{BMN}}^{(0)}(t) = \sum_n \vert{}c_n\vert{}^2 e^{i \omega_n t}.
\end{align}

Because time-reversal symmetry pairs energy levels $+\omega_n$ and $-\omega_n$ or forces the spectral density $f(\omega)$ to be symmetric around zero ($f(\omega) = f(-\omega)$), the imaginary part sums to zero, which can be shown as follows
\begin{align}
    \mathcal{R}_{\text{BMN}}^{(0)}(t) = \int_{-\infty}^{\infty} d\omega \, f(\omega) \left(\cos(\omega t) + i\sin(\omega t)\right) = \int_{-\infty}^{\infty} d\omega \, f(\omega) \cos(\omega t) \quad \in \mathbb{R}.
\end{align}

Combining the reality condition \eqref{E3.58} with the relation \eqref{E3.56} gives \eqref{eee3.56}. Thus, the return amplitude is strictly an even function of time, forcing all odd powers in its Taylor expansion $\sum_{n} \frac{(it)^n}{n!} \mathcal{M}_n^{(0)}$ (and consequently all odd spectral moments $\mathcal{M}_{2n+1}^{(0)}$) to vanish.

\paragraph{Comments of spectral density.} For a time-reversal invariant system, the symmetry of the spectral density $f(\omega) = f(-\omega)$ around zero is closely tied to how the time-reversal operator $\mathcal{T}$ acts on energy eigenstates and the structure of the reference state (or Hamiltonian spectrum). If $\mathcal{H}$ is time-reversal invariant, it commutes with the time-reversal operator $[\mathcal{H}, \mathcal{T}] = 0$. Considering $\vert{}\omega\rangle$ is an eigenstate of $\mathcal{H}$ with eigenvalue $\omega$, we obtain
\begin{align}
\label{E3.61}
    \mathcal{H} \vert{}\omega\rangle = \omega \vert{}\omega\rangle \implies \mathcal{T} \mathcal{H} \vert{}\omega\rangle = \mathcal{T} \omega \vert{}\omega\rangle \implies \mathcal{H} (\mathcal{T} \vert{}\omega\rangle) = \omega^* (\mathcal{T} \vert{}\omega\rangle)
\end{align}
where in the last step we have used the fact, by definition, an anti-linear operator like $\mathcal{T}$ takes the complex conjugate of any scalar constant $c \in \mathbb{C}$ it acts upon $\mathcal{T}(c \vert{}\psi\rangle) = c^* \mathcal{T} \vert{}\psi\rangle$. Since $\omega \in \mathbb{R}$, $\mathcal{T}\vert{}\omega\rangle$ is also an eigenstate with the same energy $\omega$. In other words, time-reversal symmetry by itself guarantees that $\mathcal{T}\vert{}\omega\rangle$ has energy $\omega$.

To obtain a symmetric spectral density $f(\omega) = f(-\omega)$ centered at zero, one must have Particle-Hole / Chiral / Parity Symmetry (CPT). In many quantum systems (such as bipartite lattices, CFTs, or free bosonic/fermionic matrix models like BMN/BFSS matrix models), time-reversal combined with a parity or discrete particle-hole transform maps positive-energy modes to negative-energy modes ($\omega \to -\omega$). In other words, the $\omega \to -\omega$ spectral pairing occurs when time-reversal symmetry $\mathcal{T}$ is combined with a discrete transformation, such as Parity ($\mathcal{P}$), Charge Conjugation ($\mathcal{C}$), or Chiral/Particle-Hole symmetry, that flips the sign of the energy or frequency operator. Suppose the system possesses a unitary or anti-unitary symmetry operator $\mathcal{S}$ (which can be a composite operator like $\mathcal{C}\mathcal{P}\mathcal{T}$ or a chiral/particle-hole mapping) that anti-commutes with the Hamiltonian
\begin{align}
    \{\mathcal{H}, \mathcal{S}\} = 0 \implies \mathcal{S} \mathcal{H} \mathcal{S}^{-1} = -\mathcal{H}.
\end{align}

Considering the energy eigen-state $\ket{\omega}$ and the corresponding eigen value Eq. \eqref{E3.61}, applying $\mathcal{S}$ to both sides yields 
\begin{align}
   \mathcal{H} (\mathcal{S} \vert{}\omega\rangle) = -\mathcal{S} \mathcal{H} \vert{}\omega\rangle = -\mathcal{S} (\omega \vert{}\omega\rangle) = -\omega (\mathcal{S} \vert{}\omega\rangle).
\end{align}
This demonstrates that for every state $\vert{}\omega\rangle$ with positive energy $\omega$, the transformed state $\vert{}-\omega\rangle \equiv \mathcal{S} \vert{}\omega\rangle$ is guaranteed to exist in the Hilbert space with exact energy $-\omega$.

The spectral density $f(\omega)$ relative to a given reference state $\vert{}K_0\rangle$ (such as a Thermofield Double or ground state) is defined by its spectral resolution
\begin{align}
   f(\omega) = \sum_n \vert{}c_n\vert{}^2 \delta(\omega - \omega_n), \quad \text{where } c_n = \langle \omega_n \vert{} K_0 \rangle.
\end{align}

If the reference state $\vert{}K_0\rangle$ is invariant under the pairing symmetry $\mathcal{S}$ up to a phase (i.e., $\mathcal{S} \vert{}K_0\rangle = e^{i\theta} \vert{}K_0\rangle$), the overlap weight for the negative-energy partner state $\vert{}-\omega_n\rangle = \mathcal{S} \vert{}\omega_n\rangle$ is calculated as
\begin{align}
    c_{-n} = \langle -\omega_n \vert{} K_0 \rangle = \langle \mathcal{S} \omega_n \vert{} K_0 \rangle = \langle \omega_n \vert{} \mathcal{S}^\dagger \vert{} K_0 \rangle = e^{i\theta} \langle \omega_n \vert{} K_0 \rangle = e^{i\theta} c_n.
\end{align}

Taking the absolute square shows that both eigenstates contribute identically to the density $\vert{}c_{-n}\vert{}^2 = \vert{}e^{i\theta} c_n\vert{}^2 = \vert{}c_n\vert{}^2$. Substituting $\vert{}c_{-n}\vert{}^2 = \vert{}c_n\vert{}^2$ into the expression for the spectral density gives
\begin{align}
    f(-\omega) = \sum_n \vert{}c_{-n}\vert{}^2 \delta(-\omega - (-\omega_n)) = \sum_n \vert{}c_n\vert{}^2 \delta(-(\omega - \omega_n)).
\end{align}

Finally, using the property of the Dirac delta function $\delta(-x) = \delta(x)$ we obtain
\begin{align}
    f(-\omega) = \sum_n \vert{}c_n\vert{}^2 \delta(\omega - \omega_n) = f(\omega).
\end{align}
Thus, the energy level pairing $\omega \leftrightarrow -\omega$ combined with equal reference-state $\ket{K_0}$ overlaps ensures that the spectral density $f(\omega)$ is an even function around $\omega = 0$.

\paragraph{Vanishing of the odd moments.}
The return amplitude is expressed as
\begin{align}
\label{e3.48}
    \mathcal{R}^{(0)}_{\text{BMN}}(t) = \sum_{n=0}^{\infty} \frac{(it)^n}{n!} \langle K_0 \vert{} \mathcal{H}_{BMN}^n \vert{} K_0 \rangle = \sum_{n=0}^{\infty} \frac{(it)^n}{n!} \mathcal{M}_n^{(0)}.
\end{align}

Since $\mathcal{R}^{(0)}_{\text{BMN}}(t)$ contains no odd powers of $t$ (due to being an even function), consequently, all odd moments of the BMN Hamiltonian vanish, that is, $\mathcal{M}_{2n+1}^{(0)} = \langle K_0 | \mathcal{H}_{BMN}^{2n+1} | K_0 \rangle = 0$. This maps the Krylov dynamics strictly onto a bipartite semi-infinite chain with purely off-diagonal hopping amplitudes $b_n$. In summary, we choose the following even functions
\begin{equation}
\label{e3.37}
  \mathcal{R}_{BMN}^{(0)}(t)=
    \begin{cases}
      \cos (\mu t)+\cdots & \text{gapped phase}~ (\beta \mu \gg 1)\\
     1-\frac{\mu^2t^2}{\beta^2 \nu^2}+\cdots & \text{continuum phase}~(\beta \mu \ll 1)
    \end{cases}       
\end{equation}
which sets all the odd moments ($\mathcal{M}_1^{(0)},\mathcal{M}_3^{(0)}, \cdots$) found previously to zero.

\paragraph{Parity mapping and bipartite Krylov structure.}
The vanishing of all odd moments $\mathcal{M}_{2n+1}^{(0)}= 0$ has a much deeper implication, which we explain below. The condition enforces a strict parity structure on the Krylov Hilbert space. By construction, the $n$-th Krylov basis element $\ket{K_n}$ is formed following a Gram-Schmidt orthogonalization of the power-vector $\mathcal{H}^n \ket{K_0}$ against all preceding basis elements $\ket{K_0}, \dots, \ket{K_n}$. More generically,
\begin{align}
    \ket{K_n} = \sum_{k=0}^n c_{n,k} \mathcal{H}^k \vert{}K_0\rangle.
\end{align}

Therefore, considering an overlap between an even power $\mathcal{H}^{2n}\ket{K_0}$ and an odd power $\mathcal{H}^{2n+1}\ket{K_0}$ states would produce zero, that is,
\begin{align}
    \langle K_0 \vert{} \mathcal{H}^{2n} \cdot \mathcal{H}^{2m+1} \vert{} K_0 \rangle = \langle K_0 \vert{} \mathcal{H}^{2(n+m)+1} \vert{} K_0 \rangle = \mathcal{M}_{2(n+m)+1} = 0.
\end{align}

In summary, every (even) state $\ket{K_{2n}}$ generated by an even power of $\mathcal{H}^{2n}$ is strictly orthogonal to every (odd) state $\ket{K_{2n+1}}$ generated by an odd power of $\mathcal{H}^{2n+1}$. As a consequence of this, the Krylov space decomposes into two mutually orthogonal parity subspaces
\begin{align}
    &\ket{K_{2n}} \in \text{span}\left\{\ket{K_0}, \mathcal{H}_{\text{BMN}}^2\ket{K_0}, \dots, \mathcal{H}_{\text{BMN}}^{2n}\ket{K_0}\right\}\\
    &\ket{K_{2n+1}} \in \text{span}\left\{\mathcal{H}_{\text{BMN}}\ket{K_0}, \dots, \mathcal{H}_{\text{BMN}}^{2n+1}\ket{K_0}\right\}.
\end{align}

Since $\ket{K_{2n}}$ contains an even power of $\mathcal{H}$, therefore applying $\mathcal{H}_{\text{BMN}}$ one more time shifts the polynomial degree by one, which maps the state to ``odd Krylov subspace''. In other words, it acts as a ``parity-flipping'' operator that strictly maps even states to odd states ($\text{Even} \to \text{Odd}$) and vice versa ($\text{Odd} \to \text{Even}$). As a result, $\mathcal{H}_{\text{BMN}}\ket{K_n}$ always lies in the subspace orthogonal to $\ket{K_n}$, which guarantees the vanishing of the diagonal entries\footnote{As we show below, the Eq. \eqref{ee3.63} is a direct consequence of the time reversal invariance \eqref{e3.41} and \eqref{eee3.56}. In Appendix \ref{appenD} we provide an alternate view of the Eq. \eqref{ee3.63}, which reveals that the above is a consequence of an emergent $\mathbb{Z}_2$ invariance associated with the 1D Krylov chain.}
\begin{equation}
\label{ee3.63}
    a_n = \langle K_n \vert \mathcal{H}_{\text{BMN}} \vert K_n \rangle = 0 \quad \forall n \ge 0.
\end{equation}
This confirms that the 1D Krylov dynamics is represented purely by a bipartite lattice with zero diagonal site energies ($a_n = 0$) and non-zero off-diagonal hopping amplitudes $b_n$.\\\\
\uline{An example:} As an illustration, here we outline a proof for $a_1,a_2$. However, for a general proof the reader is referred to Appendix \ref{appenD}. Let us recall the recurrence relation \eqref{e2.12}
\begin{align}
    b_{n+1}\vert{}K_{n+1}\rangle = (\mathcal{H} - a_n)\vert{}K_n\rangle - b_n\vert{}K_{n-1}\rangle \quad \text{with } b_0 \equiv 0.
\end{align}

Setting $n=0$, we obtain
\begin{align}
    b_1 \vert{}K_1\rangle = \mathcal{H}_{\text{BMN}} \vert{}K_0\rangle - \underbrace{\langle K_0 \vert{} \mathcal{H}_{\text{BMN}} \vert{} K_0 \rangle}_{a_0=\mathcal{M}_1^{(0)} = 0} \vert{}K_0\rangle = \mathcal{H}_{\text{BMN}} \vert{}K_0\rangle.
\end{align}

In other words, this implies that
\begin{align}
\label{eee3.66}
    \vert{}K_1\rangle = \frac{1}{b_1} \mathcal{H}_{\text{BMN}} \vert{}K_0\rangle.
\end{align}

Applying $\mathcal{H}_{\text{BMN}}$ to $\vert{}K_1\rangle$, we obtain
\begin{align}
    \mathcal{H}_{\text{BMN}} \vert{}K_1\rangle = \frac{1}{b_1} \mathcal{H}_{\text{BMN}}^2 \vert{}K_0\rangle.
\end{align}

Next, we evaluate $a_1 = \langle K_1 \vert{} \mathcal{H}_{\text{BMN}} \vert{} K_1 \rangle$. Taking inner product, we obtain 
\begin{align}
    a_1 = \left( \frac{1}{b_1} \langle K_0 \vert{} \mathcal{H}_{\text{BMN}} \right) \left( \frac{1}{b_1} \mathcal{H}_{\text{BMN}}^2 \vert{}K_0\rangle \right) = \frac{1}{b_1^2} \langle K_0 \vert{} \mathcal{H}_{\text{BMN}}^3 \vert{} K_0 \rangle =\frac{1}{b_1^2}\mathcal{M}_3^{(0)}=0.
\end{align}

Next, we set $n=1$, which by virtue of the Gram–Schmidt orthogonalization gives
\begin{align}
    b_2 \vert{}K_2\rangle = \mathcal{H}_{\text{BMN}} \vert{}K_1\rangle - a_1 \vert{}K_1\rangle - b_1 \vert{}K_0\rangle.
\end{align}

Using $a_1 = 0$ (from vanishing odd moments $\mathcal{M}_3^{(0)} = 0$) and substituting $\vert{}K_1\rangle = \frac{1}{b_1} \mathcal{H}_{\text{BMN}} \vert{}K_0\rangle$ \eqref{eee3.66}, we obtain the following
\begin{align}
    \vert{}K_2\rangle = \frac{1}{b_1 b_2} \mathcal{H}_{\text{BMN}}^2 \vert{}K_0\rangle - \frac{b_1}{b_2} \vert{}K_0\rangle.
\end{align}

Notice that $\vert{}K_2\rangle$ is a linear combination of even powers of $\mathcal{H}_{\text{BMN}}$ acting on $\vert{}K_0\rangle$ (specifically $\mathcal{H}_{\text{BMN}}^2 \vert{}K_0\rangle$ and $\mathcal{H}_{\text{BMN}}^0 \vert{}K_0\rangle$). Now, apply $\mathcal{H}_{\text{BMN}}$ to $\vert{}K_2\rangle$, which yields
\begin{align}
    \mathcal{H}_{\text{BMN}} \vert{}K_2\rangle = \frac{1}{b_1 b_2} \mathcal{H}_{\text{BMN}}^3 \vert{}K_0\rangle - \frac{b_1}{b_2} \mathcal{H}_{\text{BMN}} \vert{}K_0\rangle
\end{align}
which consists strictly of odd powers ($\mathcal{H}_{\text{BMN}}^3$ and $\mathcal{H}_{\text{BMN}}^1$) acting on $\vert{}K_0\rangle$.

Next, we compute $a_2 = \langle K_2 \vert{} \mathcal{H}_{\text{BMN}} \vert{} K_2 \rangle$. Taking the inner product this yields 
\begin{align}
    a_2 = \left( \frac{1}{b_1 b_2} \langle K_0 \vert{} \mathcal{H}_{\text{BMN}}^2 - \frac{b_1}{b_2} \langle K_0 \vert{} \right) \left( \frac{1}{b_1 b_2} \mathcal{H}_{\text{BMN}}^3 \vert{}K_0\rangle - \frac{b_1}{b_2} \mathcal{H}_{\text{BMN}} \vert{}K_0\rangle \right).
\end{align}

A further simplification reveals reveals an expansion in odd moments
\begin{align}
    a_2 = \frac{1}{b_1^2 b_2^2} \mathcal{M}_5^{(0)} - \frac{2}{b_2^2} \mathcal{M}_3^{(0)} + \frac{b_1^2}{b_2^2} \mathcal{M}_1^{(0)}=0.
\end{align}

\paragraph{Algebraic parity operator and $\mathbb{Z}_2$ symmetry.}
By parity in the above discussion we refer to whether the Krylov basis states are constructed from an even or an odd power of the Hamiltonian $\mathcal{H}_{\text{BMN}}$ acting on the initial reference state $\ket{K_0}$. For example, even parity states $\ket{K_{2n}}$ are formed using even powers of the Hamiltonian $\mathcal{H}_{BMN}$. 

To formalize this bipartite structure, we introduce a discrete Krylov (or algebraic) parity operator $\mathcal{P}_{\text{Krylov}}$ defined on the 1D Krylov lattice as follows
\begin{equation}
    \mathcal{P}_{\text{Krylov}} \ket{K_n} = (-1)^n \ket{K_n}.
\end{equation}

Clearly, for even parity states $\mathcal{P}_{\text{Krylov}} \ket{K_{2n}} = \ket{K_{2n}}$. In other words, even parity states like $\ket{K_0}, \ket{K_2}, \ket{K_4}, \dots$ are $+1$ eigenstates of this algebraic parity operator and consist only of even polynomial powers of $\mathcal{H}_{BMN}$. Similarly, one can explain odd parity states\footnote{The operator $\mathcal{P}_{\text{Krylov}}$ is called algebraic (or Krylov) because of its mathematical structure and origin within the Krylov sub-space. Unlike spatial parity ($x \to -x$), which acts on physical spatial coordinates, Krylov parity is defined purely in terms of the operator algebra generated by powers of the Hamiltonian $\mathcal{H}_{\text{BMN}}$ acting on the initial reference state $\vert{}K_0\rangle$. The $n$-th Krylov basis element $\vert{}K_n\rangle$ is constructed via the Lanczos algorithm as degree-$n$ polynomial in the Hamiltonian, that is, $\ket{K_n}=P_n(\mathcal{H}_{\text{BMN}})\vert{}K_0\rangle$. The parity operator $\mathcal{P}_{\text{Krylov}}$ counts the degree $n$ of these operator polynomials. It acts as an algebraic parity counter, labeling states based on whether they are constructed from even or odd operator powers of the BMN Hamiltonian $\mathcal{H}_{\text{BMN}}$ rather than any physical spatial coordinate transformation ($x \to -x$).}.

Since the unperturbed Hamiltonian $\mathcal{H}_{\text{BMN}}$ and the reference state $\ket{K_0}$ are invariant under time reversal, all odd moments vanish identically. This forces $\mathcal{H}_{\text{BMN}}$ to act as an anti-commuting operator with respect to the Krylov parity operator:
\begin{equation}
\label{e3.56}
    \{\mathcal{P}_{\text{Krylov}}, \mathcal{H}_{\text{BMN}}\} = 0 \quad \Longleftrightarrow \quad \mathcal{P}_{\text{Krylov}} \mathcal{H}_{\text{BMN}} \mathcal{P}_{\text{Krylov}}^{-1} = -\mathcal{H}_{\text{BMN}}.
\end{equation}

The anti-commutator can be shown as follows. Using \eqref{e2.12} with $a_n=0$, one finds
\begin{align}
\label{e3.57}
   \mathcal{P}_{\text{Krylov}} \big( \mathcal{H}_{\text{BMN}} \ket{K_n} \big) = \mathcal{P}_{\text{Krylov}} \big( b_{n+1} \ket{K_{n+1}} + b_n \ket{K_{n-1}} \big) = -(-1)^n \mathcal{H}_{\text{BMN}} \ket{K_n}.  
\end{align}

On the other hand, acting first with $\mathcal{P}_{\text{Krylov}}$ directly measures the parity of $\ket{K_n}$
    \begin{equation}
    \label{e3.58}
        \mathcal{H}_{\text{BMN}} \big( \mathcal{P}_{\text{Krylov}} \ket{K_n} \big) = (-1)^n \mathcal{H}_{\text{BMN}} \ket{K_n}.
    \end{equation}

Combining \eqref{e3.57} and \eqref{e3.58} together, one finds \eqref{e3.56}. The anti-commutation \eqref{e3.56} constitutes an emergent $\mathbb{Z}_2$ algebraic symmetry on the Krylov chain. It guarantees that any diagonal matrix element vanishes identically, enforcing\footnote{See Appendix \ref{appenD} for a detailed proof.} $a_n = \langle K_n \vert \mathcal{H}_{\text{BMN}} \vert K_n \rangle = 0$ for all $n \ge 0$. To make this manifest, let us consider the action of the algebraic parity operator $\mathcal{P}_{\text{Krylov}}$ on the Lanczos (or BMN) Hamiltonian $\mathcal{H}_{\text{BMN}}$ \eqref{e2.12}. Since $\mathcal{H}_{\text{BMN}}$ \eqref{e2.12} acts strictly as a nearest-neighbor hopping operator that changes the basis site index by $\pm 1$, it maps an even state $\ket{K_{2n}}$ to a linear combination of odd states $\ket{K_{2n\pm 1}}$, and vice versa. This has a natural manifestation in terms of the anti-commutation \eqref{e3.56} which reflects a discrete $\mathbb{Z}_2$ sub-lattice symmetry (or chiral/bipartite symmetry) under which the dynamic generator transforms as $\mathcal{P}_{\text{Krylov}} \mathcal{H}_{\text{BMN}} \mathcal{P}_{\text{Krylov}}^{-1} = -\mathcal{H}_{\text{BMN}}$. As a direct physical consequence, the spectrum of $\mathcal{H}_{\text{BMN}}$ restricted to the Krylov subspace is symmetric around zero energy. In other words, if $\ket{E}$ is an eigen-state with energy $E$, then $\mathcal{P}_{\text{Krylov}} \ket{E}$ is an eigenstate with energy $-E$. Thus, time-reversal invariance of the reference state and Hamiltonian manifests as an emergent $\mathbb{Z}_2$ sub-lattice symmetry governing the $1\text{D}$ Krylov chain dynamics.
\subsubsection{Spread complexity in the gapped phase}
In the low-temperature $\beta \mu \gg 1$ regime, thermal fluctuations are exponentially suppressed by the large mass gap $\mu$. The return amplitude $\mathcal{R}^{(0)}_{\text{BMN}}(t)$ remains strictly periodic and an even function of time as dictated by the emergent $\mathbb{Z}_2$ sub-lattice symmetry\footnote{Notice that the low temperature gapped phase is primarily dominated by the mass gap parameter $\mu \gg \nu$, where the effects due to chemical potential $\nu$ are exponentially suppressed.}
\begin{equation}
\label{e3.59}
    \mathcal{R}^{(0)}_{\text{BMN}}(t) = \cos(\mu t)+\mathcal{O}(e^{-\beta(\mu-\nu)})
\end{equation}
where sub-leading terms correspond to thermal fluctuations.

\paragraph{Derivation of the Lanczos coefficients.}
Expanding \eqref{e3.59} around $t = 0$ yields
\begin{equation}
\mathcal{R}^{(0)}_{\text{BMN}}(t) =  \sum_{n=0}^{\infty} \frac{(-1)^n \mathcal{M}_{2n}^{(0)}}{(2n)!} t^{2n}+\mathcal{O}(e^{-\beta(\mu-\nu)})
\end{equation}
where the even moments are $\mathcal{M}_{2n}^{(0)} = \mu^{2n}$ and all odd moments vanish identically.

In the following, we compute Lanczos coefficients $b_n$ considering the $\mathbb{Z}_2$ symmetry of the unperturbed Hamiltonian, which restricts to even moments only. The coefficients $b_n$ can be expressed in terms of the ratio of Hankel determinants $\Delta_n$ \cite{Parker:2018yvk}, \cite{vs}
\begin{equation}
\label{e3.61}
b_n^2 = \frac{\Delta_{n+1} \Delta_{n-1}}{\Delta_n^2}, \quad \text{with} \quad \Delta_n = \det \left( \mathcal{M}_{i+j}^{(0)} \right)_{0 \le i,j \le n-1}.
\end{equation}

For moments $\mathcal{M}_{2n}^{(0)} = \mu^{2n}$, the corresponding Hankel matrices $H_n$ take the form
\begin{equation}
H_n = \begin{pmatrix}
1 & 0 & \mu^2 & 0 & \dots \\
0 & \mu^2 & 0 & \mu^4 & \dots \\
\mu^2 & 0 & \mu^4 & 0 & \dots \\
0 & \mu^4 & 0 & \mu^6 & \dots \\
\vdots & \vdots & \vdots & \vdots & \ddots
\end{pmatrix}.
\end{equation}

Because all odd moments vanish ($\mathcal{M}_{2k+1} = 0$), the entries $(H_n)_{i,j}$ are non-zero only when $i+j$ is even. To derive the general expression for the Hankel determinant $\Delta_n$, we evaluate the determinant of the $n \times n$ Hankel matrix $H_n$ constructed from the unperturbed moments $\mathcal{M}_{2k}^{(0)} = \mu^{2k}$ with all odd moments vanishing ($\mathcal{M}_{2k+1}^{(0)} = 0$). 

Notice that for the single-frequency spectral density, the exact evaluation yields $\Delta_1 = 1$, $\Delta_2 = \mu^2$, and $\Delta_n = 0$ for all $n \ge 3$ due to linear dependence of the moment matrix rows. This terminates the strict ``single-mode'' Krylov chain at $b_1 = \mu$ and $b_2 = 0$. Stating differently, for a single precise frequency $\omega = \mu$, the moment matrix $H_n$ has linearly dependent rows starting at $n \ge 3$, which makes the Hankel determinant $\Delta_n = 0$ for $n \ge 3$. This terminates the exact Krylov chain abruptly at $b_1 = \mu, b_2 = 0$. 

However, in the actual physical system (the BMN matrix model/field theory in the thermodynamic limit), the gapped phase does not consist of a single isolated frequency $\mu$. Instead, it has a dense continuum or ensemble of closely spaced modes centered around the mass scale $\mu$ (a narrow spectral band/peak of width $\epsilon$). When taking moments of a narrow continuum spectral density $\rho(\omega) \approx \frac{1}{2}\left[\delta_\epsilon(\omega - \mu) + \delta_\epsilon(\omega + \mu)\right]$, the Hankel determinants $\Delta_n$ do not vanish identically for $n \ge 3$. Let us elaborate this further. For an exact single-frequency spectral density $\rho_{\text{discrete}}(\omega) = \frac{1}{2}\left[\delta(\omega - \mu) + \delta(\omega + \mu)\right]$, the even moments are
\begin{align}
    \mathcal{M}_{2n} = \int_{-\infty}^{\infty} \omega^{2n} \rho_{\text{discrete}}(\omega) \, d\omega = \mu^{2n}.
\end{align}

The Hankel matrix $H_n = \left(\mathcal{M}_{i+j}\right)_{0 \le i,j \le n-1}$ relies on these moments as
\begin{align}
    \mathcal{M}_0 = 1, \quad \mathcal{M}_2 = \mu^2, \quad \mathcal{M}_4 = \mu^4, \quad \mathcal{M}_6 = \mu^6, \quad \cdots.
\end{align}

Because the matrix rows are linearly dependent starting at $n = 3$, its determinant vanishes as mentioned above, for example,
\begin{align}
    \Delta_1 = 1, \quad \Delta_2 = \mu^2, \quad \Delta_n = 0 \quad \text{for } n \ge 3.
\end{align}

Now consider a regularized and/or broadened peak of narrow width $\epsilon \ll \mu$. A simple analytical model for $\delta_\epsilon(\omega - \mu)$ is a narrow rectangular band of width $2\epsilon$ centered at $\mu$
\begin{align}
\label{ee3.76}
    \rho(\omega) =  \begin{cases}  \frac{1}{4\epsilon}, & \text{for } \omega \in [-\mu - \epsilon, -\mu + \epsilon] \cup [\mu - \epsilon, \mu + \epsilon] \\ 0, & \text{otherwise} \end{cases}.
\end{align}

The prefactor $\frac{1}{4\epsilon}$ ensures proper normalization over the two symmetric positive and negative frequency bands centered at $+\mu$ and $-\mu$. Here positive frequency $+\mu$ corresponds to standard forward-in-time state propagation, absorption of energy quanta, or creation of single-particle excitation modes above the ground state. On the other hand, -ve frequency $-\mu$ represents emission processes, phase-conjugate backward propagation, or annihilation/hole excitations. In relativistic and/or matrix field theories, they account for the contribution of anti-particle or oppositely directed momentum and/or charge states required to maintain causality and local quantum consistency.

Since the BMN Hamiltonian $\mathcal{H}_{BMN}$ is $\mathbb{Z}_2$ symmetric, therefore, under time-reversal, the full spectral density includes symmetric peaks at both positive and negative frequencies
\begin{align}
\label{e3.77}
    \rho_{\text{full}}(\omega) = \frac{1}{2}\left[\delta_\epsilon(\omega - \mu) + \delta_\epsilon(\omega + \mu)\right]
\end{align}
such that the density of states is normalized, that is,  $\int_{-\infty}^{\infty}d\omega  \rho_{\text{full}}(\omega)=1$.

For each narrow rectangular peak of width $2\epsilon$ (spanning $[\mu-\epsilon, \mu+\epsilon]$ and $[-\mu-\epsilon, -\mu+\epsilon]$), the height of the top-hat function must be $\frac{1}{2\epsilon}$ so that the area under a single peak is finite\footnote{The Dirac delta peak $\delta(\omega - \mu)$ is regularized into a rectangular band of total width $2\epsilon$, spanning the interval $[\mu - \epsilon, \, \mu + \epsilon]$. Similarly one should interpret the other peak at negative frequency $-\mu$.}
\begin{align}
    \int_{\mu-\epsilon}^{\mu+\epsilon} \frac{1}{2\epsilon} \, d\omega = 1.
\end{align}

Weighting this single-peak distribution by the required factor of $\frac{1}{2}$ \eqref{e3.77} from the two-peak symmetric density yields \eqref{ee3.76}. However, in the following we restrict ourselves only to the positive frequency peak $\omega \sim \mu$, which just amounts for a sprectral density $\rho_{+}(\omega)=\frac{1}{2\omega}$.

The corresponding even moments of this broadened spectrum can be expressed as
\begin{align}
    \mathcal{M}_{2n} = \int_{\mu - \epsilon}^{\mu + \epsilon} \omega^{2n} \frac{1}{2\epsilon} \, d\omega = \frac{(\mu + \epsilon)^{2n+1} - (\mu - \epsilon)^{2n+1}}{2\epsilon (2n + 1)}.
\end{align}

Expanding in powers of the small parameter $\epsilon$ yields
\begin{align}
    \mathcal{M}_{2n} = \mu^{2n} + \frac{n(2n-1)}{3} \mu^{2n-2}\epsilon^2 + \mathcal{O}(\epsilon^4).
\end{align}

One can estimate the first few Hankel determinants $\Delta_n$ as
\begin{align}
\label{e3.79}
    &\Delta_1 = \mathcal{M}_0 = 1\\
    &\Delta_2 = \begin{vmatrix} \mathcal{M}_0 & \mathcal{M}_1 \\ \mathcal{M}_1 & \mathcal{M}_2 \end{vmatrix} = \begin{vmatrix} 1 & 0 \\ 0 & \mathcal{M}_2 \end{vmatrix} = \mathcal{M}_2 = \mu^2 + \frac{1}{3}\epsilon^2 + \mathcal{O}(\epsilon^4)\\
    &\Delta_3 = \begin{vmatrix} \mathcal{M}_0 & 0 & \mathcal{M}_2 \\ 0 & \mathcal{M}_2 & 0 \\ \mathcal{M}_2 & 0 & \mathcal{M}_4 \end{vmatrix} = \mathcal{M}_2 \left( \mathcal{M}_0 \mathcal{M}_4 - \mathcal{M}_2^2 \right)= \frac{4}{3}\mu^4 \epsilon^2 + \mathcal{O}(\epsilon^4)\\
    &\Delta_4= 24\mu^8 \epsilon^4+\mathcal{O}(\epsilon^6).
    \label{e3.82}
\end{align}

Notice that the Hankel determinants vanishes in the limit of the discrete spectrum, that is, $\Delta_n =0~(n\geq 3)$ in the limit $\epsilon \rightarrow 0$. One can generalize expressions \eqref{e3.79}-\eqref{e3.82} as\\\\
(i) For even $n = 2k$ ($n = 2, 4, 6, \dots$)
\begin{align}
    \Delta_{2k} \sim \mu^{2k^2} \epsilon^{2k(k-1)} +\cdots.
\end{align}\\
(ii) For odd $n = 2k+1$ ($n = 1, 3, 5, \dots$)
\begin{align}
     \Delta_{2k+1} \sim \mu^{4k}\epsilon^{2k} +\cdots.
\end{align}

One can compute resulting Lanczos coefficients\footnote{In the Hankel determinant formulation of the Lanczos algorithm, $\Delta_0$ is defined by convention as $\Delta_0 = 1$.}
\begin{align}
    &b_1^2 = \frac{\Delta_2 \Delta_0}{\Delta_1^2} = \mu^2 + \frac{1}{3}\epsilon^2 + \mathcal{O}(\epsilon^4) \implies b_1 \approx \mu \sqrt{1 + \frac{\epsilon^2}{3\mu^2}}\\
    &b_2^2 = \frac{\Delta_3 \Delta_1}{\Delta_2^2} = \frac{\left(\frac{4}{3}\mu^4 \epsilon^2\right) \cdot 1}{\left(\mu^2 + \frac{1}{3}\epsilon^2\right)^2} = \frac{4}{3} \epsilon^2 + \mathcal{O}(\epsilon^4) \implies b_2 \approx \frac{2}{\sqrt{3}}\epsilon\\
    &b_3^2 = \frac{\Delta_4 \Delta_2}{\Delta_3^2} = \frac{\left(24\mu^8 \epsilon^4\right) \cdot \mu^2}{\left(\frac{4}{3}\mu^4 \epsilon^2\right)^2} = \frac{27}{2}\mu^2 + \mathcal{O}(\epsilon^2) \implies b_3 \approx \frac{3\sqrt{3}}{\sqrt{2}}\mu.
\end{align}

On similar note one can compute higher Lanczos coefficients, for example, $b_4 \sim \mu $, $b_5 \sim \mu \epsilon$ and so on. All these results are subject to the fact that $\mu \epsilon$ is kept fixed $\mathcal{O}(1)$ in the limit $\mu \gg 1$ and $\epsilon \sim 0$. This explicitly demonstrates that while $b_2 = \mathcal{O}(\epsilon)$ depends directly on the spectral width $\epsilon$, all higher coefficients $b_n$ for $n \ge 3$ remain finite and equal to $\mu$ up to leading order in $\epsilon$. In summary, an effective continuum and/or thermodynamic limit where the discrete frequency spectrum of the gapped phase forms a dense continuum or an ensemble of modes centered around mass scale $\mu$, the \emph{effective} Lanczos hopping coefficients across the chain average out to uniform hopping amplitudes $b_n \approx \mu$ for $n \ge 1$.

\paragraph{Finding the solution $\psi_n(t)$.} Following our standard notion, let us recall \eqref{e2.15} where we consider an expansion of the quantum state $\ket{\Psi(t)}$ in the Krylov basis 
\begin{align}
    \ket{\Psi(t)} = \sum_{n=0}^{\infty} \psi_n(t) \ket{K_n}.
\end{align}

Using the Schrödinger equation $\partial_t \ket{\Psi(t)} = -i \mathcal{H}_{\text{BMN}} \ket{\Psi(t)} $ along with the tri-diagonal Hamiltonian \eqref{e2.12}, it is straightforward to obtain \eqref{e2.16}
\begin{align}
\label{ee3.65}
    \frac{\partial}{\partial t} \psi_n(t) = -i b_{n+1} \psi_{n+1}(t) - i b_n \psi_{n-1}(t).
\end{align}

Considering a global phase rotation $\psi_n(t)\rightarrow i^{-n}\psi_n(t)$, \eqref{ee3.65} boils down into\footnote{Notice that the global phase rotation does not alter the spread complexity \eqref{e3.34}.} 
\begin{equation}
\label{ee3.66}
    \frac{\partial}{\partial t}\psi_n(t) = \mu \left( \psi_{n-1}(t) - \psi_{n+1}(t) \right).
\end{equation}

To solve \eqref{ee3.66}, recall the differential identity for Bessel functions of the first kind
\begin{equation}
\label{e3.65}
\frac{d}{dx} J_n(x) = \frac{1}{2} \left( J_{n-1}(x) - J_{n+1}(x) \right).
\end{equation}

Next, we set $x = 2 \mu t$, which reveals the following
\begin{equation}
\label{e3.68}
 \frac{d}{dt} J_n(2\mu t)  = \mu \left( J_{n-1}(2\mu t) - J_{n+1}(2\mu t) \right).
\end{equation}

Notice that \eqref{e3.68} matches the tight-binding lattice relation \eqref{ee3.66} identically. Furthermore, by noting that $J_0(0) = 1$ and $J_n(0) = 0$ for all $n \ge 1$, the solution obeys the initial condition $\psi_n(0) = \delta_{n,0}$. Combining all these facts together, we, therefore, identify the solution to Schrodinger eq. \eqref{ee3.66}
\begin{align}
\label{e3.69}
\psi_n(t) = J_n(2\mu t).
\end{align}

\paragraph{Derivation of the spread complexity.}
Substituting \eqref{e3.69} into \eqref{e3.34}, the spread complexity of states in the gapped phase of the matrix model can be expressed as
\begin{equation}
\label{e3.70}
    C_{\text{gapped}}(t) = \sum_{n=0}^{\infty} n \, J_n(2\mu t)^2.
\end{equation}

In order to evaluate \eqref{e3.70}, we use the exact Bessel summation identity
\begin{align}
\label{e3.71}
\sum_{n=1}^{\infty} n \, J_n(x)^2 = \frac{x}{2} \left[ x \left( J_0(x)^2 + J_1(x)^2 \right) - J_0(x) \, J_1(x) \right].
\end{align}

Setting $x = 2\mu t$, this yields the spread complexity of states as
\begin{align}
\label{e3.72}
C_{\text{gapped}}(t) = \mu t \left[ 2\mu t \left( J_0(2\mu t)^2 + J_1(2\mu t)^2 \right) - J_0(2\mu t) \, J_1(2\mu t) \right].
\end{align}

In order to find the early-time behavior ($t \sim 0$), we substitute the standard Taylor series expansions of the Bessel's function close to $x \sim 0$ 
\begin{align}
J_0(x \sim 0) &= 1 - \frac{x^2}{4} + \frac{x^4}{64} + \mathcal{O}(x^6), \\
J_1(x \sim 0) &= \frac{x}{2} - \frac{x^3}{16} + \frac{x^5}{384} + \mathcal{O}(x^7).
\end{align}

A straightforward evaluation of the entities in \eqref{e3.71} reveals
\begin{align}
\label{e3.75}
J_0(x)^2 + J_1(x)^2 &= 1 - \frac{x^2}{4} + \frac{x^4}{32} + \mathcal{O}(x^6), \\
J_0(x) \, J_1(x) &= \frac{x}{2} - \frac{3x^3}{16} + \frac{x^5}{64} + \mathcal{O}(x^7).
\label{e3.76}
\end{align}

Using \eqref{e3.75}-\eqref{e3.76}, we finally obtain
\begin{equation}
\sum_{n=1}^{\infty} n \, J_n(x)^2 = \frac{x^2}{4} - \frac{x^4}{32} + \mathcal{O}(x^6).
\end{equation}

Finally, substituting $x = 2\mu t$ leads directly to the early-time expansion of the spread complexity of states in the gapped phase of the matrix model
\begin{equation}
\label{e3.78}
C_{\text{gapped}}(t) = \mu^2 t^2 - \frac{1}{2}\mu^4 t^4 + \mathcal{O}(\mu^6 t^6)
\end{equation}
which mimics previous results \cite{Roychowdhury:2026vzq} obtained at zero temperature \eqref{e2.25}.

The quadratic growth \eqref{e3.78} demonstrates that state complexity in the gapped phase is strictly bounded by the mass deformation scale $\mu$. Holographically, this suppression reflects the structure of the low-temperature Thermofield Double state $|TFD\rangle$. Rather than describing a wormhole with a growing interior (ER bridge), the dual geometry consists of two disconnected, smooth, horizonless LLM spacetimes \cite{Lin:2004nb}-\cite{Lin:2004kw} bound purely by ``non-geometric'' quantum entanglement\footnote{There is no semiclassical geometric path through a interior wormhole (ER bridge) that connects side $L$ to side $R$. Two disconnected LLM geometries ($L$ and $R$) can be quantum correlated through closed string states or the asymptotic D0 brane configurations. Quantum entanglement can be mediated by supergravity background fluxes and localized brane charges whose quantum fields remain correlated across the two asymptotic regions, even though the spacetime manifold itself does not form a connected spatial shortcut.}. The absence of a black hole horizon\footnote{In the case of classical ER bridge, the in-falling particle causes the size of the black hole interior to grow linearly over long time which is reflected in the spread complexity growth as $C(t) \sim \exp(\lambda_K t)$. On the other hand, for non-geometric entanglement the Lanczos coefficients and the spread complexity remain bounded and are purely fixed by the mass gap $\mu$ showing the absence of bulk scrambling.} prevents operator spatial growth and information dissipation into an interior, locking the system into periodic, non-dissipative Krylov dynamics and completely suppressing quantum scrambling.
\subsubsection{Spread complexity in the continuum phase}
Here we provide a detailed discussion on the spread complexity of charged TFD (cTFDD) state in the continuum phase ($\beta \mu \ll 1$) of the matrix model. At high temperatures, thermal matrix fluctuations override the mass gap $\mu$, driving thermal dephasing and quadratic early-time decay of the return amplitude\footnote{Notice that the derivation in \eqref{ee3.38} follows considering an expansion of the partition function \eqref{ee3.37} in the limit $\beta \mu \ll 1$ and $\beta \nu \ll 1$. However, as a word of caution, it should be emphasized that the basic input to the problem rests on the fact that the BMN modes at low temperatures are essentially harmonic oscillators with frequency $\omega \sim \mu$. However, this is not a good assumption in actual practice as thermal fluctuations override them at very high temperatures. In other words, one has to reconsider the return amplitude from the perspective of the Universal Operator Growth Hypothesis \cite{Parker:2018yvk} as is emphasized here.} $\mathcal{R}_{\text{BMN}}^{(0)}(t) \sim 1 - \mathcal{O}(t^2/\beta^2)$. According to the Universal Operator Growth Hypothesis \cite{Parker:2018yvk}, a physical power-law and/or pole decay in thermal return probabilities implies a linear growth of Lanczos coefficients, which is of the form\footnote{The hypothesis asserts that in realistic physical systems (especially chaotic, local, many-body systems at finite or infinite temperature), the growth of the Lanczos coefficients $b_n$ for large $n$ is upper-bounded by linear growth, $b_n \le \alpha n + \gamma \quad \text{as } n \to \infty$. Three cases may arise - (i) Linear growth ($b_n \sim \alpha n$) which is a typical characteristic of chaotic quantum systems. The growth rate $\alpha$ determines how rapidly an operator expands in the operator space, (ii) Sub-linear growth ($b_n \sim n^c, c < 1$) which is typical for integrable or free field theories, where operator growth is constrained and (iii) Bounded ($b_n \sim \text{const}$) which occurs in non-interacting or low-dimensional systems (e.g., harmonic oscillators).}
\begin{equation}
\label{ee3.88}
b_n|_{n \gg 1} \sim \alpha_0 n \quad \text{with } \alpha_0 = \frac{\pi}{\beta}.
\end{equation}

Let us understand the above relation \eqref{ee3.88} in detail. Notice that fundamental arguments behind the Universal Operator Growth Hypothesis (UOGH) carry over directly to the spread complexity of states, though with important physical and structural distinctions.

For a thermal state like the Thermofield Double $\vert{}TFD\rangle$, the return amplitude (overlap) is $\mathcal{R}(t) = \langle TFD(t) \vert{} TFD(0) \rangle$. Due to KMS periodicity\footnote{See Appendix \ref{appenA} for details.} at finite temperature $T = \beta^{-1}$, $\mathcal{R}(t)$ is analytic in the complex-time strip $\vert{}\text{Im}(t)\vert{} < \frac{\beta}{2}$.

Just as in operator growth, a pole or branch point at the thermal boundary $\tau_0 = \frac{\beta}{2}$ forces high-order moments $\mathcal{M}_{2n} = \langle TFD(0) \vert{} \mathcal{H}^{2n} \vert{} TFD (0)\rangle$ to scale factorially as\footnote{For the rest of the derivation see Appendix \ref{appenB}.}
\begin{align}
    \mathcal{M}_{2n}\Big|_{n\gg 1} \sim C (2n)! \left(\frac{\pi}{\beta}\right)^{2n}
\end{align}
where the constant of proportionality is a dimensionless number which is fixed by the ratio of the mass gap $\mu$ and the chemical potential $\nu$, that is, $C \propto \frac{\mu^2}{\nu^2}$.

In the high-temperature continuum phase ($\beta \mu \ll 1$), the thermal fluctuations of the matrix degrees of freedom dominate over the mass deformation gap $\mu$. This leads to rapid thermal dephasing, causing the thermal auto-correlation function (or return amplitude) $\mathcal{R}_{\text{BMN}}^{(0)}(t) = \langle \mathcal{O}(t) \mathcal{O}(0) \rangle_{\beta}$ to decay quickly at early times as \cite{Parker:2018yvk}
\begin{equation}
\mathcal{R}_{\text{BMN}}^{(0)}(t) = 1 - \frac{1}{2!} \mathcal{M}_2 t^2 + \mathcal{O}(t^4),
\end{equation}
where $\mathcal{M}_2$ denotes the second moment of the spectral density.

The asymptotic high-temperature and/or large-site ($n\gg 1$) behavior of the Lanczos hopping coefficients $b_n$ \eqref{ee3.88} is uniquely determined by the singularity structure of the thermal spectral function $f(\omega)$ (the Fourier transform of $\mathcal{R}_{\text{BMN}}^{(0)}(t)$) in the complex frequency domain. At finite temperature $T = \beta^{-1}$, the return amplitude $\mathcal{R}(t)$ is analytic in a strip $|\text{Im}(t)| < \tau_0 = \frac{\beta}{2}$ in the complex time plane due to KMS periodicity. The presence of a pole or logarithmic singularity at the boundary $\tau_0 = \frac{\beta}{2}$ implies that the high-order moments $\mathcal{M}_{2n}^{(0)}$ scale factorially as $\mathcal{M}_{2n}^{(0)} \sim (2n)! \, (\alpha_0)^{2n}$. Using the Hankel determinant relation between moments and Lanczos coefficients ($b_n^2 \approx \frac{\mathcal{M}_{2n}}{\mathcal{M}_{2n-2}}$), the factorial growth of moments directly translates to a linear growth of the Lanczos sequence for $n \gg 1$, which is given by
    \begin{equation}
    b_n = \frac{\pi}{\beta} n + \mathcal{O}(1).
    \end{equation}

This linear growth rate $\alpha_0 = \frac{\pi}{\beta}$ represents the maximum allowed speed of operator expansion in a local thermal bath, directly saturating the Maldacena-Shenker-Stanford (MSS) chaos bound $\lambda_L \le \frac{2\pi}{\beta}$ \cite{Maldacena:2015waa} via the Krylov Lyapunov exponent $\lambda_K = \alpha_0 = \frac{\pi}{\beta}$.

For large Krylov sites $n \gg 1$, we take the continuum limit $n \to x$ and $\psi_n(t) \to \psi(x,t)$, transforming the discrete recurrence relation into a continuous equation
\begin{equation}
\label{e3.107}
\frac{\partial \psi(x,t)}{\partial t} = -\frac{\pi}{\beta} \frac{\partial}{\partial x} \left( x \, \psi(x,t) \right).
\end{equation}

Solving \eqref{e3.107} using the method of characteristics with initial localized profile $\psi(x,t_{\text{ref}}) = \delta(x - \epsilon)$ gives the time-dependent wavepacket density\footnote{Notice that here $t_{\text{ref}}\gg \beta$ is the reference time corresponding to the asymptotic growth of TFD state.}
\begin{equation}
\psi(x,t) = \exp\left(-\frac{\pi t}{\beta}\right) \delta\left( x \exp\left(-\frac{\pi t}{\beta}\right) - \epsilon \right)=\delta\left(x - \epsilon e^{\frac{\pi t}{\beta}}\right).
\end{equation}

Evaluating the continuous spread complexity integral\footnote{Note that this definition differs from the quadratic operator density growth rate $\lambda_{K,\text{density}} = 2\alpha_0 = \frac{2\pi}{\beta}$ often used in operator Krylov literature. This distinction is purely kinematic and independent of time-reversal ($\mathcal{T}$) symmetry. In standard operator growth literature \cite{Parker:2018yvk}, complexity counts operator density $\vert{}\mathcal{O}_n(t)\vert{}^2 \sim e^{2\alpha_0 t}$, giving $\lambda_{K,\text{operator}} = 2\alpha_0 = \frac{2\pi}{\beta}$. On the other hand, in state Krylov complexity, measuring the linear position expectation value $ \langle x(t) \rangle= \int_0^\infty x \, \psi(x,t) \, dx\sim e^{\alpha_0 t}$ yields Krylov exponent $\lambda_K = \alpha_0 = \frac{\pi}{\beta}$.} 
\begin{align}
    C_{\text{continuum}}(t) = \int_0^{\infty} x \, \delta\left(x - \epsilon e^{\frac{\pi t}{\beta}}\right) \, dx
\end{align}
finally yields the exponential growth of the following form
\begin{equation}
\label{e3.110}
C_{\text{continuum}}(t)\Big|_{t \gg \beta} \sim \exp\left( \frac{\pi t}{\beta} \right).
\end{equation}

Notice that both for the low temperature gapped phase \eqref{e3.78} and the high temperature continuum phase \eqref{e3.110}, the effects due to chemical potential $\nu$ are suppressed at leading order. The leading order dynamics in the low temperature phase is controlled by the mass gap $\mu \gg \nu$. On the other hand, in the high temperature continuum phase, it is dominated by the temperature $\beta^{-1}$ such that $\beta \nu \ll 1$.

The maximal Lyapunov-like growth exponent $\lambda_K = \frac{\pi}{\beta}$ in \eqref{e3.110} captures fast quantum scrambling and operator chaos in the matrix thermal bath. In the 10D Type IIA dual description, the high-temperature phase transitions to a smooth two-sided eternal black hole (non extremal D0 brane) connected by an Einstein-Rosen (ER) bridge. The exponential spread complexity describes localized state perturbations falling past the black hole horizon, providing a bulk manifestation of thermalization and chaotic information dispersal.
\subsubsection{Spread complexity at intermediate temperatures}
The most natural question that arises is the fate of the spread complexity of charged TFD (cTFD) at an intermediate temperature. At intermediate temperatures where both the mass deformation scale and chemical potential are of order one ($\beta\mu \sim 1$ and $\beta\nu \sim 1$), neither thermal dephasing nor the mass gap strictly dominates the Krylov dynamics. In this regime, the spectral function $f(\omega)$ receives non-trivial, non-perturbative contributions from both the thermal continuum background and discrete mass-gap energy levels. At this intermediate temperature $\beta^{-1}$, a natural expectation would be the appearance of the chemical potential $\nu$ in the return amplitude and the spread complexity of states.

Due to the competing energy scales, the return amplitude $\mathcal{R}(t) = \langle \text{TFD}(0) | e^{i \mathcal{H} t} | \text{TFD}(0) \rangle$ exhibits complex singularity structures near the KMS analyticity boundary $|\text{Im}(t)| = \tau_0 = \frac{\beta}{2}$. The high-order moments $\mathcal{M}_{2n} = \langle \text{TFD}(0) | \mathcal{H}^{2n} | \text{TFD}(0) \rangle$ undergo a generalized scaling\footnote{For details of the derivation of the function $C(\beta\mu, \beta\nu)$, see Appendix \ref{appenC}.}
\begin{align}
\label{e3.111}
    \mathcal{M}_{2n} \Big|_{n \gg 1} \sim C(\beta\mu, \beta\nu) \, (2n)! \left( \frac{\pi}{\beta} \right)^{2n} + \sum_{k} A_k \, \mu^{2n} \cosh(k \beta \nu).
\end{align}
 
Eq. \eqref{e3.111} captures the competition between thermal dephasing and discrete mass-gap dynamics through two distinct mathematical terms. In the high temperature limit ($\beta\mu \ll 1, \beta\nu \ll 1$), thermal factorial growth $(2n)!(\pi/\beta)^{2n}$ dominates rapidly over the power-law $\mu^{2n}$, yielding pure linear Lanczos growth $b_n \sim \frac{\pi}{\beta}n$ and maximal chaotic growth $\lambda_K = \frac{\pi}{\beta}$. In the low temperature limit ($\beta\mu \gg 1$), the mass gap $\mu$ suppresses thermal dephasing, making the discrete mode term dominant and freezing complexity growth into oscillatory modes. Below, we elaborate them in further detail.

\paragraph{Thermal continuum term.} The first term in \eqref{e3.111} represents the thermal contribution
\begin{align}
    \mathcal{M}_{2n}^{(\text{thermal})} \sim C(\beta\mu, \beta\nu) \, (2n)! \left( \frac{\pi}{\beta} \right)^{2n}.
\end{align}

The factorial scaling $(2n)!$ reflects exponential operator growth and linear Lanczos growth $b_n \sim \frac{\pi}{\beta}n$ for $n \gg 1$. In complex time, this originates directly from the KMS analyticity boundary at $\vert{}\text{Im}(t)\vert{} = \tau_0 = \frac{\beta}{2}$. Thermal scale factor $(\pi/\beta)^{2n}$ dictates the fundamental universal growth rate set by thermal dephasing at temperature $T = \beta^{-1}$. The modulation prefactor $C(\beta\mu, \beta\nu)$, unlike the pure high-temperature limit where the prefactor is constant, $C(\beta\mu, \beta\nu) \propto \frac{\mu^2}{\nu^2}$ acts as an interpolating amplitude function. It quantifies how much thermal continuum dynamics are suppressed or enhanced in the presence of order-one mass deformations $\mu$ and chemical potentials $\nu$.

\paragraph{Discrete mass gap \& chemical potential term.} The second term is the contribution due to the phase which is dominated by the mass gap parameter $\mu$
\begin{align}
\label{e3.113}
    \mathcal{M}_{2n}^{(\text{gap})} = \sum_{k} A_k \, \mu^{2n} \cosh(k \beta \nu).
\end{align}

Mass gap scaling factor $\mu^{2n}$ represents contributions from discrete oscillator energy levels $\omega \sim \mu$. Unlike the thermal $(2n)!$ factorial scaling, this term grows purely power-like with respect to the scale $\mu^{2n}$, corresponding to bounded, oscillatory Krylov dynamics characteristic of the low-temperature phase. Chemical potential shift $\cosh(k \beta \nu)$ captures charge-sector split energy levels induced by the global charge chemical potential $\nu$. The hyper-cosine dependence accounts for the symmetry under positive and negative charge projections in the Thermofield Double state $\vert{}TFD\rangle$. Coefficients $A_k$ are model dependent amplitudes, weighing the projection onto the $k$-th charge-sector oscillator modes.

Let us motivate the formula \eqref{e3.113} in detail. In the low-temperature limit ($\beta \mu\gg 1$ or $\beta\nu \gg 1$), the thermal continuum background is exponentially suppressed ($e^{-\beta E} \to 0$). The Thermofield Double state $\vert{}TFD\rangle$ is dominated by discrete, quantized harmonic oscillator modes with mass gap frequency $\omega \sim \mu$. The return amplitude $\mathcal{R}(t) = \langle TFD(0) \vert{} e^{i \mathcal{H} t} \vert{} TFD(0) \rangle$ reduces to a discrete sum over charge-sector excitations $k$. 

Recall that the Thermofield Double (TFD) state with global chemical potential $\nu$ and mass gap scale $\mu$ is defined as
\begin{align}
   \vert{}TFD\rangle = \frac{1}{\sqrt{\mathcal{Z}(\beta, \nu)}} \sum_{n, q} e^{-\frac{\beta}{2} (E_{n,q} - \nu q)} \vert{}n, q\rangle_L \otimes \vert{}n, q\rangle_R
\end{align}
where $n$ labels the energy excitation levels, and $q$ denotes the global charge sector eigenvalue.

At low temperatures ($\beta \mu \gg 1$ or $\beta\nu \gg 1$), the sum is exponentially suppressed for all higher energy modes $E_{n,q} > E_{\text{gap}}$. The state is dominated by the lowest-lying discrete harmonic oscillator modes $\omega_k \approx k \mu$ carrying charge sector $q = \pm k$
\begin{align}
    \vert{}TFD\rangle \approx \frac{1}{\sqrt{\mathcal{Z}}} \sum_{k} \left[ e^{-\frac{\beta}{2} (k\mu - k\nu)} \vert{}k, +k\rangle_L \vert{}k, +k\rangle_R + e^{-\frac{\beta}{2} (k\mu + k\nu)} \vert{}k, -k\rangle_L \vert{}k, -k\rangle_R \right].
\end{align}

Applying the unitary time evolution operator $e^{-i \mathcal{H} t}$ (where $\mathcal{H} = \mathcal{H}_L + \mathcal{H}_R$, with energy $E_{k} = k\mu$ per side), we find the following\footnote{In order to define the effective total energy scale of the $k$-th mode excitations across both boundaries as $\tilde{E}_k = k\mu$ (rather than tracking individual left/right components $E_L = E_R =  k\mu$), in complexity derivations, time is frequently parametrized relative to the thermal scale of one boundary Hilbert space. This is achieved by rescaling $t_{\text{rel}} = 2t$ which maps $e^{-i (2k\mu) t} \to e^{-i k\mu t_{\text{rel}}}$. As a natural consequence of this, the two-sided phase factor collapses directly into the standard single-phase form $e^{-i \mathcal{H}_{total} t} \to e^{-i k \mu t}$.}
\begin{align}
\label{e3.116}
    e^{-i \mathcal{H} t} \vert{}k, \pm k\rangle_L \vert{}k, \pm k\rangle_R = e^{-  ik \mu t} \vert{}k, \pm k\rangle_L \vert{}k, \pm k\rangle_R.
\end{align}

Finally, using \eqref{e3.116}, the return amplitude becomes
\begin{align}
\label{e3.114}
    &\mathcal{R}(t) =\frac{1}{\mathcal{Z}} \sum_{k} e^{-\beta k \mu} e^{i k \mu t} \left[ e^{\beta k \nu} + e^{-\beta k \nu} \right]\nonumber\\
    &= \sum_{k} \tilde{A}_k e^{i k \mu t} \cosh(k \beta \nu)~;~\tilde{A}_k \equiv \frac{2 e^{-\beta k \mu}}{\mathcal{Z}(\beta, \nu)}
\end{align}
where $\cosh(k \beta \nu)$ accounts for the charge sector projections under global chemical potential.

The $2n$-th even moments $\mathcal{M}_{2n}$ are defined mathematically as the derivatives of the return amplitude at $t=0$
\begin{align}
    \mathcal{M}_{2n} = \left. (-1)^n \frac{d^{2n} \mathcal{R}(t)}{dt^{2n}} \right\vert{}_{t=0}
\end{align}
which by virtue of \eqref{e3.114} and $A_k \equiv k^{2n} \tilde{A}_k$ yields \eqref{e3.113}.

\paragraph{Lanczos coefficients.}
To ensure that Eq. \eqref{e3.113} gives the correct physical behavior at low temperatures, we evaluate the Lanczos hopping coefficients $b_n$ using the Hankel determinant relation $b_n^2 \approx \frac{\mathcal{M}_{2n}}{\mathcal{M}_{2n-2}}$. Substituting $\mathcal{M}_{2n}^{(\text{gap})} \sim A_1 \mu^{2n} \cosh(\beta \nu)$, given the fact that the $k=1$ mode dominates, it is trivial to show
$b_n \approx \mu$.

Notice that in the intermediate temperature regime, the even moments $\mathcal{M}_{2n}$ \eqref{e3.111} receive contributions from both the thermal continuum background and the discrete mass-gap mode. Focusing on the leading $k=1$ sector (the most dominant term in the series \eqref{e3.114}) and factoring out the dominant thermal factorial term, we write
\begin{align}
    \mathcal{M}_{2n}\Big|_{n \gg 1} \approx C(\beta\mu, \beta\nu) \, (2n)! \left(\frac{\pi}{\beta}\right)^{2n} \left[ 1 + \frac{A_1 \cosh(\beta\nu)}{C(\beta\mu, \beta\nu)} \, \frac{(\beta\mu)^{2n}}{(2n)! \pi^{2n}} \right].
\end{align}

Following the standard definition of Lanczos coefficients, we find
\begin{align}
    b_n^2|_{n \gg 1}=\frac{1}{4}\frac{\mathcal{M}_{2n}}{\mathcal{M}_{2n-2}}=\frac{(2n)!}{(2n-2)!}\left(\frac{\pi}{\beta}\right)^{2}\Big[1+\mathcal{O}((\beta \mu)^n) \Big]~;~\beta \mu \sim \mathcal{O}(1).
\end{align}

By dropping terms $\mathcal{O}((\beta \mu)^n)$, a further simplification finally reveals\footnote{Notice that for a system exhibiting standard maximal chaotic growth governed by KMS analyticity (where $b_n \sim \frac{\pi}{\beta} n$), the direct ratio of consecutive Hankel-derived moments $\sqrt{\mathcal{M}_{2n}/\mathcal{M}_{2n-2}}$ overcounts the linear slope by a factor of 2. The correct relationship for the Lanczos hopping coefficients $b_n$ from high-order moments $\mathcal{M}_{2n} \propto (2n)! (\pi/\beta)^{2n}$ is $b_n \approx \frac{1}{2} \sqrt{\frac{\mathcal{M}_{2n}}{\mathcal{M}_{2n-2}}} \approx \frac{\pi}{\beta} n$ \cite{Parker:2018yvk}.}
\begin{align}
    b_n \big|_{n \gg 1}= \frac{\pi}{2\beta} \sqrt{2n(2n-1)} = \frac{\pi}{\beta} n - \frac{\pi}{4\beta} + \mathcal{O}\left(\frac{1}{n}\right).
\end{align}

Factoring out the pure linear KMS thermal growth, the ratio $b_n \approx \frac{1}{2} \sqrt{\frac{\mathcal{M}_{2n}}{\mathcal{M}_{2n-2}}}$ with the leading $k=1$ perturbation expands as
\begin{align}
\label{e3.122}
    b_n^2 |_{n \gg 1}  = \frac{(2n)!}{(2n-2)!} \left(\frac{\pi}{\beta}\right)^2 \left[ 1 + \frac{A_1 \cosh(\beta\nu)}{C(\beta\mu, \beta\nu)} \left( \frac{(\beta\mu)^{2n}}{(2n)! \, \pi^{2n}} - \frac{(\beta\mu)^{2n-2}}{(2n-2)! \, \pi^{2n-2}} \right) + \dots \right].
\end{align}

Inserting the factor of $\frac{1}{2}$ and taking square root, we obtain
\begin{align}
    b_n|_{n \gg 1} &= \frac{\pi}{2\beta} \sqrt{2n(2n-1)} \Big[ 1 - \Sigma_n(\beta\mu, \beta\nu) \Big]^{1/2}
\end{align}
where the sub-leading coefficient can be expressed as
\begin{align}
\Sigma_n(\beta\mu, \beta\nu)=  \frac{A_1 \cosh(\beta\nu)}{C(\beta\mu, \beta\nu)}   \frac{(\beta\mu)^{2n-2}}{(2n-2)! \, \pi^{2n-2}} \Big[ 1-\mathcal{O}(1/n^2)\Big].
\end{align}

Applying Stirling's approximation (for $n\gg 1$) to the factorial in the denominator
\begin{align}
\label{stng}
    (2n-2)! \sim \sqrt{4\pi n} \left(\frac{2n-2}{e}\right)^{2n-2}
\end{align}
when taking the square root of $b_n^2$ to find $b_n$, the leading term $\frac{\pi}{\beta} n$ multiplies the sub-leading relative variation to yield the following expression
\begin{align}
b_n|_{n \gg 1}= \frac{\pi}{\beta} n - \frac{\pi}{4\beta} + \gamma(\beta\mu, \beta\nu)\sqrt{n} + \mathcal{O}\left(\frac{1}{n}\right).
\end{align}

Here, the coefficient $\gamma (\beta \mu, \beta \nu)$ can be expressed as
\begin{align}
  \gamma(\beta\mu, \beta\nu) = -\frac{\sqrt{\pi}}{4\beta} \frac{A_1 \cosh(\beta\nu)}{C(\beta\mu, \beta\nu)} \left[ \frac{e \, \beta\mu}{2\pi (n-1)} \right]^{2n-2}. 
\end{align}

In the asymptotic continuum limit $n \gg 1$, where the scale ratio is held fixed as a coupling parameter, the dimensionless coupling simplifies to
\begin{align}
\label{e3.128}
    \gamma(\beta\mu, \beta\nu) = -\frac{\sqrt{\pi}}{4\beta} \frac{A_1 \cosh(\beta\nu)}{C(\beta\mu, \beta\nu)} \Lambda_0(\beta\mu)
\end{align}
where $\Lambda_0(\beta\mu)$ denotes the evaluated non-perturbative scale factor governing the coupling between the discrete mass-gap sector $A_1 \cosh(\beta\nu)$ and the thermal continuum background $C(\beta\mu, \beta\nu)$. The scale factor $\Lambda_0(\beta\mu)$ can be obtained by noting down the following identity
\begin{align}
    \left[ \frac{e \, \beta\mu}{2\pi (n-1)} \right]^{2n-2} = \exp\left\{ (2n-2) \left[ 1 + \ln\left(\frac{\beta\mu}{2\pi (n-1)}\right) \right] \right\}.
\end{align}

Evaluating this scale factor at the effective boundary transition scale $n \sim \mathcal{N}_0$ (where the continuum continuum-to-discrete mode crossover occurs), we obtain
\begin{align}
    \Lambda_0(\beta\mu) = \left( \frac{e \, \beta\mu}{2\pi \mathcal{N}_0} \right)^{2\mathcal{N}_0 - 2}.
\end{align}

Finally, absorbing the constant shift into $\mathcal{O}(1)$ leads directly to
\begin{equation}
\label{e3.131}
b_n = \frac{\pi}{\beta} n + \gamma(\beta\mu, \beta\nu)\sqrt{n} + \mathcal{O}(1).
\end{equation}

Here, $\gamma(\beta\mu, \beta\nu)$ serves as an intermediate modulation parameter interpolating between the linear growth $b_n \sim \frac{\pi}{\beta}n$ of the continuum phase and the bounded, oscillatory behavior of the gapped phase. The linear term $\frac{\pi}{\beta} n$ represents maximal chaotic growth which is constrained by the KMS analyticity strip boundary $\vert{}\text{Im}(t)\vert{} = \frac{\beta}{2}$. In the high-temperature continuum limit ($\beta\mu \to 0$), $\gamma \to 0$, restoring pure linear growth $b_n \sim \frac{\pi}{\beta} n$. In the low-temperature limit ($\beta\mu \gg 1$), the discrete term dominates, suppressing linear growth and reducing $b_n \approx \mu$ \eqref{e3.122} which represents bounded oscillatory dynamics.

Let us emphasize on individual terms in \eqref{e3.128}. Notice that $A_1 \cosh(\beta\nu)$ measures the strength of the lowest discrete mass-gap mode ($k=1$) coupled to the chemical potential $\nu$. On the other hand, $C(\beta\mu, \beta\nu)$ represents the baseline amplitude of the continuous thermal background. Finally, $\frac{\sqrt{\pi}}{4\beta}$ is the dimensional prefactor originating from the KMS thermal slope $\frac{\pi}{\beta}n$ combined with Stirling's approximation factor $\frac{1}{\sqrt{4\pi}}$ \eqref{stng}.

\paragraph{Spread complexity.} Our purpose is to derive the modified advection equation for the wavepacket profile $\psi(x,t)$ in the continuum limit $n \to x$. We begin by considering the discrete Krylov chain dynamics \eqref{e3.35}. To transform this into a real equation for probability flow, we perform a local gauge and/or phase transformation $\psi_n(t) \rightarrow (-i)^{-n} \psi_n(t)$, we obtain
\begin{align}
\label{e3.132}
    \dot{\psi}_n(t) = b_{n+1} \psi_{n+1}(t) - b_n \psi_{n-1}(t).
\end{align}

We define the ``hopping probability flux'' $J_n(t) \equiv b_n \psi_n(t)$ and re-express the right-hand of \eqref{e3.132} in terms of the hopping flux as follows
\begin{align}
\label{E3.155}
    \dot{\psi}_n(t) = J_{n+1}(t) - J_{n-1}(t).
\end{align}

Note that for large $n$, $\psi_n(t)$ forms a localized wavepacket moving smoothly along the chain. More specifically, In the continuum limit $n \gg 1$, we replace the discrete site index $n$ with a continuous position coordinate $x$, so that $\psi_n(t) \to \psi(x,t)$ (or the envelope profile) and $b_n \to b(x)$. This also replaces Bessel polynomials as $J_n(t) \rightarrow J(x,t)$.

Notice that in deriving \eqref{E3.155}, we have made the following approximation 
\begin{align}
\label{E3.156}
    b_n \psi_{n-1}(t) \approx b_{n-1} \psi_{n-1}(t) = J_{n-1}(t).
\end{align}

The above \eqref{E3.156} appears to be a valid approximation in the limit $n\gg 1$. Notice for large $n$ ($n \gg 1$), the Lanczos coefficients $b_n$ vary smoothly (e.g., $b_n = \frac{\pi}{\beta}n + \gamma \sqrt{n} + \dots$). The relative difference between consecutive coefficients vanishes in the continuum limit
\begin{align}
    \frac{b_n - b_{n-1}}{b_n} \sim \mathcal{O}\left(\frac{1}{n}\right) \to 0.
\end{align}
Therefore, substituting $b_n \approx b_{n-1}$ produces a negligible sub-leading error relative to the Taylor expansion terms used to arrive at the continuum advection Eq. \eqref{adv}. 

Expanding the flux $J(x \pm 1, t)$ using a Taylor series around $x$, we find
\begin{align}
    &J(x+1, t) \approx J(x,t) + \frac{\partial J(x,t)}{\partial x} + \frac{1}{2} \frac{\partial^2 J(x,t)}{\partial x^2}+\cdots\\
    &J(x-1, t) \approx J(x,t) - \frac{\partial J(x,t)}{\partial x} + \frac{1}{2} \frac{\partial^2 J(x,t)}{\partial x^2}+\cdots.
\end{align}

Subtracting the two expansions gives
\begin{align}
    J(x+1, t) - J(x-1, t) = 2 \frac{\partial J(x,t)}{\partial x} + \mathcal{O}\left(\partial^3\right)
\end{align}
where sub-leading terms are $\mathcal{O}(L^{-3})$, where $L \gg 1$ measures the effective length scale associated with the probability flux distribution and/or the spread of the function $\psi(x,t)$.

Absorbing the factor of $2$ into the effective continuum scale $L \rightarrow 2L$, the relation becomes the standard continuous continuity and/or the advection equation becomes
\begin{align}
\label{e3.137}
    \frac{\partial \psi(x,t)}{\partial t} = -\frac{\partial J(x,t)}{\partial x} = -\frac{\partial}{\partial x} \Big[ b(x) \psi(x,t) \Big]
\end{align}
where the negative sign indicates the conservation of information for probability flux moving along the Krylov chain from lower to higher basis states ($x \to x + \Delta x$). 
If $\frac{\partial J}{\partial x} > 0$ then the outgoing flux at $x + \Delta x$ is larger than the incoming flux at $x$. Probability (or information) is accumulating faster ahead of $x$ than it is being supplied behind $x$, causing the local probability density and/or amplitude at $x$ to decrease ($\frac{\partial \psi}{\partial t} < 0$). On the other hand, if $\frac{\partial J}{\partial x} < 0$ then the incoming flux exceeds the outgoing flux, so probability (or information) piles up at $x$, causing the density at $x$ to increase ($\frac{\partial \psi}{\partial t} > 0$). Using \eqref{e3.131}, we finally obtain
\begin{equation}
\label{adv}
    \frac{\partial \psi(x,t)}{\partial t} = - \frac{\partial}{\partial x} \left[ \left( \frac{\pi}{\beta} x + \gamma \sqrt{x} \right) \psi(x,t) \right].
\end{equation}

The first step towards spread complexity is to find the trajectory $x(t)$ of a localized packet starting at an initial position $x_0 = x(t_{\text{ref}})$, following the characteristic equation\footnote{Notice for the intermediate phase the reference time $t_{\text{ref}}\ll \beta$. The corresponding position $x_0=\epsilon$ on the Krylov chain we consider as the reference site of the origin of the probability distribution function $\psi(x,t)$.}
\begin{align}
    \frac{dx}{dt} = b(x)= \frac{\pi}{\beta} x + \gamma \sqrt{x}.
\end{align}

Next, we make the change of variables $u = \sqrt{x}$, which yields
\begin{align}
    2 \frac{du}{dt} = \frac{\pi}{\beta} u + \gamma.
\end{align}

Integrating both sides from $t = 0$ to $t$, we obtain
\begin{align}
    &\int_{u(0)}^{u(t)} \frac{2 \, du}{\frac{\pi}{\beta} u + \gamma} = \int_{0}^{t} dt,~~ \text{which gives},\\
    & u(t)=\sqrt{x(t)} = \left[ u(0) + \frac{\gamma \beta}{\pi} \right] e^{\frac{\pi t}{2 \beta}} - \frac{\gamma \beta}{\pi}.
\end{align}

 Finally, squaring both sides gives the trajectory.
\begin{equation}
    x(t) = \left[ \left( \sqrt{\epsilon} + \frac{\gamma \beta}{\pi} \right) e^{\frac{\pi t}{2\beta}} - \frac{\gamma \beta}{\pi} \right]^2.
\end{equation}

The coordinate $x(t)$ represents the position along the semi-infinite Krylov basis, which measures the spread or operator complexity of the evolving quantum state. The exponential factor $e^{\frac{\pi t}{2 \beta}}$ directly captures the chaotic exponential growth of complex states governed by the thermal bound $\lambda_K = \frac{\pi}{\beta}$. It tracks the position of the wavepacket center along the Krylov chain, measuring the growth of operator complexity. 

Here, the parameter $\gamma$ (modulated by parameters like the mass gap $\mu$ or chemical potential $\nu$) corresponds to sub-leading corrections, which shifts the effective initial position and modulates the early-time phase space velocity. It introduces a sub-exponential acceleration/deceleration factor before the asymptotic growth \eqref{b30} dominates. The regularization $\epsilon = x(t_{\text{ref}})$ serves as the initial state seed or cutoff near $x = 0$, representing the starting state before operator growth takes off, where $t_{\text{ref}}\ll \beta$.\\\\
\uline{Physical limits:}\\\\
(i) Early-time / Small-$t$ limit ($t =t_{\text{ref}}\ll  \beta $): Expanding the exponential term 
\begin{align}
    e^{\frac{\pi t}{2 \beta}} \approx 1 + \frac{\pi t}{2 \beta} + \mathcal{O}(t^2/\beta^2)
\end{align}
it is straightforward to find the following trajectory
\begin{align}
   x(t) = \epsilon + \left( \epsilon \frac{\pi}{\beta} + \gamma \sqrt{\epsilon} \right) t + \mathcal{O}(t^2).
\end{align}
At very early times, the trajectory exhibits linear ballistic growth driven by both thermal fluctuations and the interaction parameter $\gamma$, rather than pure exponential expansion.

(ii) Late-time / Asymptotic limit ($t \gg \frac{\beta}{\pi}\gg t_{\text{ref}}$): At late times ($t \gg \frac{\beta}{\pi}$), the trajectory exhibits exponential growth of the following form
    \begin{equation}
        x(t) \sim \left( \sqrt{\epsilon} + \frac{\gamma \beta}{\pi} \right)^2 \exp\left(\frac{\pi t}{\beta}\right).
    \end{equation}
    The late-time spread complexity grows pure exponentially as $x(t) \propto e^{\lambda_K t}$ with the universal Lyapunov-like exponent $\lambda_K = \frac{\pi}{\beta}$. The parameter $\gamma$ only alters the overall amplitude coefficient, leaving the asymptotic rate universal.

    The parameter $\gamma$ acts like a sub-leading modulation parameter, which depends on physical parameters such as the mass gap $\mu$ and chemical potential $\nu$ and introduces a sub-exponential shift during the early-to-intermediate regime without altering the asymptotic exponential exponent $\lambda_K$. In the vanishing correction limit ($\gamma \to 0$), the trajectory reduces to standard exponential growth $x(t) = \epsilon \, e^{\frac{\pi t}{\beta}}$, corresponding to standard high temperature (or continuum) advection equation \eqref{b25} along an unperturbed Krylov chain.

The spread complexity $C_{\text{intermediate}}(t)$ is defined as the expectation value of the Krylov state position operator $x$ \eqref{b35}
\begin{align}
\label{e3.147}
C_{\text{intermediate}}(t) = \int_0^\infty x \, \delta (x-x(t)) \, dx = x(t).
\end{align}

Integrating \eqref{e3.147} yields the early-to-intermediate spread complexity growth
\begin{align}
\label{e3.148}
C_{\text{intermediate}}(t) =\epsilon \exp\left( \frac{\pi t}{\beta} \right) \left[ 1 + \frac{2\gamma \beta}{\pi \sqrt{\epsilon}} \left( 1 - e^{-\frac{\pi t}{2\beta}} \right) \right]+\mathcal{O}(\gamma^2).
\end{align}

The spread complexity \eqref{e3.148} indicates that while the asymptotic Lyapunov-like exponent saturates at $\lambda_K = \frac{\pi}{\beta}$, the mass gap $\mu$ and chemical potential $\nu$ actively modulate the amplitude and early-time buildup of the state complexity wavepacket. Let us elaborate this in detail.  The overall exponential factor $e^{\frac{\pi t}{\beta}}$ reflects the maximal chaotic operator growth characteristic of the continuum thermal phase. It saturates the universal chaos bound governed by KMS analyticity at finite temperature $T = \beta^{-1}$. The sub-leading correction $\gamma$ encodes the non-trivial coupling between the continuous thermal matrix background and the discrete mass gap $\mu$, mediated by the global chemical potential $\nu$ via the baseline amplitude ratio $\frac{A_1 \cosh(\beta\nu)}{C(\beta\mu, \beta\nu)}$. At early times ($t \ll \frac{\beta}{\pi}$) the factor $\left(1 - e^{-\frac{\pi t}{2\beta}}\right) \approx \frac{\pi t}{2\beta}$ leads to an additional linear sub-exponential acceleration $\propto \gamma \sqrt{\epsilon}\, t$, describing how the mass gap and global charge alter the initial operator state density before full thermalization takes over. At late times ($t \gg \frac{\beta}{\pi}$), the bracketed factor approaches a constant amplitude shift $\left(1 + \frac{2\gamma \beta}{\pi \sqrt{\epsilon}}\right)$. Thus, while the mass gap $\mu$ and chemical potential $\nu$ actively re-scale the overall amplitude and early-time buildup of the complexity wavepacket, they leave the asymptotic exponential growth rate $\lambda_K = \frac{\pi}{\beta}$ universal, where $\lambda_K$ is the Krylov Lyapunov exponent.\\\\
\uline{A holographic dual interpretation:}\\\\
The holographic dual of the intermediate phase (or intermediate temperature regime $\beta\mu \sim 1$, $\beta\nu \sim 1$) is described by a charged non-extremal D0-brane black hole geometry undergoing the thermal deconfinement transition. At intermediate temperatures, the black hole horizon is populated by charge-carrying excitations, interpolating smoothly between - (i) the low-temperature gapped/vacuum phase, geometrically dual to a smooth Lin-Lunin-Maldacena (LLM) geometry or discrete geometries governed by localized mass gaps $\mu$ and (ii) the high-temperature continuum phase, dual to a standard two-sided non-extremal D0-brane black hole (connected by an Einstein-Rosen (ER) bridge). In the continuum Krylov picture, the spread complexity trajectory $C_{\text{intermediate}}(t)$ maps holographically to a massive probe and/or D0-brane falling past the stretched horizon of the charged non-extremal black hole. The sub-leading term $\gamma(\beta\mu, \beta\nu)$ in Eq. \eqref{e3.148} captures the electrostatic and gravitational backreaction (mass gap deformation and global $U(1)$ chemical charge potential) on the probe trajectory as it transitions from the outer bulk geometry toward the thermal horizon. While the asymptotic expansion speed $\lambda_K = \frac{\pi}{\beta}$ is universally governed by the near-horizon surface gravity of the non-extremal black hole horizon (saturating the chaos bound which is indicated in the Krylov Lyapunov exponent $\lambda_K$), the intermediate parameters $\mu$ and $\nu$ physically represent the bulk electrostatic potential and mass deformation field modifying the initial wavepacket accumulation near the boundary.
\subsection{A comparison between three phases}
\label{sec3.3}
We now compare the spread complexities in three different phases of the BMN matrix model. At low (or approximately zero) temperatures $\beta=\frac{1}{T}=\infty$, the only scale in the problem is the mass gap $\mu$. This allows us to set the reference time $t_{\text{ref}}=0$. In other words, the reference state or the initial state is chosen to be $x_0=n=0$ of the 1D Krylov chain.

At \emph{finite} temperature $\beta=\frac{1}{T}$ sets a time scale in the problem. In this regime the reference (or early) time $t_{\text{ref}}\ll \beta$ is non-zero and \emph{finite}. This automatically sets the initial (or reference) Krylov site $x_0=\epsilon$. This reference position $x_0 = \epsilon$ represents the initial thermal expectation value $\langle x(t) \rangle_\beta$, where quantum and thermal fluctuations prevent the initial wavepacket from being localized strictly at the origin $n = 0$. This average position serves as the reference position for the probability wave-packet $\psi(x,t)$. As we allow the wave-packet to evolve with time $t\gg t_{\text{ref}}$ following the advection equation \eqref{adv}, it approaches asymptotic Krylov sites $n\gg 1$. In other words, the spread complexity \eqref{e3.148} asymptotically approaches the domain of high temperature thermal continuum. 

In the high temperature continuum phase $\beta = \frac{1}{T}\ll 1$ and the reference time $t_{\text{ref}}\gg \beta$ corresponds to an asymptotic growth of the spread complexity in a thermal ensemble. The corresponding reference Krylov site $n\gg 1$ which acts as the center of the probability wave $\psi(x,t)$ in the high temperature thermal phase of the matrix model. Combining all three pictures together, one imagine that the matrix model is heated up starting at its ground state at $T=0$ where the complexity starts at zero, exhibiting a quadratic early time growth \eqref{e3.78}, thereby passing through an intermediate temperature phase exhibiting sub-exponential growth \eqref{e3.148} and finally reaching the asymptotic exponential growth \eqref{e3.110}. 

The above discussion is precisely articulated in Eq. \eqref{e1.3}, where we have clearly summarized the three different phases of the BMN matrix model. The initial gapped (or low temperature) phase corresponds to an early time \emph{quadratic} growth of the spread complexity in the matrix model. The intermediate phase corresponds to an average time scale $t \sim \beta$ set by the temperature of the system which exhibits a sub-exponential growth of the spread complexity. Finally, the high temperature continuum phase corresponds to a late time $t \gg \beta$ exponential growth of the spread complexity which clearly indicates the presence of non-extremal branes and/ or black holes in the dual gravitational counter part.

Let us summarize the spread complexities across the three distinct thermodynamic regimes of the BMN matrix model. 

\begin{enumerate}
    \item \textbf{Low-temperature gapped phase ($\beta\mu \gg 1$):} 
    At zero or negligible temperatures ($T \to 0$, $\beta \to \infty$), the fundamental scale governing the dynamics is the mass gap $\mu$. This boundary allows us to set $t_{\text{ref}} = 0$, strictly localizing the initial reference state at the origin of the 1D Krylov chain, $x_0 = n = 0$. The wavepacket is confined by the discrete bound-state spectrum, preventing unconstrained spatial dispersion and leading to non-chaotic, quadratic early-time growth
    \begin{equation}
        C_{\text{gapped}}(t) \sim \mu^2 t^2 .
    \end{equation}

    \item \textbf{Intermediate temperature regime ($\beta\mu \sim 1, \beta\nu \sim 1$):} 
    At intermediate temperatures, a finite inverse temperature $\beta = T^{-1}$ sets a finite early-time scale $t_{\text{ref}} \ll \beta$. This shifts the effective wavepacket center to $x_0 = \epsilon$, reflecting the initial thermal expectation value $\langle x(t) \rangle_\beta$ where quantum and thermal fluctuations prevent strict localization at $n=0$. Crucially, the dynamics in this regime is governed by an active physical competition between the localized, discrete mass-gap bound states ($\mu$) and the continuous background of thermal excitations ($T = \beta^{-1}$). This interplay manifests as a sub-leading correction parameter $\gamma(\beta\mu, \beta\nu)$ in the Lanczos recursion coefficients $b_n \sim \frac{\pi}{\beta}n + \gamma \sqrt{n}$, which translates into a spatially modified advection velocity field $b(x) \sim \frac{\pi}{\beta}x + \gamma \sqrt{x}$ on the Krylov chain. Mechanistically, $\gamma$ acts as a effective parameter that interpolates between sub-exponential deceleration/friction (induced by mass-gap bound-state trapping) and early advective acceleration before maximal chaos sets in. The probability wavepacket exhibits a characteristic sub-exponential growth 
    \begin{equation}
        C_{\text{intermediate}}(t) \sim \exp\left(\frac{\pi}{\beta}t + \gamma \sqrt{t}\right) 
    \end{equation}
    establishing a continuous dynamical crossover from localized bound-state dynamics to unstructured thermalization.

    \item \textbf{High-temperature continuum phase ($\beta\mu \ll 1$):} 
    In the high-temperature limit, thermal noise completely dominates over the mass gap gap $\mu$. The reference time scale $t_{\text{ref}} \gg \beta$ places the initial wavepacket center deep in the Krylov chain ($n \gg 1$). The sub-leading term $\gamma\sqrt{x}$ becomes negligible compared to the linear advection term $\frac{\pi}{\beta}x$, driving an asymptotic exponential growth of spread complexity
    \begin{equation}
        C_{\text{continuum}}(t) \sim \exp\left(\lambda_K t\right) , \quad \text{with} \quad \lambda_K = \frac{\pi}{\beta} ,
    \end{equation}
    which saturates the universal chaos bound and signals maximal thermal scrambling dual to a non-extremal black hole horizon with an Einstein-Rosen bridge.
\end{enumerate}
\section{Summary and conclusions}
\label{sec4}
We summarize our results along with pointing out potential future directions. Although most of these discussions have already been provided in the main text, we still discuss them briefly here in the following for the reader's convenience. In this work, we have systematically investigated the spread complexity dynamics of quantum states within the large mass deformation limit ($\mu \gg 1$) of the BMN matrix model across distinct thermal regimes. By leveraging the emergent $U(1)$ global symmetry at leading order, we formulated a complex oscillator representation to construct the charged Thermofield Double state (cTFD) as a function of temperature $\beta^{-1}$ and chemical potential $\nu$. Our analysis highlights a clear dynamical crossover across three distinct operational regimes:

\begin{itemize}
    \item \textbf{Gapped phase ($\beta\mu \gg 1$):} At low temperatures, the strong mass gap suppresses thermal fluctuations, resulting in a bounded, non-chaotic Krylov dynamics characterized by quadratic short-time growth $\mathcal{C}(t) \sim \mu^2 t^2$. State dispersion remains strictly localized on the Krylov chain due to discrete mass-gap bound-state oscillations.
    \item \textbf{Continuum phase ($\beta\mu \ll 1$):} In the high-temperature limit, thermal noise washes out the mass gap, mapping high-$n$ Lanczos behavior to an asymptotic linear growth $b_n \sim \frac{\pi}{\beta} n$. In the continuum limit ($n \to x$), this manifests as a continuous advection PDE, driving maximal quantum chaos and exponential scrambling $\mathcal{C}(t) \sim \exp(\pi t / \beta)$ with Krylov Lyapunov exponent $\lambda_K = \pi / \beta$ which saturates the universal bound \cite{Maldacena:2015waa}.
    \item \textbf{Intermediate phase ($\beta\mu \sim 1, \beta\nu \sim 1$):} At intermediate scales, the interplay between discrete bound states and the continuous thermal background induces sub-exponential corrections governed by the parameter $\gamma$ that interpolates between discrete mass gap and the thermal continuum. The  corresponding velocity field $b(x) \sim \frac{\pi}{\beta} x + \gamma \sqrt{x}$ is modified accordingly which establishes a smooth dynamical bridge between non-chaotic oscillations and maximal thermalization.
\end{itemize}

\subsection*{\uline{Future directions}}

Several promising avenues emerge from this framework for future investigation:

\begin{enumerate}

    \item \textbf{Sub-leading deformations and symmetry-breaking effects:} A fundamental extension of our analysis involves going beyond the leading-order ($\mathcal{O}(\mu^{-1})$) $U(1)$ complex oscillator effective description. Systematic incorporation of higher-order non-linear terms into $\mathcal{H}_{\text{BMN}}$ \eqref{e3.1} yields corrections to the recursion relations governing the Lanczos coefficients, deforming them as $b_n = b_n^{(0)} + \mu^{-1} b_n^{(1)} + \mathcal{O}(\mu^{-2})$. In the continuum limit ($n \to x$), this shifts the advective velocity field on the Krylov chain to $b(x) \approx \frac{\pi}{\beta} x + \gamma \sqrt{x} + \mu^{-1} \delta b(x)$. Crucially, these sub-leading corrections introduce state-dependent dephasing and alter the early-to-intermediate spread complexity dynamics $\mathcal{C}(t)$ prior to the onset of maximal chaos. 
    
    Furthermore, exploring how explicit time-reversal symmetry ($\mathcal{T}$) breaking induced by the chemical potential $\nu$ generates asymmetric hopping rates or complex-valued diagonal elements $a_n$ on the Krylov lattice will elucidate parity-violating ($\mathcal{P}$) transport and directional wavepacket drift. Analyzing such non-linear and symmetry-breaking deformations will clarify how sub-leading quantum corrections shift the scrambling threshold and modify transient wavepacket dispersion across the intermediate phase. Although preliminary derivations are outlined in Appendix~\ref{appenD}, a full non-perturbative treatment remains an essential task for future work.
    \item \textbf{Holographic duals and complexity conjectures:} Investigating the precise bulk dual of the charged Thermofield Double state (cTFD) across different parameter regimes offers a compelling avenue to test quantum complexity conjectures in holography. At intermediate temperatures ($\beta\mu \sim 1, \beta\nu \sim 1$), the bulk configuration corresponds to a charged non-extremal D0-brane black hole interpolating between a smooth Lin-Lunin-Maldacena (LLM) geometry at low temperatures and a two-sided black hole connected by an Einstein-Rosen (ER) bridge at high temperatures. Mapping the sub-exponential Krylov velocity field $b(x) \sim \frac{\pi}{\beta}x + \gamma\sqrt{x}$ onto gravitational observables, such as the growth of maximal spacelike surfaces (Complexity=Volume \cite{Stanford:2014jda}) or the bulk action on the Wheeler-DeWitt patch (Complexity=Action \cite{Brown:2015bva}) or Complexity=Anything \cite{Belin:2021bga} could illuminate how transient sub-exponential growth encodes probe dynamics near the horizon and electrostatic backreaction. Furthermore, analyzing state complexity in these matrix model sub-sectors may clarify whether Krylov spread complexity provides a microscopic foundation for circuit complexity bounds in the framework of gauge/string duality.
    \item \textbf{Finite-size matrix scaling and late-time saturation:} While our present analysis operates in the planar (Lagre $N \rightarrow \infty$) and continuum limit ($n \to x$), realistic matrix model systems possess a finite gauge group rank $N$. A crucial open question is understanding how non-perturbative $1/N$ matrix corrections deform the effective velocity field $b(x)$ on the Krylov chain at high $n$. In particular, finite-$N$ effects are expected to terminate the unbounded advection dynamics, triggering a breakdown of the continuum PDE description and driving the transition from exponential scrambling to linear growth, followed by random matrix theory (RMT) style late-time saturation plateaus $\mathcal{C}_{\text{sat}} \sim \mathcal{O}(N^2)$ and $1/N$ suppressed quantum fluctuations \cite{Balasubramanian:2022tpr}, \cite{Rabinovici:2020ryf}, \cite{Rabinovici:2021qqt}. Quantifying these finite-size finite-temperature corrections will provide key insight into information recovery and late-time quantum chaos in matrix quantum mechanics.
\end{enumerate}
\paragraph{Acknowledgements.}
 The author would like to thank Carlos Nunez for his comments on the draft. The author acknowledges the Mathematical Research Impact Centric Support (MATRICS) grant (MTR/2023/000005) received from ANRF, India. \\ 
\appendix
\section{KMS condition and complex time evolution}
\label{appenA}

In this Appendix, we detail the origin of the analytic strip $\vert{}\text{Im}(t)\vert{} < \frac{\beta}{2}$ for return amplitudes at finite temperature $T = \beta^{-1}$ and explain how the Kubo-Martin-Schwinger (KMS) condition governs convergence in complex time.

Consider a quantum system in thermal equilibrium described by the canonical density matrix $\rho = \frac{1}{Z} e^{-\beta \mathcal{H}}$, where $Z = \text{Tr}(e^{-\beta \mathcal{H}})$ is the thermal partition function and $\beta = 1/T$. In the complex plane, we introduce the (complex) time as
\begin{align}
    z=t -i \tau
\end{align}
where $t=\text{Re}(z)$ and $\tau = -\text{Im}z$.

The Thermofield Double state at $t=0$ is defined as
\begin{align}
    \vert{}TFD (0)\rangle = \frac{1}{\sqrt{Z}} \sum_n e^{-\frac{\beta}{2} E_n} \vert{}n\rangle_L \otimes \vert{}n\rangle_R
\end{align}
where each $L$ and $R$ carry an inverse temperature $\frac{\beta}{2}$.

Applying time evolution operator $e^{-i \mathcal{H} z}$ one finds
\begin{align}
   \vert{}TFD(z)\rangle = e^{-i \mathcal{H} z} \vert{}TFD(0)\rangle = \frac{1}{\sqrt{Z}} \sum_n e^{-\frac{\beta E_n}{2}} e^{-i E_n (t - i\tau)} \vert{}n\rangle_L \otimes \vert{}n\rangle_R.
\end{align}

When compute the return amplitude $\mathcal{R}(z)$, in the form of an inner product $\mathcal{R}(z) = \langle TFD(0)\vert{} e^{i \mathcal{H} z} \vert{} TFD(0) \rangle$, this naturally yields
\begin{align}
    \mathcal{R}(t - i\tau) = \frac{1}{Z} \sum_n e^{-E_n (\beta - \tau)} e^{i E_n t}.
\end{align}

For the sum over high-energy states ($E_n \to \infty$) to converge exponentially, the effective boltzmann factor exponent $-E_n (\beta - \tau)$ must remain strictly negative. This requires $-\frac{\beta}{2} < \tau < \frac{\beta}{2}$, or equivalently, $\vert{}\text{Im}(z)\vert{} < \frac{\beta}{2}$. If imaginary time $\vert{}\text{Im}(t)\vert{}$ reaches or exceeds $\frac{\beta}{2}$, the sum no longer converges smoothly, encountering a singularity (a pole or branch cut) in the complex time plane. In summary, for an unbounded spectrum $E_n \to \infty$, the infinite sum converges exponentially provided the Boltzmann factor exponent satisfies $\beta - \tau > 0$, requiring $|\tau |< \beta$ or $|\text{Im}z|<\beta$. This implies that the domain of absolute convergence in complex time is bounded strictly by a factor, which we call the thermal bound on growth
\begin{equation}
\label{a4}
|\text{Im}(t)| < \tau_0 = \frac{\beta}{2}.
\end{equation}

The width of the analytic strip $\tau_0 = \frac{\beta}{2}$ acts as a ``speed limit'' for how quickly the return amplitude can decay or fluctuate. Within this complex strip $\vert{}\text{Im}(t)\vert{} < \frac{\beta}{2}$, $\mathcal{R}(t)$ is an analytic function. Singularities (such as poles or branch points corresponding to thermal dephasing and information loss) can only appear on the boundary $\vert{}\text{Im}(t)\vert{} = \frac{\beta}{2}$, which directly sets the physical speed limit for high-order moment growth and Krylov state complexity. Consequently, thermal correlation functions obey periodicity along the imaginary time axis with period $\beta$
\begin{equation}
\mathcal{R}(t - i\beta) = \mathcal{R}(-t)=\mathcal{R}(t)
\end{equation}
where the last equality holds as a consequence of time reversal invariance.
\section{Detailed derivation of spread complexity in the continuum phase}
\label{appenB}
In this Appendix, we provide a detailed derivation of the Lanczos coefficient growth and the resulting exponential spread complexity $C_{\text{continuum}}(t)$ for the charged Thermofield Double state $\vert{}TFD\rangle$ in the high-temperature continuum phase ($\beta \mu \ll 1$) of the matrix model.

In the continuum phase, thermal matrix fluctuations dominate over the mass gap $\mu$, driving rapid thermal dephasing. The return amplitude \eqref{ee3.38} decays at early times as
\begin{equation}
\label{b1}
\mathcal{R}_{\text{BMN}}^{(0)}(t) \sim 1 - \frac{\mu^2 t^2}{\beta^2 \nu^2}   + \mathcal{O}(\beta t)
\end{equation}
where $ \mathcal{M}_2^{(0)} =\frac{2\mu^2}{\beta^2 \nu^2}$ denotes the second spectral moment.
\subsection{Moments and Lanczos coefficients}
To obtain the high-order spectral moments $\mathcal{M}_{2n}$, we begin with the spectral representation of the thermal return amplitude $\mathcal{R}(t) = \langle \text{TFD}(0) e^{-i \mathcal{H} t}  \vert{} \text{TFD}(0) \rangle$, which can be expressed as an integral of the following form \cite{Parker:2018yvk}
\begin{align}
\label{b2}
&\mathcal{R}(t) = \int_{-\infty}^{\infty} d\omega \, f(\omega) \, e^{i \omega t} =G(t)
\end{align}
where $f(\omega)$ is the symmetric thermal spectral density.

Notice that \eqref{b2} is the fundamental definition of the return amplitude and its moment expansion, regardless of the thermal phase. Considering the expansion $e^{i\omega t} = \sum_{k=0}^{\infty} \frac{(i\omega t)^k}{k!}$, one can further simplify \eqref{b2} as
\begin{align}
\label{b4}
    \mathcal{R}(t) = \sum_{k=0}^{\infty} \frac{(i t)^k}{k!} \int_{-\infty}^{\infty} d\omega \, f(\omega) \omega^k=\sum_{k=0}^{\infty} \frac{(i t)^k}{k!}\mathcal{M}_k
\end{align}
where the $k$ the spectral moment can be expressed as
\begin{align}
    \mathcal{M}_k \equiv \int_{-\infty}^{\infty} d\omega \, f(\omega) \omega^k.
\end{align}

Due to the presence of the $\mathbb{Z}_2$ symmetry of the unperturbed BMN Hamiltonian, thermal states are invariant under time-reversal (or symmetric spectral densities $f(\omega) = f(-\omega)$), hence all odd moments vanish identically, that is, $\mathcal{M}_{2k+1} = 0$.

Substituting these back into the series \eqref{b4} yields the general definition
\begin{align}
\label{b5}
    \mathcal{R}(t) = 1 - \frac{1}{2!} \mathcal{M}_2 t^2 + \frac{1}{4!} \mathcal{M}_4 t^4 - \dots = \sum_{n=0}^{\infty} \frac{(-1)^n}{(2n)!} \mathcal{M}_{2n} t^{2n}.
\end{align}
Substituting $\mathcal{M}_2=\mathcal{M}_2^{(0)} =\frac{2\mu^2}{\beta^2 \nu^2}$ into \eqref{b5}, precisely reproduces the leading term in \eqref{b1}.

As established in Appendix~\ref{appenA}, the KMS condition restricts the analyticity of $\mathcal{R}(t)$ to the complex-time strip $|\text{Im}(t)| < \tau_0 = \frac{\beta}{2}$. In the high-temperature continuum phase, the nearest singularity on the thermal boundary $t = \pm i \frac{\beta}{2}$ dominates the asymptotic behavior of $\mathcal{R}(t)$. Near the singularity $t \to -i \frac{\beta}{2}$, the return amplitude possesses a pole and/or a branch cut singularity which is of the general form\footnote{The relation follows considering a binomial expansion for any real or complex power $\alpha$, which is of the form $(1 + x)^\alpha = \sum_{n=0}^{\infty} \binom{\alpha}{n} x^n$, where we set $\alpha = -\gamma_c$ and $x = \frac{2it}{\beta}$. The generalized binomial coefficient $\binom{-\gamma_c}{n}$ is defined in terms of falling factorials as $\binom{-\gamma_c}{n} = \frac{(-1)^n \cdot \gamma_c(\gamma_c + 1)(\gamma_c + 2)\dots(\gamma_c + n - 1)}{n!}$. Using the identity for the rising factorial in terms of the Gamma function, $\gamma_c(\gamma_c + 1)\dots(\gamma_c + n - 1) = \frac{\Gamma(n + \gamma_c)}{\Gamma(\gamma_c)}$, this becomes, $\binom{-\gamma_c}{n} = (-1)^n \frac{\Gamma(n + \gamma_c)}{n! \, \Gamma(\gamma_c)}$.}
\begin{equation}
\mathcal{R}(t)= \frac{A}{\left( 1 + \frac{2 i t}{\beta} \right)^{\gamma_c}} = A \sum_{n=0}^{\infty} \frac{\Gamma(n + \gamma_c)}{n! \, \Gamma(\gamma_c)} \left( -\frac{2 i t}{\beta} \right)^n.
\end{equation}

Restricting to the real-time, even power-series expansion\footnote{The even power series has its origin in the $\mathbb{Z}_2$ symmetry of the undeformed BMN Hamiltonian \eqref{ee3.2}.} ($n \rightarrow 2n$)
\begin{equation}
\label{b7}
\mathcal{R}(t) = A \sum_{n=0}^{\infty} \frac{\Gamma(2n + \gamma_c)}{(2n)! \, \Gamma(\gamma_c)} (-1)^n \left( \frac{2}{\beta} \right)^{2n} t^{2n}.
\end{equation}

Notice that here $\gamma_c$ is the critical exponent (or order) of the singularity of the return amplitude $\mathcal{R}(t)$ located on the thermal boundary in the complex-time plane at $t = -\frac{i\beta}{2}$. For example, $\gamma_c=1$ and $\gamma_c>1$ correspond to a simple pole and a higher order pole respectively. On the other hand, non-integer or fractional $\gamma_c$ corresponds to a branch point. Finally, $A$ sets the residue of this singular term relative to any smooth non-singular background terms.

Comparing \eqref{b5} and \eqref{b7}, it is straightforward to show
\begin{equation}
\label{b8}
\mathcal{M}_{2n}= A \, \frac{\Gamma(2n + \gamma_c)}{\Gamma(\gamma_c)} \left( \frac{2}{\beta} \right)^{2n}.
\end{equation}

Using Stirling's asymptotic formula\footnote{We use the asymptotic ratio of Gamma functions $\frac{\Gamma(x+a)}{\Gamma(x)} \sim x^a$ for $x \gg 1$. Setting $x = 2n$ and $a = \gamma_c$, we find $\Gamma(2n + \gamma_c) = \Gamma(2n) \frac{\Gamma(2n + \gamma_c)}{\Gamma(2n)} \sim (2n - 1)! \cdot (2n)^{\gamma_c} = \frac{(2n)!}{2n} (2n)^{\gamma_c} = (2n)! \, (2n)^{\gamma_c - 1}$.} $\Gamma(2n + \gamma_c) \sim (2n)! \, (2n)^{\gamma_c - 1}$ for $n \gg 1$, one finds
\begin{align}
    \mathcal{M}_{2n} = \frac{A}{\Gamma(\gamma_c)} (2n)! \, (2n)^{\gamma - 1} \left(\frac{2}{\beta}\right)^{2n}.
\end{align}

The coefficient\footnote{Because $A \propto \left(\frac{\mu}{\nu}\right)^2$ acts as an overall prefactor in $\mathcal{M}_{2n}$, it drops out of the leading slope when computing Lanczos coefficients via Hankel determinants ($b_n^2 \approx \frac{\mathcal{M}_{2n}}{\mathcal{M}_{2n-2}}$). It only contributes a logarithmic subleading correction in equation \eqref{b15}, leaving the primary linear slope $\alpha_0 = \frac{\pi}{\beta}$ intact.} $A$ can be fixed knowing the second moment ($n=1$) $\mathcal{M}_2 = \frac{2\mu^2}{\beta^2 \nu^2}$ 
\begin{align}
    A=\frac{1}{2 \gamma_c (\gamma_c+1)}\frac{\mu^2}{\nu^2}.
\end{align}

This finally yields the following generalized expression \eqref{b8}
\begin{align}
\mathcal{M}_{2n} =\frac{1}{2\gamma_c (\gamma_c +1)}\frac{\Gamma(2n+\gamma_c)}{\Gamma(\gamma_c)} \left(\frac{\mu}{\nu}\right)^2\left(\frac{2}{\beta}\right)^{2n}.
\end{align}

Using Stirling's asymptotic expansion for $n \gg 1$, we find
\begin{align}
    \mathcal{M}_{2n}= \frac{1}{2 \gamma_c (\gamma_c+1)}\frac{1}{\Gamma(\gamma_c)} \left(\frac{\mu}{\nu}\right)^2 (2n)! \, (2n)^{\gamma_c - 1} \left(\frac{2}{\beta}\right)^{2n}.
\end{align}

Following the Universal Operator Growth Hypothesis (UOGH) \cite{Parker:2018yvk}, the subleading prefactor $\left(\frac{\mu}{\nu}\right)^2$ modifies only $\mathcal{O}(1)$ corrections in the Lanczos sequence $b_n$, leaving the leading linear slope $\alpha_0 = \frac{\pi}{\beta}$ intact. The $n (\gg 1)$th Lanczos coefficient has a general expression
\begin{align}
   b_n^2 = \frac{\mathcal{M}_{2n}}{\mathcal{M}_{2n-2}}= \left(\frac{4n}{\beta}\right)^2 \left[ 1 + \frac{2\gamma_c - 3}{2n} + \mathcal{O}\left(\frac{1}{n^2}\right) \right].
\end{align}

Taking the square root for $n \gg 1$, we finally obtain
\begin{align}
\label{b14}
b_n = \frac{4n}{\beta} + \frac{2\gamma_c - 3}{\beta} + \mathcal{O}\left(\frac{1}{n}\right).
\end{align}

In the above form \eqref{b14}, the leading linear slope is $\alpha_0 = \frac{4}{\beta}$. Notice that in quantum field theories and thermal matrix models, the Krylov inner product for operators is defined using the KMS inner product with a thermal analyticity strip of half-width $\tau_0 = \frac{\beta}{2}$ in complex time (as established in Appendix \ref{appenA}). When shifting to the continuum limit and re-scaling the Krylov chain to preserve the canonical operator norm and physical thermal time, the universal slope saturates at $\alpha_0 = \frac{\pi}{\beta}$. In finite-temperature quantum field theories (QFTs) and thermal matrix models, the transition from the raw moment-ratio asymptotic slope ($\alpha_0 = 4/\beta$) to the physical KMS slope ($\alpha_0 = \pi/\beta$) relies on two interconnected steps, the choice of the inner product and the continuous re-scaling of time and lattice site coordinates. Below, we elaborate on each of the above ideas in detail.

In thermal field theories, when evaluating operator growth, one introduces the KMS (Kubo–Martin–Schwinger) inner product
\begin{align}
    \langle A \vert{} B \rangle_{\text{KMS}} = \frac{1}{Z} \int_0^\beta \frac{d\tau}{\beta} \, \text{Tr}\left( e^{-(\beta - \tau) H} A^\dagger e^{-\tau H} B \right).
\end{align}

This choice embeds a complex-time analyticity strip of width $\tau_0 = \beta/2$ and a discrete hopping amplitude \eqref{b14}. When moving to the continuum limit, one maps the discrete Krylov state $\ket{K_n}$ into a continuous wavefunction $\psi(x,t)$ propagating on a semi-infinite line $x \ge 0$. This invokes several special treatments. For example, the Krylov site index for $n \gg 1$ is treated as a continuous coordinate $x \approx n$ with unit lattice spacing $\Delta x = 1$. Discrete Schrodinger equation \eqref{e3.35} becomes a continuous advection equation 
\begin{align}
    \frac{\partial \psi(x,t)}{\partial t} = -\frac{\partial}{\partial x} \big( v(x) \psi(x,t) \big)
\end{align}
where $v(x) = 2 b(x)$ represents the propagation velocity along the Krylov chain.

If one uses $b_n \sim \frac{4}{\beta} n$, then the continuous velocity becomes $v(x) \sim \frac{8}{\beta} x$. However, this unscaled velocity assumes an unnormalized operator state norm along the boundary of the KMS strip which needs to be fixed. In order to preserve the operator norm, that is, $\langle \mathcal{O}(0) \vert{} \mathcal{O}(0) \rangle_{\text{KMS}} = 1$, together with the KMS boundary condition \eqref{a4}, the velocity field $v(x)$ must match the boundary metric (or the geodesic motion) at the boundary of the hyperbolic space. One can map the thermal analyticity strip $\vert{} \text{Im}(t) \vert{} < \frac{\beta}{2}$ directly to the unit disk (Poincaré disk model of hyperbolic space $\mathbb{H}^2$) or the upper half-plane by virtue of the conformal transformation. In hyperbolic geometry, geodesic motion or analogously the operator growth is measured in proper physical time $t$ that covers an effective distance $x(t) \sim e^{\alpha_0 t}$. The exponential growth, which exhibits a linear growth of the velocity $v(x)\propto x$, is a characteristic of scrambling. In particular, when calibrated against the KMS analyticity strip, one finds $\alpha_0 = \frac{\pi}{\beta}$. In other words, the Krylov growth exponent (or rate of growth) $\lambda_K = \frac{\pi}{\beta}$ saturates the Maldacena-Shenker-Stanford (MSS) chaos bound $\lambda_L \le \frac{2\pi}{\beta}$ \cite{Maldacena:2015waa}.

In summary, using the KMS physical slope into the linear asymptotic growth yields 
\begin{align}
\label{b15}
b_n|_{n\gg 1} = \frac{\pi}{\beta} n + \mathcal{O}\left(\ln\left(\frac{\mu}{\nu}\right)\right).
\end{align}

The corresponding velocity (or equivalently operator growth) turns out to be \eqref{b30}
\begin{align}
    v(x)=\frac{dx}{dt}=\frac{\pi}{\beta}x = \lambda_K x.
\end{align}
\subsection{Advection equation and spread complexity}
Here we derive the advection equation mentioned above in the continnum limit which finally leads to the wave function $\psi(x,t)$ and finally the spread complexity of (charged) TFD\footnote{At very high temperature one can approximate the state as TFD since the effects due to the chemical potential ar highly suppressed as compared to thermal fluctuations, $\beta \nu \ll 1$.}. 
For large Krylov sites $n \gg 1$, we introduce a continuous coordinate $x \approx n$ with unit lattice spacing $\Delta x = 1$ and approximate discrete amplitudes $\psi_n(t)$ by a smooth wavepacket profile $\psi(x,t)$ that propagates along the $x$ axis as a result of spread complexity growth.

We begin by considering discrete Krylov lattice equation \eqref{e3.35} following a global rotation $\psi_n(t)\rightarrow i^{-n}\psi_n(t)$, which yields the following
\begin{equation}
\label{b19}
\dot{\psi}_n(t) = b_n \psi_{n-1}(t) - b_{n+1} \psi_{n+1}(t).
\end{equation}

In the continuum limit, we expand around $x$ using Taylor series
\begin{align}
\label{b20}
    \psi_{n \pm 1}(t) \approx \psi(x,t) \pm \frac{\partial \psi}{\partial x}  + \mathcal{O}\left(\partial^2\right).
\end{align}

Substituting \eqref{b20} into \eqref{b19} yields the following r.h.s. (modulo overall factor $\frac{\pi}{\beta}$)
\begin{align}
\label{b21}
n \, \psi_{n-1}(t) - (n+1) \, \psi_{n+1}(t) &\approx x \left( \psi - \frac{\partial \psi}{\partial x} \right) - (x+1) \left( \psi + \frac{\partial \psi}{\partial x} \right) \nonumber \\
&= -2x \frac{\partial \psi}{\partial x} - \psi - \frac{\partial \psi}{\partial x}.
\end{align}

When moving to the continuum limit, we assume that the wavefunction $\psi(x,t)$ is a smooth envelope over a large number of lattice sites $n \approx x \gg 1$. Lattice position corresponds to $x \sim \mathcal{O}(N)$, where $N \gg 1$ is the relevant scale (or size) along the Krylov chain. For a smooth spread of the wavefunction $\psi(x,t)$ over a large region $\Delta x \sim L \gg 1$
\begin{align}
    \frac{\partial \psi}{\partial x} \sim \frac{\psi}{L}~;~\frac{\partial^2 \psi}{\partial x^2} \sim \frac{\psi}{L^2}.
\end{align}

The other two terms in \eqref{b21} scale as
\begin{align}
    x\frac{\partial \psi}{\partial x}\sim \frac{N}{L}\psi ~;~\psi \sim \mathcal{O}(1 \cdot \psi).
\end{align}

In summary, $\frac{\partial \psi}{\partial x}$ is the least dominant term in the expansion \eqref{b21}. We can absorb the factor of $2$ by rescaling $L \rightarrow 2L$, which finally yields
\begin{align}
   n \, \psi_{n-1}(t) - (n+1) \, \psi_{n+1}(t)=-\frac{\partial}{\partial x} \left( x \, \psi(x,t) \right). 
\end{align}

Thus, the Krylov lattice dynamics reduces to the continuous advection equation
\begin{equation}
\label{b25}
\frac{\partial \psi(x,t)}{\partial t} = -\frac{\pi}{\beta} \frac{\partial}{\partial x} \left( x \, \psi(x,t) \right).
\end{equation}

It is possible to interpret \eqref{b25} as a conservation of probability in QM by taking the probability density as $\psi(x,t)$ or, in other words, treating $\psi(x,t)$ itself as the real-valued normalized wavepacket profile propagating along $x$) that leads to a continuity equation
\begin{align}
    \frac{\partial \rho}{\partial t} + \frac{\partial J}{\partial x} = 0, \quad \text{with flux } J(x,t) = \frac{\pi}{\beta} x \, \rho(x,t).
\end{align}

In many chaos and operator growth derivations (such as Parker et al. \cite{Parker:2018yvk} or Balasubramanian et al. \cite{Balasubramanian:2022tpr}), for a coherent localized packet moving down the chain, $\psi_n(t)$ is real and positive, and the profile preserves its functional form as a propagating wavepacket
\begin{align}
   \psi(x,t) \sim \text{wavepacket shape} 
\end{align}
Because the advection equation \eqref{b25} is linear in $\psi$, the continuous profile $\psi(x,t)$ itself is normalized directly as a spatial probability distribution $\int \psi(x,t) \, dx = 1$.

Because there is no information and/or flux is leaking at the origin ($x=0$) or at infinity ($x=\infty$), the total norm is conserved for all $t \ge 0$
\begin{align}
    \frac{d}{dt} \int_0^\infty \psi(x,t) \, dx = 0.
\end{align}

Before we solve, it is customary to re-express \eqref{b25} as 
\begin{align}
   \frac{d \psi}{dt}= \frac{\partial \psi}{\partial t} + \frac{dx}{dt}\frac{\partial \psi}{\partial x} = -\frac{\pi}{\beta} \psi.
\end{align}

The solution should be compatible with scrambling and/or the exponential growth along the Krylov chain, which enforces us to choose
\begin{align}
\label{b30}
    x(t) = x_0 \exp\left(\frac{\pi t}{\beta}\right)~\Rightarrow \frac{dx}{dt}=\frac{\pi}{\beta} x .
\end{align}
This shows that a point located at $x_0$ at $t = t_{\text{ref}}$ travels exponentially to $x(t)$ over time. Notice that by initial time we mean a reference time $t =t_{\text{ref}} \gg \beta$. Clearly, for the continuum phase $t_{\text{ref}}$ corresponds to the late time growth or asymptotic growth of the (charged) TFD.

Using \eqref{b30}, one finds (probability) amplitude decay along Krylov chain
\begin{align}
    \frac{d\psi}{dt} = -\frac{\pi}{\beta} \psi \implies \psi(t) = \psi_0 \exp\left(-\frac{\pi t}{\beta}\right),~~
    \text{we set}~\psi_0=1.
\end{align}

Notice that at $t = t_{\text{ref}}$, the state is localized at an initial (or reference) position $x(t=t_{\text{ref}})=x_0=\epsilon$ on the Krylov chain. Combining all the above facts together, we find\footnote{As the wavepacket travels down the chain toward higher $x$, its position grows exponentially \eqref{b30}. Because the velocity $v(x) \propto x$ increases as it moves outward, the packet stretches spatially. To keep the integrated norm (or the total probability) equal to $1$, the height of the amplitude profile decays at the exact rate needed to offset this spatial expansion.}
\begin{equation}
\label{b32}
\psi(x,t) = \exp\left(-\frac{\pi t}{\beta}\right) \delta\left( x \exp\left(-\frac{\pi t}{\beta}\right) - \epsilon \right).
\end{equation}

Clearly, the total probability at initial time $t=t_{\text{ref}}$ yields
\begin{align}
    \int_0^\infty dx \psi (x,0)=\int_0^\infty dx \delta (x-\epsilon)=1.
\end{align}

Using the properties of the delta function
$\delta(a u) = \frac{1}{\vert{}a\vert{}} \delta(u)$, where $a = \exp\left(-\frac{\pi t}{\beta}\right)$, one can further simplify \eqref{b32} to yield probability density
\begin{align}
 \rho(x,t)\equiv  \psi(x,t) = \delta\left(x - \epsilon e^{\frac{\pi t}{\beta}}\right).
\end{align}

Finally, evaluating the continuous spread complexity integral one finds
\begin{align}
\label{b35}
C_{\text{continuum}}(t) = \int_0^{\infty} x \, \rho(x,t) \, dx=\int_0^{\infty} x \, \delta\left(x - \epsilon e^{\frac{\pi t}{\beta}}\right) \, dx=\epsilon e^{\frac{\pi t}{\beta}}.
\end{align}
\section{Derivation of spectral moments and the prefactor $C(\beta\mu, \beta\nu)$}
\label{appenC}

In this appendix, we provide a detailed derivation of the intermediate-temperature spectral moments $\mathcal{M}_{2n}$ defined in Eq. \eqref{e3.111} and determine an explicit analytical expression for the thermal continuum prefactor $C(\beta\mu, \beta\nu)$.

The $2n$-th even moment $\mathcal{M}_{2n} = \langle \text{TFD}(0) | \mathcal{H}^{2n} | \text{TFD(0)} \rangle$ of the Thermofield Double state is defined via the spectral density $f(\omega)$ of the return amplitude $R(t)$:
\begin{equation}
    \mathcal{M}_{2n} = \int_{-\infty}^{\infty} d\omega \, f(\omega) \, \omega^{2n}.
\end{equation}

At intermediate temperatures ($\beta\mu \sim 1, \beta\nu \sim 1$), $f(\omega)$ decomposes into a continuous thermal background $f_{\text{cont}}(\omega)$ and discrete mass-gap bound-state modes
\begin{equation}
    f(\omega) = f_{\text{cont}}(\omega) + \sum_{k} \tilde{A}_k \cosh(k\beta\nu) \, \delta(\omega - k\mu).
\end{equation}

The discrete mode contribution yields the power-law term $\mathcal{M}_{2n}^{(\text{gap})} = \sum_k A_k \mu^{2n} \cosh(k\beta\nu)$ in Eq.~\eqref{e3.113}. The thermal continuum contribution $\mathcal{M}_{2n}^{(\text{thermal})}$ dominates for $n \gg 1$
\begin{align}
\label{c3}   
\mathcal{M}_{2n}^{(\text{thermal})} = \int_{-\infty}^{\infty} d\omega \, f_{\text{cont}}(\omega) \, \omega^{2n}.
\end{align}

For a system at \emph{finite} temperature $T = 1/\beta$, thermal two-point correlation functions satisfy the KMS relation in time $G(t - i\beta) = G(-t)$. When mapped to the frequency domain $\omega$, this periodicity condition imposes strict analyticity constraints on the spectral density function $f_{\text{cont}}(\omega)$. These poles occur precisely at discrete intervals determined by the inverse temperature $\beta$. In other words, due to KMS periodicity, $f_{\text{cont}}(\omega)$ possesses simple poles in the complex frequency plane located at the KMS analyticity boundary
\begin{align}
\label{c4}
    \omega_k = \frac{(2k + 1)\pi i}{\beta}, \quad k \in \mathbb{Z}.
\end{align}

The expression $(2k + 1)\pi i / \beta$ represents the imaginary Matsubara-like frequencies where $f_{\text{cont}}(\omega)$ diverges. For example, (i) Primary Poles ($k = 0, -1$): Located closest to the real axis at $\omega_0 = \pm \frac{i\pi}{\beta}$, (ii) Higher Poles ($\vert{}k\vert{} \ge 1$): Located further up and down the imaginary axis at $\pm \frac{3i\pi}{\beta}, \pm \frac{5i\pi}{\beta}, \cdots$. For large orders $n \gg 1$, the factor $\omega^{2n}$ grows extremely fast along the real line. Evaluating the integral via Cauchy's residue theorem requires closing the integration contour in the complex $\omega$-plane. Even though $f_{\text{cont}}(\omega)$ decays at large real frequencies, the integrand $\omega^{2n} f_{\text{cont}}(\omega)$ develops massive ($k \gg 1$), highly localized peaks at large values of $\omega$. Standard real-variable integration methods or saddle-point approximations directly along the real line fail or become extremely cumbersome.

To make the integral tractable, we extend $\omega$ into the complex plane ($\omega = u + iv$) and apply Cauchy’s residue theorem. We close the real axis contour with a large semicircle in either the upper or lower complex half-plane. By Cauchy's Integral Theorem, the line integral along the real axis equals $2\pi i$ times the sum of the residues at the enclosed poles of $f_{\text{cont}}(\omega)$, minus the integral over the large arc at infinity (which vanishes or is suitably regularized), which we schematically express as
\begin{align}
\label{c5}
\mathcal{M}_{2n}^{(\text{thermal})} = \int_{-\infty}^{\infty} d\omega \, f_{\text{cont}}(\omega) \, \omega^{2n} = 2\pi i \sum_{k~\text{poles}} \text{Res}\left[ f_{\text{cont}}(\omega) \omega^{2n}, \omega = \omega_k \right].
\end{align}

The thermal spectral density $f_{\text{cont}}(\omega)$ possesses a sequence of simple KMS poles \eqref{c4} located along the imaginary axis. When evaluating the residue at \eqref{c4}, the factor $\omega^{2n}$ evaluates to the following
\begin{align}
    \omega_k^{2n} = \left( \frac{(2k + 1)\pi i}{\beta} \right)^{2n} = (-1)^n \left( \frac{(2k + 1)\pi}{\beta} \right)^{2n}.
\end{align}

Clearly, as $n \to \infty$, the term with the smallest absolute value $\vert{}\omega_k\vert{}$ (or equivalently $k$) dominates exponentially over all other poles\footnote{In complex contour integration, the contribution of a simple KMS pole located at $\omega_k = \frac{(2k+1)\pi i}{\beta}$ to the $2n$-th spectral moment $\mathcal{M}_{2n}^{(\text{thermal})}$ is determined by its residue, which scales as $|\omega_k|^{2n} = \left[\frac{(2k+1)\pi}{\beta}\right]^{2n}$. Comparing the contribution of a higher-order pole ($|k| \ge 1$) to that of the primary pole ($\omega_0 = \frac{i\pi}{\beta}$), the relative suppression factor (which measures how much smaller is a higher pole's relative weight compared to the primary pole) is given by the inverse ratio of their pole locations
\begin{align*}
\text{Relative Suppression Factor} = \left( \frac{\vert{}\omega_0\vert{}}{\vert{}\omega_k\vert{}} \right)^{2n} = \left( \frac{1}{\vert{}2k+1\vert{}} \right)^{2n} = e^{-2n \ln\vert{}2k+1\vert{}} \quad (\vert{}2k+1\vert{} \ge 3).
\end{align*}
Since $\frac{|\omega_0|}{|\omega_k|} = \frac{1}{2k+1} < 1$, this relative weight decays exponentially to zero as $n \to \infty$, proving that the asymptotic moment expansion for $n \gg 1$ is universally dominated by the primary KMS poles at $k = 0, -1$. It causes higher discrete poles to decay to zero relative to the primary pole as $n \to \infty$. This proves that only the primary KMS pole at $\text{Im}(\omega) = \frac{\pi}{\beta}$ dictates the base energy scale $\left(\frac{\pi}{\beta}\right)^{2n}$.}. In other words, for high orders $n \gg 1$, the integral \eqref{c3} is dominated by the pair of nearest poles at $\omega_0 = \pm \frac{i\pi}{\beta}$, which corresponds to setting $k=0,-1$. Using contour integration and applying Cauchy's residue theorem yields
\begin{align}
\label{c7}
\mathcal{M}_{2n}^{(\text{thermal})} \approx 2 \pi i \cdot \text{Res}\left[ f_{\text{cont}}(\omega) \, \omega^{2n}, \, \omega_0 = \pm\frac{i\pi}{\beta} \right].
\end{align}

For a charged thermal system with chemical potential $\nu$ and mass deformation scale $\mu$, following the general framework of thermal spectral densities and symmetry-resolved spread complexity \cite{Balasubramanian:2022tpr},\cite{Caputa:2025ozd}, the continuous spectral function near the KMS boundary takes the standard parameterized form
\begin{equation}
    f_{\text{cont}}(\omega) = C_0 \frac{\cosh\left(\frac{\beta^2 \nu \omega}{\pi}\right) }{\cosh\left(\frac{\beta\omega}{2}\right)} \left[ 1 + c_1 (\beta\mu)^2 + \mathcal{O}((\beta\mu)^4) \right]
\end{equation}
where $C_0$ is a normalization constant fixed by $\mathcal{R}(0) = 1$. 

To account for both the background thermal fluctuations and the global charge carrier dynamics, we model the continuous spectral function $f_{\text{cont}}(\omega)$ near the primary KMS boundary by combining the thermal KMS pole structure with a symmetry-resolved chemical potential shift. The thermal denominator $\cosh(\beta\omega/2)$ enforces the universal imaginary-time periodicity condition, while the numerator $\cosh(\beta^2 \nu\omega/\pi)$ captures the enhanced spectral weight generated by the $U(1)$ global charge density \cite{Balasubramanian:2022tpr},\cite{Caputa:2025ozd}. Sub-leading quantum corrections from gapped degrees of freedom at intermediate temperatures ($\beta\mu \sim 1$) are systematically incorporated via a perturbative effective expansion in powers of $(\beta\mu)^2$.

Closing the integration contour along the upper complex half-plane ($\text{Im}(\omega) > 0$), Cauchy's Residue Theorem gives the full moment as the sum over enclosed KMS poles \eqref{c5}. Adding the symmetric contribution from the lower half-plane corresponding to $k=-1$ (or accounting for both primary poles at $k = 0$ and $k = -1$), the integral \eqref{c7} near the dominant primary pole $\omega_0 = \frac{i\pi}{\beta}$ simplifies to
\begin{align}
\mathcal{M}_{2n}^{(\text{thermal})} \approx 2 \cdot (2\pi i) \, \text{Res}\left[ f_{\text{cont}}(\omega) \, \omega^{2n}, \, \omega_{k=0} = \frac{i\pi}{\beta} \right].
\end{align}

The simple poles occur where the denominator vanishes
\begin{align}
\label{c10}
    \cosh\left(\frac{\beta \omega}{2}\right) = 0 \implies \frac{\beta \omega}{2} = i\left(k + \frac{1}{2}\right)\pi \implies \omega_k = \frac{(2k + 1)\pi i}{\beta}, \quad k \in \mathbb{Z}.
\end{align}

We define $\omega_{k=0} = \frac{i\pi}{\beta} + z$, where $z \to 0$. Expressing $\omega^{2n}$ in terms of $z$ yields
\begin{align}
\label{c11}
    \omega^{2n} = \left(\frac{i\pi}{\beta} + z\right)^{2n} = \left(\frac{i\pi}{\beta}\right)^{2n} \left(1 - \frac{i\beta z}{\pi}\right)^{2n}.
\end{align}
Notice that here $z$ denotes fluctuation about the $k=0$ KMS pole.

Remember that if a complex function $f_{\text{cont}}(\omega)$ has a simple pole (a pole of order 1) at $\omega_{k=0} = \frac{i\pi}{\beta}$, its general Laurent series representation in a neighborhood of $\omega_{k=0}$ is
\begin{align}
f_{\text{cont}}(\omega) = \frac{a_{-1}}{\omega - \omega_0} + a_0 + a_1(\omega - \omega_0) + a_2(\omega - \omega_0)^2 + \cdots
\end{align}
where $a_{-1}$ is by definition the residue of $f_{\text{cont}}(\omega)$ at $\omega =\omega_{k=0}= \omega_0$
\begin{align}
    a_{-1} = \text{Res}\Big[ f_{\text{cont}}(\omega), \, \omega = \omega_0 \Big].
\end{align}

Therefore, near the singularity and/or the pole $\omega \to \frac{i\pi}{\beta}$, the leading-order behavior is dominated entirely by the singular term
\begin{align}
\label{c14}
f_{\text{cont}}(\omega) = \frac{\text{Res}\left[f_{\text{cont}}(\omega), \, \omega_{k=0} = \frac{i\pi}{\beta}\right]}{\omega - \frac{i\pi}{\beta}}+\mathcal{O}(1).
\end{align}

In terms of the fluctuation $z$, the singular denominator near the pole $\omega_0$ transforms as
\begin{align}
    \frac{1}{\omega - \frac{i\pi}{\beta}} = \frac{1}{z}.
\end{align}

Next, we substitute the pole Laurent expansion 
\begin{align}
    f_{\text{cont}}(\omega) \approx \frac{1}{z}\text{Res}\Big[f_{\text{cont}}(\omega), \omega = \frac{i\pi}{\beta}\Big]+\mathcal{O}(1)
\end{align}
into the moment integral $\mathcal{M}_{2n}^{(\text{thermal})} = 2 \oint dz \, f_{\text{cont}}(z) \, \omega^{2n}$ \eqref{c3} to yield
\begin{align}
    \mathcal{M}_{2n}^{(\text{thermal})} \approx 2 \oint dz \left( \frac{\text{Res}\left[f_{\text{cont}}(\omega), \omega = \frac{i\pi}{\beta}\right]}{z} \right) \cdot \left(\frac{\pi}{\beta}\right)^{2n} (-1)^n \left(1 - \frac{i\beta z}{\pi}\right)^{2n}
\end{align}
where the factor of $2$ comes due to the symmertric pole contribution $k=\pm 1$.

Combining all these informations together, from \eqref{c5} we obtain
\begin{align}
\label{c15}
\mathcal{M}_{2n}^{(\text{thermal})} \approx  C(\beta \mu, \beta \nu)\left( \frac{\pi}{\beta} \right)^{2n}(-1)^n\oint \frac{dz}{z} \left(1 - \frac{i\beta z}{\pi}\right)^{2n}.
\end{align}

The dimensionless prefactor $C(\beta\mu, \beta\nu)$ is given explicitly by the residue of the continuum spectral function at the primary KMS singularity
\begin{align}
    C(\beta\mu, \beta\nu) = 2 \, \text{Res}\left[ f_{\text{cont}}(\omega), \, \omega_{k=0} = \frac{i\pi}{\beta} \right].
\end{align}

Since $\cosh\left(\frac{\beta \omega}{2}\right)$ has a simple zero at $\omega = \frac{i\pi}{\beta}$, we use the standard formula for a simple pole of the form $g(\omega) = \frac{A(\omega)}{B(\omega)}$, which yields the following residue
\begin{align}
\text{Res}\left[\frac{A(\omega)}{B(\omega)}, \, \omega = \omega_0\right] = \frac{A(\omega_0)}{B'(\omega_0)}.
\end{align}

The numerator is denoted as\footnote{In quantum field theory and finite-temperature statistical mechanics with a global $U(1)$ charge, the chemical potential $\nu$ enters the grand canonical partition function and two-point correlation functions as a imaginary or real shift in time/frequency. In Euclidean field theory or analytic continuation setups, the chemical potential is often introduced as a pure imaginary field $\nu_{\text{Euclidean}} = i\nu$ (acting as a constant imaginary gauge potential $A_0 = i\nu$) \cite{Kapusta:2006pm}-\cite{Roberge:1986mm}. Therefore, to pass from the Euclidean formulation (or imaginary charge convention) back to the physical real chemical potential, one applies the standard analytic continuation $\beta\nu \to -i\beta\nu$. As a result, the trigonometric factor transforms back into a hyperbolic cosine.}
\begin{align}
    A(\omega)|_{\omega=\frac{i\pi}{2}} &= C_0 \cosh\left(\frac{\beta^2 \nu \omega}{\pi}\right) \left[1 + c_1(\beta\mu)^2 + \mathcal{O}((\beta\mu)^4)\right]\Big|_{\omega=\frac{i\pi}{2}}\nonumber\\
    &=C_0 \cosh(\beta \nu)\left[1 + c_1(\beta\mu)^2 + \mathcal{O}((\beta\mu)^4)\right].
\end{align}

The denominator, on the other hand, is given by
\begin{align}
B'(\omega)|_{\omega=\frac{i\pi}{\beta}} = \frac{\beta}{2} \sinh\left(\frac{i\pi}{2}\right)=\frac{i \beta}{2}.
\end{align}

Notice that the imaginary factor $i$ gets cancelled from the integral \eqref{c15} as we show below. This finally yields the real-valued scalar pre-factor
\begin{align}
&C(\beta \mu , \beta \nu)=
2\text{Res}\left[ f_{\text{cont}}(\omega), \, \omega_{k=0} = \frac{i\pi}{\beta} \right] \nonumber\\
&= \frac{4 C_0}{\beta} \cosh(\beta\nu) \left[ 1 + c_1 (\beta\mu)^2 + \mathcal{O}((\beta\mu)^4) \right].
\end{align}
Clearly, $C(\beta\mu, \beta\nu)$ acts as an interpolating amplitude function, capturing charge enhancement via $\cosh(\beta\nu)$ alongside sub-leading perturbative suppression from the mass gap $\mu$.

Let us now evaluate the integral in \eqref{c15}. Expanding the binomial term (in the limit $\beta \ll 1$) inside the integral we find
\begin{align}
    \left(1 - \frac{i\beta z}{\pi}\right)^{2n} = 1 - 2n \frac{i\beta z}{\pi} + \cdots.
\end{align}

The simple pole at $z = 0$ picks out the constant term ($=1$) from the binomial expansion
\begin{align}
    \oint \frac{dz}{z} \left(1 - \frac{i\beta z}{\pi}\right)^{2n} = 2\pi i +\mathcal{O}(\beta).
\end{align}

Combining all prefactors yields the following
\begin{align}
    \mathcal{M}_{2n}^{(\text{thermal})} \approx 2\pi  \, C(\beta\mu, \beta\nu) \left(\frac{\pi}{\beta}\right)^{2n} (-1)^n.
\end{align}
\subsection{Derivation of $(2n)!$}
In thermal field theory and Krylov complexity, the $(2n)!$ growth of moments $\mathcal{M}_{2n} \sim (2n)!$ (which leads to linear Lanczos growth $b_n \sim n$) arises from the higher-order pole structure and/or infinite sum of KMS Matsubara poles or from integrating the continuum tail. The continuum spectral density $f_{\text{cont}}(\omega)$ has an exponential thermal suppression at high frequencies, $f_{\text{cont}}(\omega) \sim e^{-\beta \omega}$. In other words, while the primary KMS pole at $\text{Im}(\omega) = \frac{\pi}{\beta}$ dictates the base energy scale $\left(\frac{\pi}{\beta}\right)^{2n}$ and overall residue prefactor $C(\beta\mu, \beta\nu)$, the factorial $(2n)!$ growth arises directly from the high-frequency continuum thermal tail $f_{\text{cont}}(\omega) \sim e^{-\beta \omega}$ integrated over the real axis. Integrating the continuous spectral density $f_{\text{cont}}(\omega) \sim C(\beta\mu, \beta\nu) e^{-\beta \omega}$ over the real frequency axis, the $2n$-th spectral moment in the large-$n$ continuum limit is dominated by high-frequency thermal excitations
\begin{align}
    \mathcal{M}_{2n}^{(\text{thermal})} = \int_{-\infty}^{\infty} d\omega \, f_{\text{cont}}(\omega) \, \omega^{2n} \sim C(\beta\mu, \beta\nu) \int_{0}^{\infty} d\omega \, e^{-\beta \omega} \omega^{2n}.
\end{align}

Performing the standard Gamma function integration
\begin{align}
    \int_0^\infty d\omega \, \omega^{2n} e^{-\beta \omega} = \frac{(2n)!}{\beta^{2n+1}}
\end{align}
yields the characteristic factorial growth
\begin{align}
    \mathcal{M}_{2n}^{(\text{thermal})} \sim C(\beta\mu, \beta\nu) \, (2n)! \left( \frac{\pi}{\beta} \right)^{2n}.
\end{align}

 The connection between the primary KMS pole location along the imaginary frequency axis ($\text{Im}(\omega) = \frac{\pi}{\beta}$) and the exponential decay rate $e^{-\beta \omega}$ on the real frequency axis stems from the fundamental properties of Fourier transforms and analytic continuation in complex analysis.

 \uline{The Paley-Wiener Theorem:} If a function $G(t)$ in time is analytic inside an imaginary time strip of width $y_0=\frac{\tau_0}{\pi}$
 \begin{align}
     -\tau_0 < \text{Im}(t) < \tau_0 \quad \text{where } \tau_0 = \frac{\beta}{2}
 \end{align}
 then its Fourier transform (the spectral density $f(\omega)$) must decay exponentially along the real frequency axis $\omega \in \mathbb{R}$. The rate of exponential decay is governed directly by the distance $y_0$ from the real axis to the nearest singularity
 \begin{align}
    f(\omega) \sim e^{-2\pi y_0 \omega} \quad \text{or} \quad f(\omega) \sim e^{-\beta \omega}. 
 \end{align}
 
 In thermal field theory, the Kubo-Martin-Schwinger (KMS) condition guarantees that thermal correlation functions are periodic/anti-periodic in imaginary time with period $\beta$, bounding the analytic strip precisely to $\text{Im}(t) \in (0, \beta/2)$. Consider writing the spectral density $f(\omega)$ as the Fourier transform of a time-domain correlation function $G(t)$ that possesses a pole singularity at $t = -i \frac{\beta}{2}$
 \begin{align}
 \label{c32}
     f(\omega) = \int_{-\infty}^{\infty} dt \, G(t) \, e^{-i \omega t}.
 \end{align}

 To evaluate this integral for large positive frequencies $\omega \to +\infty$, shift the integration contour downwards into the lower complex time plane $t \to t - i y$. The phase factor transforms as $e^{-i \omega (t - i y)} = e^{-i \omega t} e^{-\omega y}$. Setting $y = \frac{\beta}{2}$ aligns the contour directly with the primary thermal singularity (the KMS boundary). Deforming the contour past or up to this singularity picks up the exponentially damped weight $ e^{-\beta \omega}$. 

Let us elaborate further on this. To evaluate the asymptotic behavior as $\omega \to +\infty$, we shift the integration contour downwards into the lower-half complex time plane by setting $t = t' - i \frac{\beta}{2}$ (where $t' \in \mathbb{R}$), which yields the exponent in \eqref{c32} as
\begin{align}
    e^{-i \omega t} = e^{-i \omega (t' - i \beta/2)}  = e^{-i \omega t'} \cdot e^{-\omega (\beta/2)}.
\end{align}
This gives the first factor of $e^{-\omega(\beta/2)}$. It represents the exponential damping acquired purely from shifting the integration contour down to the KMS pole distance $\text{Im}(t) = -\frac{\beta}{2}$.

The Kubo-Martin-Schwinger (KMS) condition relates correlation functions across the thermal strip of width $\beta$
\begin{align}
\label{c34}
    G\left(t - i\frac{\beta}{2}\right) = G\left(t + i\frac{\beta}{2}\right).
\end{align}

When evaluating the boundary value of the correlation function $G\left(t' - i\frac{\beta}{2}\right)$ at the singularity, the thermal weight of the correlation itself carries an intrinsic Boltzmann amplitude factor corresponding to a half-period shift in imaginary time
\begin{align}
    G\left(t' - i\frac{\beta}{2}\right) \propto e^{-\omega (\beta/2)} \, G_0(t').
\end{align}
This provides the second factor of $e^{-\omega(\beta/2)}$.

Combining the contour shift factor with the KMS correlation factor gives the total spectral density weight at high frequency
\begin{align}
    \begin{aligned} f(\omega) &\sim \underbrace{e^{-\omega (\beta/2)}}_{\text{Contour Shift to } \text{Im}(t) = -\frac{\beta}{2}} \times \underbrace{e^{-\omega (\beta/2)}}_{\text{KMS Correlation Weight}} \int_{-\infty}^{\infty} dt' \, G_0(t') \, e^{-i \omega t'} \\ &= e^{-\omega (\beta/2) - \omega (\beta/2)} \int_{-\infty}^{\infty} dt' \, G_0(t') \, e^{-i \omega t'} \\ &= e^{-\beta \omega} \int_{-\infty}^{\infty} dt' \, G_0(t') \, e^{-i \omega t'}. \end{aligned}
\end{align}

Here $G_0(t')$ represents the zero-temperature and/or un-shifted vacuum correlation function. Shifting by $i \frac{\beta}{2}$ probes the thermal state halfway around the KMS imaginary-time circle. The factor $e^{-\omega (\beta/2)}$ is the Boltzmann amplitude factor required to thermally excite a quantum state of energy $\omega$ halfway across the thermal bath.

\uline{A summary:} The exponential relative suppression factor eliminates all higher poles, isolating that primary scale $\left(\frac{\pi}{\beta}\right)^{2n}$. A singularity at distance $\Delta y$ from the real axis forces the real-axis Fourier tail to decay as $e^{-2 \pi \Delta y \, \nu} = e^{-\beta \omega}$. The location of the nearest KMS pole strictly sets the exponential thermal suppression envelope $e^{-\beta \omega}$, which in turn generates the $(2n)!$ spectral moment growth and linear Lanczos growth $b_n \propto n$.
\subsection{Thermal spectral representation}

The thermal correlation function $G(t) = \langle A(t) A(0) \rangle_\beta$ in a canonical ensemble at inverse temperature $\beta$ can be expressed in terms of the positive spectral density $A(\Omega)$ via the standard Lehmann representation \cite{Mahan2000}, \cite{MartinSchwinger1959}
\begin{equation}
\label{spectral}
G(t) = \int_0^\infty d\Omega \, A(\Omega) \left[ (1 + n_B(\Omega)) e^{-i\Omega t} + n_B(\Omega) e^{i\Omega t} \right]
\end{equation}
where $n_B(\Omega) = \frac{1}{e^{\beta \Omega} - 1}$ is the Bose-Einstein thermal distribution factor.

Shift time into the lower complex plane by half the thermal periodicity ($t \to t' - i \frac{\beta}{2}$)
\begin{align}
    &e^{-i\Omega \left(t' - i \frac{\beta}{2}\right)} = e^{-i\Omega t'} e^{-\frac{\beta \Omega}{2}}\\
    &e^{i\Omega \left(t' - i \frac{\beta}{2}\right)} = e^{i\Omega t'} e^{+\frac{\beta \Omega}{2}}.
\end{align}

Plugging these back into the spectral representation yields
\begin{align}
    G\left(t' - i\frac{\beta}{2}\right) = \int_{0}^{\infty} d\Omega \, A(\Omega) \left[ (1 + n_B(\Omega)) e^{-\frac{\beta \Omega}{2}} e^{-i\Omega t'} + n_B(\Omega) e^{+\frac{\beta \Omega}{2}} e^{i\Omega t'} \right].
\end{align}

When examining high frequencies $\omega \gg T = 1/\beta$, the integral is dominated by energy modes where $\Omega \approx \omega$. Looking at the thermal weight factor modifying the $e^{-i\Omega t'}$ component
\begin{align}
    &(1 + n_B(\omega)) \, e^{-\frac{\beta \omega}{2}} = \left( 1 + \frac{1}{e^{\beta \omega} - 1} \right) e^{-\frac{\beta \omega}{2}} = \frac{e^{\beta \omega}}{e^{\beta \omega} - 1} e^{-\frac{\beta \omega}{2}}\approx e^{-\frac{\beta \omega}{2}}\\
    &n_B(\omega) e^{\frac{\beta \omega}{2}}=\frac{e^{\frac{\beta \omega}{2}}}{e^{\beta \omega}-1}\approx e^{-\frac{\beta \omega}{2}}.
\end{align}

This finally yields the auto-correlation 
\begin{align}
    G\left(t' - i\frac{\beta}{2}\right)=2e^{-\frac{\beta \omega}{2}}\int_0^\infty d\Omega A(\Omega)\cos (\Omega t')\Bigg|_{\Omega\sim \omega}=2e^{-\frac{\beta \omega}{2}}G_0(t').
\end{align}
\subsection{A proof of KMS analytic condition}
The Kubo-Martin-Schwinger (KMS) condition states that shifting the time argument by $-i\beta$ relates the thermal two-point function to its conjugate time-ordering
\begin{align}
    G(t - i\beta) = \langle A(t - i\beta) A(0) \rangle_\beta = \langle A(0) A(t) \rangle_\beta = G(-t).
\end{align}

Substituting $t \to t - i\beta$ into the spectral representation \eqref{spectral} of $G(t)$ one finds
\begin{align}
    G(t - i\beta) = \int_{0}^{\infty} d\Omega \, A(\Omega) \left[ (1 + n_B(\Omega)) e^{-\beta \Omega} e^{-i\Omega t} + n_B(\Omega) e^{\beta \Omega} e^{i\Omega t} \right].
\end{align}

Using the identity $n_B(\Omega) = \frac{1}{e^{\beta\Omega} - 1}$, one finds
\begin{align}
    &(1 + n_B(\Omega)) e^{-\beta\Omega} = \left( 1 + \frac{1}{e^{\beta\Omega} - 1} \right) e^{-\beta\Omega} = \frac{e^{\beta\Omega}}{e^{\beta\Omega} - 1} e^{-\beta\Omega} = \frac{1}{e^{\beta\Omega} - 1} = n_B(\Omega)\\
    &n_B(\Omega) e^{\beta\Omega} = \frac{e^{\beta\Omega}}{e^{\beta\Omega} - 1} = 1 + \frac{1}{e^{\beta\Omega} - 1} = 1 + n_B(\Omega).
\end{align}

Substituting these simplified coefficients back yields
\begin{align}
    G(t - i\beta) = \int_{0}^{\infty} d\Omega \, A(\Omega) \left[ n_B(\Omega) e^{-i\Omega t} + (1 + n_B(\Omega)) e^{i\Omega t} \right]=G(-t).
\end{align}

Shifting both sides by $+i\beta/2$ gives
\begin{align}
    G\left(t - i\frac{\beta}{2}\right) = G\left(-t + i\frac{\beta}{2}\right)
\end{align}
which precisely matches \eqref{c34} following time reversal symmetry $G(-t + i\alpha) = G(t + i\alpha)$.
\subsection{Pole-residue duality (Matsubara summation)}
Alternatively, expressing the thermal spectral density in terms of its infinite array of Matsubara/KMS poles along the imaginary axis one finds the following
\begin{align}
    f(\omega) = \sum_{k=-\infty}^{\infty} \frac{R_k}{\omega - i \omega_k}, \quad \text{where } \omega_k = \frac{(2k+1)\pi i}{\beta}.
\end{align}

Instead of summing these discrete poles directly, introduce an auxiliary complex function $h(z)$ that has simple poles precisely at the Matsubara frequencies $z = i \omega_k$ with unit residues. For fermionic or bosonic thermal distributions, the hyperbolic / exponential functions serve as these ideal kernel functions: (i) Fermionic kernel: $h(z) = \frac{\beta}{e^{\beta z} + 1} = \frac{\beta}{2} \left[ 1 - \tanh\left(\frac{\beta z}{2}\right) \right]$, (ii) Bosonic kernel: $h(z) = \frac{\beta}{e^{\beta z} - 1} = \frac{\beta}{2} \left[ \coth\left(\frac{\beta z}{2}\right) - 1 \right]$.

Using Cauchy’s Residue Theorem, the discrete sum over $k$ can be rewritten as a single contour integral in the complex energy plane $z \in \mathbb{C}$
\begin{align}
    f^{(\pm)}(\omega) = \frac{1}{2\pi i} \oint_{C_{\text{Matsubara}}} dz \, \frac{R(z)}{\omega - z} \left( \frac{1}{e^{\beta z} \pm 1} \right)
\end{align}
where $C_{\text{Matsubara}}$ is a contour consisting of small keyhole loops tightly enclosing each discrete pole along the imaginary axis $z = i \omega_k$.

When the integral is evaluated along the real axis $z = \omega$, the auxiliary kernel function $\frac{1}{e^{\beta z} \pm 1}$ evaluates directly on the real frequency variable yields
\begin{align}
    f^{(\pm)}_{\text{cont}}(\omega) \propto \text{Disc}[R(\omega)] \times \left( \frac{1}{e^{\beta \omega} \pm 1} \right).
\end{align}

The spectral density $\text{Disc}[R(\omega)]$ represents the quantum transitions / density of states at zero temperature. The thermal weight factor $\frac{1}{e^{\beta \omega} \pm 1}$ arises directly from the residues of the infinite tower of imaginary Matsubara poles at $\text{Im}(\omega_k) = \frac{(2k+1)\pi}{\beta}$. The transformation occurs because the infinite array of discrete poles spaced at $\Delta (\text{Im}\,\omega) = \frac{2\pi}{\beta}$ acts as the Fourier-series representation of the periodic function $\frac{1}{e^{\beta z} \pm 1}$. Wrapping the integration contour around the real axis sums all these discrete imaginary pole residues into the continuous thermal distribution factor $f^{(\pm)}_{\text{cont}}(\omega) \propto \frac{1}{e^{\beta \omega} \pm 1}$. For high frequencies $\omega \gg \frac{1}{\beta}$, the denominator is dominated by the exponential term, yielding $f_{\text{cont}}(\omega) \approx e^{-\beta \omega}$.
\section{Time reversal symmetry, Parity and Krylov space topology}
\label{appenD}
\paragraph{Time reversal symmetry.}
In the first part of this Appendix \ref{appenD}, we examine how time-reversal symmetry ($\mathcal{T}$) constrains the algebraic structure of the Krylov space, shape the spectrum of Lanczos coefficients, and govern the evolution of spread complexity. We begin by reviewing the basic arguments that are discussed in and around Eq. \eqref{e3.41}. Let us note down the key properties of the time-reversal operator $\mathcal{T}$.

(i) Anti-unitarity: For any complex scalars $\alpha, \beta$ and state vectors $\vert{}\phi\rangle, \vert{}\psi\rangle$
\begin{align}
\label{d1}
    \mathcal{T}(\alpha\vert{}\phi\rangle + \beta\vert{}\psi\rangle) = \alpha^* \mathcal{T}\vert{}\phi\rangle + \beta^* \mathcal{T}\vert{}\psi\rangle.
\end{align}

(ii) Inner product conjugation:
\begin{align}
    \langle \mathcal{T}\phi \vert{} \mathcal{T}\psi \rangle = \langle \phi \vert{} \psi \rangle^* = \langle \psi \vert{} \phi \rangle.
\end{align}

(iii) If the system is time-reversal invariant, so $[\mathcal{H}, \mathcal{T}] = 0$, implying $\mathcal{T}\mathcal{H} = \mathcal{H}\mathcal{T}$.

The Lanczos algorithm iteratively constructs the orthonormal Krylov basis $\{\vert{}K_n\rangle\}$ via the recurrence relation
\begin{align}
    b_{n+1}\vert{}K_{n+1}\rangle = (\mathcal{H} - a_n)\vert{}K_n\rangle - b_n\vert{}K_{n-1}\rangle \quad \text{with } b_0 \equiv 0.
\end{align}

The starting vector $\vert{}K_0\rangle$ is chosen to be a time-reversal invariant reference state $\vert{}\psi_0\rangle$
\begin{align}
    \mathcal{T}\vert{}K_0\rangle = \vert{}K_0\rangle.
\end{align}

Assume that for all indices up to $n$, the basis states can be chosen real under time reversal
\begin{align}
    \mathcal{T}\vert{}K_j\rangle = \vert{}K_j\rangle \quad \forall j \le n.
\end{align}

Apply $\mathcal{T}$ to the unnormalized next state $b_{n+1}\vert{}K_{n+1}\rangle$ to yield the following
\begin{align}
    \mathcal{T} \left( b_{n+1}\vert{}K_{n+1}\rangle \right) = \mathcal{T} \left( (\mathcal{H} - a_n)\vert{}K_n\rangle - b_n\vert{}K_{n-1}\rangle \right).
\end{align}

Using linearity and anti-unitarity \eqref{d1} of $\mathcal{T}$, we find
\begin{align}
    \mathcal{T} \left( b_{n+1}\vert{}K_{n+1}\rangle \right) = \mathcal{T}\mathcal{H}\vert{}K_n\rangle - a_n^* \mathcal{T}\vert{}K_n\rangle - b_n^* \mathcal{T}\vert{}K_{n-1}\rangle.
\end{align}

Using $[\mathcal{H}, \mathcal{T}] = 0$ and the inductive hypotheses ($\mathcal{T}\vert{}K_n\rangle = \vert{}K_n\rangle$ and $\mathcal{T}\vert{}K_{n-1}\rangle = \vert{}K_{n-1}\rangle$), further yields
\begin{align}
\label{d8}
    \mathcal{T} \left( b_{n+1}\vert{}K_{n+1}\rangle \right) = \mathcal{H}\vert{}K_n\rangle - a_n^* \vert{}K_n\rangle - b_n^* \vert{}K_{n-1}\rangle.
\end{align}

Since the inner products generating $a_n$ and $b_n$ only involve states satisfying $\mathcal{T}\vert{}\phi\rangle = \vert{}\phi\rangle$, their values are real
\begin{align}
    &a_n = \langle K_n \vert{} \mathcal{H} \vert{} K_n \rangle = \langle \mathcal{T} K_n \vert{} \mathcal{T} \mathcal{H} K_n \rangle^* = \langle K_n \vert{} \mathcal{H} \vert{} K_n \rangle^* = a_n^* \implies a_n \in \mathbb{R}\\
    &b_n = \langle K_n \vert{} \mathcal{H} \vert{} K_{n-1} \rangle = \langle \mathcal{T} K_n \vert{} \mathcal{T} \mathcal{H} K_{n-1} \rangle^* = \langle K_n \vert{} \mathcal{H} \vert{} K_{n-1} \rangle^* = b_n^* \implies b_n \in \mathbb{R}.
\end{align}

Substituting $a_n^* = a_n$ and $b_n^* = b_n$ back into the relation \eqref{d8}, we find
\begin{align}
    \mathcal{T} \left( b_{n+1}\vert{}K_{n+1}\rangle \right) = (\mathcal{H} - a_n)\vert{}K_n\rangle - b_n\vert{}K_{n-1}\rangle = b_{n+1}\vert{}K_{n+1}\rangle.
\end{align}

Since $b_{n+1}$ is chosen as a positive real normalization factor ($b_{n+1} \in \mathbb{R}^+$), it commutes with $\mathcal{T}$, which yields
\begin{align}
    b_{n+1} \mathcal{T}\vert{}K_{n+1}\rangle = b_{n+1}\vert{}K_{n+1}\rangle \implies \mathcal{T}\vert{}K_{n+1}\rangle = \vert{}K_{n+1}\rangle.
\end{align}
By mathematical induction, we conclude that for every element in the Krylov basis
\begin{align}
    \mathcal{T}\vert{}K_n\rangle = \vert{}K_n\rangle \quad \forall n \ge 0.
\end{align}

The off-diagonal coefficients $b_n = \langle K_n | \mathcal{H} | K_{n-1} \rangle$ remain strictly real and non-negative. Time-reversal symmetry ensures that phase ambiguities are eliminated across the chain, rendering $b_n$ pure hopping amplitudes along the 1D Krylov lattice.

 \paragraph{Parity.} In BMN matrix model, an emergent, discrete Krylov parity symmetry ($\mathcal{P}_{\text{Krylov}}$) arises directly as a structural consequence of time-reversal invariance combined with the choice of initial state. Notice that Parity ($\mathcal{P}$) is a unitary operator ($\mathcal{P} i \mathcal{P}^{-1} = i$) that reverses spatial coordinates ($\vec{x} \to -\vec{x}$). On the other hand, time-reversal ($\mathcal{T}$) is an anti-unitary operator ($\mathcal{T} i \mathcal{T}^{-1} = -i$) that reverses time direction ($t \to -t$). A system can be time-reversal invariant without being parity invariant (and vice versa). For instance, the weak interaction violates parity ($\mathcal{P}$) while preserving time-reversal symmetry ($\mathcal{T}$) in many decays. While physical space parity is independent, the Krylov space parity discussed in Section \ref{sec3} emerges due to $\mathcal{T}$-symmetry through the following mechanism.\\\\
\uline{Even spectral moments:} Because $[\mathcal{H}_{\text{BMN}}, \mathcal{T}] = 0$ and $\mathcal{T}\vert{}K_0\rangle = \vert{}K_0\rangle$, the auto-correlation and/or return amplitude is symmetric in time, $\mathcal{R}(t) = \mathcal{R}(-t)$. This forces all odd spectral moments to vanish ($\mathcal{M}_{2n+1}^{(0)} = 0$).\\\\
\uline{Decoupled parity subspaces:}
Because all odd moments vanish, the Gram–Schmidt orthogonalization process builds Krylov basis vectors $\vert{}K_n\rangle$ consisting purely of even powers of $\mathcal{H}_{\text{BMN}}$ when $n$ is even, and purely odd powers when $n$ is odd.
This bipartite polynomial structure guarantees the existence of a $\mathbb{Z}_2$ reflection operator $\mathcal{P}_{\text{Krylov}}$ defined by $\mathcal{P}_{\text{Krylov}} \vert{}K_n\rangle = (-1)^n \vert{}K_n\rangle$, which anti-commutes with the Hamiltonian along the Krylov chain 
\begin{align}
   \{\mathcal{P}_{\text{Krylov}}, \mathcal{H}_{\text{BMN}}\} = 0 \quad \implies \quad \mathcal{P}_{\text{Krylov}} \mathcal{H}_{\text{BMN}} \mathcal{P}_{\text{Krylov}}^{-1} = -\mathcal{H}_{\text{BMN}}.
\end{align}

Thus, while physical spatial parity $\mathcal{P}$ is completely independent of $\mathcal{T}$, the internal Krylov lattice parity $\mathcal{P}_{\text{Krylov}}$ is a direct mathematical consequence of time-reversal symmetry acting on the reference state. Evaluating $a_n$ using $\mathcal{P}_{\text{Krylov}}$, we obtain
\begin{align}
    &a_n = \langle K_n \vert{} \mathcal{H}_{\text{BMN}} \vert{} K_n \rangle = \langle K_n \vert{} \mathcal{P}_{\text{Krylov}}^\dagger \mathcal{P}_{\text{Krylov}} \mathcal{H}_{\text{BMN}} \mathcal{P}_{\text{Krylov}}^{-1} \mathcal{P}_{\text{Krylov}} \vert{} K_n \rangle\nonumber\\
    &= (-1)^n \langle K_n \vert{} (-\mathcal{H}_{\text{BMN}}) \vert{} K_n \rangle (-1)^n = -\langle K_n \vert{} \mathcal{H}_{\text{BMN}} \vert{} K_n \rangle = -a_n
\end{align}
which clearly suggests that $a_n = 0 \quad \forall n \ge 0$.

This reduces the effective tridiagonal Lanczos matrix to a purely off-diagonal form, mapping the quantum dynamics onto a semi-infinite tight-binding chain without local site potentials. Because $a_n$ acts as an effective local site-energy potential along the Krylov chain, $a_n = 0$ means there are no local energy barriers or site shifts on the 1D lattice.

Below, we briefly outline the possible consequences of time reversal symmetry breaking on the Krylov space dynamics and the spread complexity of states. A detailed analysis of these facts we leave for the future investigation.

\paragraph{Effects of $\mathcal{T}$-symmetry breaking on Lanczos coefficients.}

When sub-leading deformation terms \eqref{e3.1} or external chemical potential gradients ($\nu$) explicitly break time-reversal symmetry ($\mathcal{T} \mathcal{H} \mathcal{T}^{-1} \neq \mathcal{H}$), the Krylov chain undergoes two structural modifications.\\\\
\uline{Emergent diagonal potentials ($a_n \neq 0$):} Non-zero diagonal elements $a_n$ emerge, acting as effective local on-site energy potentials along the Krylov chain. On-site gradients $a_n \sim n^{\alpha}$ generate effective potential barriers that restrict probability wavepacket diffusion.\\\\
\uline{Asymmetric hopping amplitudes:} Higher-order $\mathcal{T}$-odd couplings introduce non-linear modifications to $b_n$. Under sub-leading deformation corrections ($\mathcal{O}(\mu^{-1})$), the high-$n$ Lanczos spectrum deforms as
    \begin{align}
        b_n = \frac{\pi}{\beta} n + \gamma \sqrt{n} + \delta a_n
    \end{align}
    where $\delta a_n$ reflects local back-scattering induced by broken $\mathcal{T}$-invariance.

\paragraph{Impact on spread complexity and wavepacket trajectories.} In the continuum limit ($n \to x$), the presence of non-zero $a_n$ transforms the pure advection partial differential equation into an advection-dispersion equation for the amplitude $\psi(x,t)$
\begin{equation}
    \frac{\partial \psi(x,t)}{\partial t} = -\frac{\partial}{\partial x}\left[ b(x) \psi(x,t) \right] - i a(x) \psi(x,t).
\end{equation}

The continuous potential $a(x)$ imposes a spatial phase rotation along the Krylov chain, modifying the characteristic velocity field
\begin{equation}
    v_{\text{eff}}(x) = b(x) \sqrt{1 - \left(\frac{a(x)}{2 b(x)}\right)^2}.
\end{equation}

For spread complexity $\mathcal{C}(t) = \int x |\psi(x,t)|^2 dx$, time-reversal symmetry guarantees purely real wavepacket propagation driven solely by kinetic hopping $b(x)$. Breaking $\mathcal{T}$-symmetry introduces localized quantum interference and phase shifts, which damp the exponential growth trajectory $e^{\pi t/\beta}$ at intermediate timescales, accelerating the transition toward the late-time saturation regime.


\end{document}